# Onsager Coefficients for Liquid Metal Flow in a Conduit under a Magnetic Field

Sindu E. Shanmugadas and Haim H. Bau*

Dept. Mechanical Engineering and Applied Mechanics, University of Pennsylvania, Philadelphia, PA 1904-6315. USA

***Corresponding author:** *bau@seas.upenn.edu*

**Abstract:**

We analyze the flow of room and near room-temperature liquid metals in shallow, long rectangular conduits with two insulating and two perfectly conducting walls under a uniform magnetic field perpendicular to the flow direction and the insulating surfaces, focusing on moderate Hartmann numbers. A pressure gradient and Lorentz body forces may drive or oppose the flow. We derive explicit expressions for the Onsager coefficients that relate the flow rate and electric current on the one hand to the potential difference across electrodes and the pressure gradient on the other hand. We further demonstrate that these coefficients satisfy Onsager-Casimir reciprocity. These simplified expressions provide a convenient framework for analyzing, optimizing, and controlling magnetohydrodynamic (MHD) machines operating with liquid metals in applications such as power conversion, energy harvesting, pumping, actuation, valving, breaking, and sensing without moving components.

**Nomenclature**

(Variables without specified units are dimensionless)

**Latin Symbols**

| | |
|---|---|
| $a$ | Conduit's half height ($m$). |
| $b$ | Conduit's half width ($m$). |
| $c$ | Numerical coefficient. |
| $B_0$ | Uniform magnetic field in the $\boldsymbol{y}$-direction ($T$). |
| $h$ | Normalized induced current stream function. |
| $G$ | Pressure gradient $\left(= -\frac{dp}{dz}\right)$ ($Pa\ m^{-1}$). |
| $H_z$ | Current stream function ($A\ m^{-1}$). |
| $\boldsymbol{J}$ | Electric current density ($A\ m^{-2}$). |
| $K_{EE}$ | Onsager coefficient - electrical conductance ($\Omega^{-1} m^{-1}$). |
| $K_{EH}$ | Cross Onsager coefficient, the electrical current generated by a $1\ Pa{\cdot}m^{-1}$ pressure gradient ($m^2 A\ N^{-1}$). |
| $K_{HE}$ | Cross Onsager coefficient, the flow rate generated by a $1V$ potential difference ($m^2 A\ N^{-1}$). |
| $K_{HH}$ | Onsager coefficient, hydrodynamic conductance ($m^6 N^{-1} s^{-1}$). |
| $L$ | Half conduit length ($m$). |
| $p$ | Pressure ($Pa$) |
| $P$ | Body force scale $(\frac{B_0 \Delta H}{2a} + G)$ ($N\ m^{-3}$) |
| $Q$ | Flow rate ($m^3 s^{-1}$) |
| $r$ | Resistance ratio |
| $R_{ex}$ | External resistance ($\Omega\ m$) |
| $R_{in}$ | Internal resistance ($\Omega\ m$) |
| $u$ | Normalized axial velocity $(= \frac{\mu}{a^2 P} v_z)$ |
| $v_z$ | Axial velocity ($m\ s^{-1}$) |

$\bar{v}_z$ Axial average velocity $=(4ab)^{-1}Q$ ($m\ s^{-1}$)

$x, y, z$ Cartesian coordinates ($m$)

**Subscripts**

$c$ Core (outer) region

$i$ Boundary layer (inner) region

$m$ maximum

**Greek Symbols**

$\alpha, \beta$ Numerical coefficients.

$\varepsilon$ Conversion efficiency

$\delta$ Length scale for the lateral boundary layer

$\phi$ Potential (V)

$\tilde{\varphi}$ Normalized potential (equation 7)

$\mu$ Dynamic viscosity ($Pa\ s$)

$\nu$ Kinematic viscosity ($m^2 s^{-1}$)

$\sigma$ Electric conductivity ($S\ m^{-1}$)

$\xi$ Normalized $x$ coordinate ($=a^{-1}x$)

$\eta$ Normalized $y$ coordinate ($=b^{-1}y$)

$\zeta$ Stretched boundary layer coordinate in the $\boldsymbol{x}$-direction

$\Psi$ Potential difference across electrodes ($V$)

**Dimensionless Groups**

$A$ Aspect ratio $(a/b)$

$H_a$ Hartmann Number $(=aB_0 \frac{\sigma}{\mu})$

$Re$ Reynolds Number $\frac{\bar{v}_z a}{\nu}$

$S$ Ratio of pressure gradient and Lorentz body force $\frac{B_0 \Delta H}{2aG}$

**Introduction:**

In recent years, there has been a growing interest in room-temperature and near-room-temperature liquid metals, particularly gallium-based eutectic alloys, across various fields, including flexible, stretchable, reconfigurable, and self-healing electronics; soft robotics; energy harvesting and conversion; on-demand thermal control; wearable sensors; and microfluidics [1]. Gallium-based alloys are attractive due to their high electrical and thermal conductivities, low toxicity, and biocompatibility.

Earlier, microfluidics researchers leveraged magnetic fields and electric currents to induce Lorentz body forces in aqueous solutions, driving liquid flow [2, 3, 4], stirring liquids [5, 6, 7], and controlling fluidic networks [8] without any moving members or mechanical components. However, aqueous solutions have relatively low electric conductivity and are susceptible to electrochemical reactions at electrode surfaces, significantly limiting the strength of the Lorentz body forces that can be produced. In contrast, liquid metals, such as gallium-based alloys, offer a significant advantage due to their electrical conductivity being orders of magnitude higher ($\sim 10^6\ S\ m^{-1}$) compared to aqueous solutions ($\sim 10\ S\ m^{-1}$), enabling the generation of substantial Lorentz body forces even under moderate magnetic fields, such as available from small rare-earth, permanent magnets.

The use of eutectic gallium alloys comes with several challenges. A primary concern is that gallium and its alloys readily form a surface oxide layer when exposed to oxygen. This oxide layer reduces surface energy, hindering the flow and wetting behavior of the liquid metal and adversely affecting contact electrical resistance. Strategies to mitigate the formation and effects of this oxide layer include operating in controlled environments, such as in inert (oxygen-free) atmospheres; using acidic conditions to dissolve the oxide layer [9]; using reducing electrochemistry to reduce gallium oxide; and applying coatings to the liquid metal surface to prevent gallium's exposure to oxygen [10, 11, 12]. Additionally, gallium is corrosive to certain metals, such as aluminum and copper. Gallium infiltrates the grain boundaries of these metals, leading to embrittlement and structural degradation over time. Protective coatings, such as barrier layers, can be applied to susceptible metals, preventing direct contact with gallium [10, 11, 12].

Despite these challenges, applications of room-temperature and near-room-temperature liquid metals MHD are expanding. Although liquid metals have 10-fold lower electrical conductivity than solid copper and cannot compete with solid copper in conventional electromagnetic machines, they are compatible with flexible, reconfigurable, compact devices. Applications include, among other things, harvesting energy from ocean waves [13], low-temperature heat sources [14], motion [15, 16], and body locomotion [17]; on-demand thermal control [18, 19]; actuation, sensing, and assisting in fabricating soft robotics and wearable devices.

Many minute systems that utilize liquid metal MHD comprise shallow, long conduits with rectangular cross-sections ($-b \leq x \leq b, -a \leq y \leq a$) with aspect ratio: $A = \frac{a}{b} < 1$; electrodes at $x = \pm b$; and uniform magnetic field ($B_0$) in the $y$-direction (**Fig. 1**). Permanent, rare-earth magnets often provide the magnetic field. Flows in these devices are usually laminar ($Re << 10^3$). Since the magnetic Reynolds number and the current flux are small, distortions to the externally applied magnetic field $B_0$ are insignificant [20]. The Hartmann number $H_a = aB_0\frac{\sigma}{\mu}$, gauging the relative importance of induced Lorentz and viscous forces, is *O(10).* In the above, $\sigma$ and $\mu$ are, respectively, the liquid metal electric conductivity and dynamic viscosity.

Except for the asymptotic limits of small [21] and large [22] Hartmann numbers, the MHD equations must be solved numerically. The analysis, design, optimization, and control of magnetohydrodynamic systems, such as generators and energy harvesters, pumps, actuators, breaks, and sensors, would benefit from simple relationships between control inputs, such as potential across electrodes and flow assisting/retarding pressure gradient on the one hand, and outputs, such as flow rate and current, on the other hand. Our paper aims to provide such relationships.

Our paper is organized as follows. Section 2 presents the mathematical model for fully developed MHD flow in a long conduit with a rectangular cross-section and solves the MHD equations by finite elements. Section 3 carries asymptotic analysis in the limit of aspect ratio: $A \to 0$. The result of this analysis enables us to obtain relationships between flow rate and current on the one hand and potential difference across the electrodes and pressure gradient on the other hand in the form of Onsager relations that satisfy Onsager-Casimir reciprocity. Section 4 takes advantage of these simple relations to analyze the performance of various MHD machines.

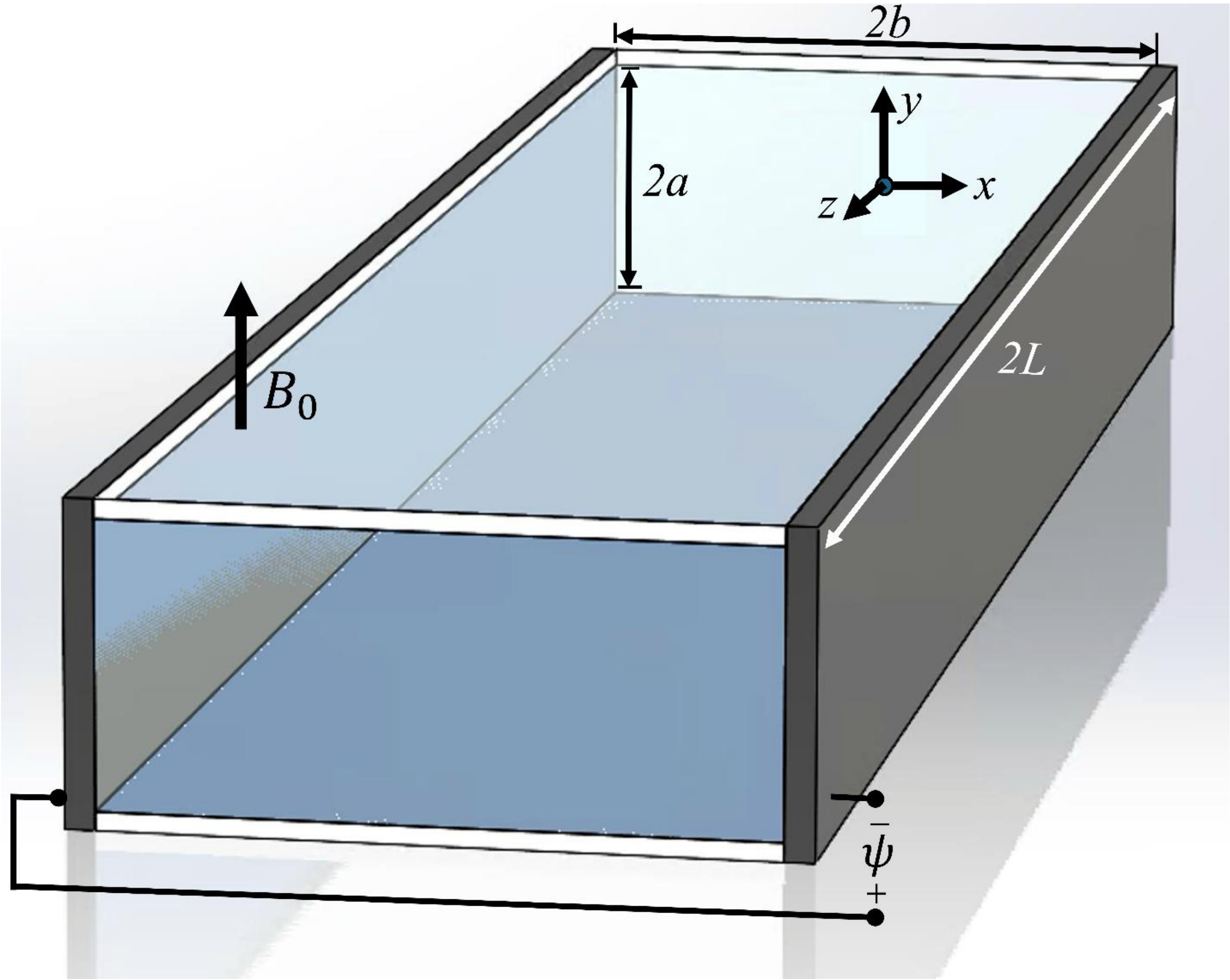

**Fig. 1:** Our long flow conduit $-L < z < L$ with the uniform, rectangular cross-section $-b \le x \le b$, $-a \le y \le a$. The uniform magnetic field $B_0$ is in the positive $\boldsymbol{y}$-direction. The surfaces $x = \pm b$ are perfect conductors (electrodes). The surfaces $y = \pm a$ are insulating.

## 2. Mathematical Model

Consider a long duct ($-L < z < L$) with the rectangular cross-section $-b \le x \le b$, $-a \le y \le a$; (**Fig. 1**). The coordinate $z$ aligns with the conduit's axis. The conduit's surfaces $x = \pm b$ are perfectly conducting electrodes, each at a uniform potential and in perfect contact with the liquid metal. The surfaces $y = \pm a$ are perfectly insulating (blocking electrical current). The duct is filled with a conductive liquid with uniform electric conductivity $\sigma$ and viscosity $\mu$. The conduit is subjected to a uniform magnetic field $B_0$ in the $\boldsymbol{y}$-direction, unaffected by induced magnetic fields. Since the magnetic Reynolds number and the current in the duct are small, so is the magnetic induction. Thus, the externally applied magnetic field remains unaffected. The steady flow in the conduit is assumed to be nearly isothermal due to the liquid metal's high thermal conductivity and the efficient heat exchange with the duct walls, facilitated by the duct's shallow geometry. We modeled the MHD flow using the time-independent Navier-Stokes equation with non-slip boundary conditions at all solid surfaces. $L \gg Max(a, b)$. At a distance *O(a Re)* from the inlet, the flow is fully-developed, such that $\frac{\partial v_z}{\partial z} = 0$. When the flow is fully-developed, the momentum equation in the $\boldsymbol{z}$-direction reduces to:

$$0 = -\frac{dp}{dz} + \mu\left(\frac{\partial^2 v_z}{\partial x^2} + \frac{\partial^2 v_z}{\partial y^2}\right) + J_x B_0, \tag{1}$$

where $p$ is the pressure, $J_x$ the $\boldsymbol{x}$-direction component of the current density, and $v_z(x, y)$ the axial velocity. The pressure gradient $G = -\frac{dp}{dz}$ is constant.

The electric current density accounts for both Ohmic and induced currents.

$$J_x = \sigma\left(-\frac{\partial \phi}{\partial x} - v_z B_0\right), \qquad J_y = \sigma\left(-\frac{\partial \phi}{\partial y}\right), \tag{2}$$

where $\phi(x, y)$ is the electric potential. Charge conservation requires the current density $\boldsymbol{J}$ to be divergence-free ( $\boldsymbol{\nabla} \cdot \boldsymbol{J} = 0$). We use bold letters to describe vectors and regular font for scalar quantities. Following Hunt-Stewartson (hereafter, **HS)** [22], we introduce the current stream function $H_z(x, y)$

$$J_x = \frac{\partial H_z}{\partial y}, \qquad J_y = -\frac{\partial H_z}{\partial x}, \tag{3}$$

which satisfies charge conservation. With the introduction of $H_z$, equations (1) and (2) reduce to two coupled, linear *pdes* with $v_z$ and $H_z$ as the dependent variables.

$$0 = -\frac{dp}{dz} + \mu\left(\frac{\partial^2 v_z}{\partial x^2} + \frac{\partial^2 v_z}{\partial y^2}\right) + B_0 \frac{\partial H_z}{\partial y}. \tag{4}$$

$$0 = \frac{1}{\sigma}\left(\frac{\partial^2 H_z}{\partial x^2} + \frac{\partial^2 H_z}{\partial y^2}\right) + B_0 \frac{\partial v_z}{\partial y}. \tag{5}$$

At the insulating surfaces, $H_z(x, a) = H_1$, $H_z(x, -a) = H_2$, and $\Delta H = H_1 - H_2$ is the net current flow between the opposing electrodes per conduit's unit length. $\Delta H$ is known a priori only when operating under current control.

We use $b$ as the length scale in the $\boldsymbol{x}$-direction ($\xi = \frac{x}{b}$) and $a$ as the length scale in the $\boldsymbol{y}$-direction ($\eta = \frac{y}{a}$). $P = \frac{B_0 \Delta H}{2a} + G$ ($N\,m^{-3}$) scales the body force. $\frac{B_0 \Delta H}{2a}$ is the cross-section average Lorentz body force and $G = -\frac{dp}{dz}$ is the pressure-induced body force. To assess their relative importance, we define the dimensionless group: $S = \frac{B_0 \Delta H}{2aG}$.

Next, following HS, we define the dimensionless velocity $u(\xi, \eta)$ and the alternation in the current stream function resulting from the flow $h(\xi, \eta)$. $v_z = \frac{a^2 P}{\mu} u(\xi, \eta)$ and $H_z = \frac{1}{2}(H_1 + H_2) + \frac{\Delta H}{2}\eta + a^2 P \sqrt{\frac{\sigma}{\mu}} h(\xi, \eta)$. Since only differences in $H_z$ matter, we set $H_1 = \frac{1}{2}\Delta H$ and $H_2 = -\frac{1}{2}\Delta H$, where $\Delta H = H_1 - H_2$. The resulting dimensionless equations are:

$$\begin{pmatrix} A^2 \frac{\partial^2}{\partial \xi^2} + \frac{\partial^2}{\partial \eta^2} & H_a \frac{\partial}{\partial \eta} \\ H_a \frac{\partial}{\partial \eta} & A^2 \frac{\partial^2}{\partial \xi^2} + \frac{\partial^2}{\partial \eta^2} \end{pmatrix} \begin{pmatrix} u \\ h \end{pmatrix} = \begin{pmatrix} -1 \\ 0 \end{pmatrix} \tag{6}$$

with the boundary conditions $u(\pm 1, \eta) = u(\xi, \pm 1) = \frac{\partial h(\pm 1, \eta)}{\partial \xi} = h(\xi, \pm 1) = 0$. In the above, $H_a = aB_0 \sqrt{\frac{\sigma}{\mu}}$ is the Hartman number and $A$ is the conduit cross-section's aspect ratio.

It is apparent from equation (6) and the boundary conditions that the velocity is an even function of both $\xi$ and $\eta$: $u(-\xi, \eta) = u(\xi, \eta)$ and $u(\xi, -\eta) = u(\xi, \eta)$, while $h$ is odd in $\eta$ and even in $\xi$: $h(\xi, -\eta) = -h(\xi, \eta)$, $h(-\xi, \eta) = h(\xi, \eta)$, and $h(\xi, 0) = 0$.

The velocity ($u$) satisfying the Dirichlet boundary condition, while ($h$) satisfies the Neumann boundary condition at $\xi = \pm 1$ complicates the analytical solution of equations (6). For small $A$ and moderate $H_a$, equations (6) can be readily solved numerically, *e.g.,* with finite elements (Supplement S1). When the Hartmann number is large, numerical solutions are challenged by thin Hartmann boundary layers of thickness $H_a^{-1}$ next to $\eta = \pm 1$ and Shercliff boundary layers of thickness $H_a^{-1/2}$ next to $\xi = \pm 1$. A fine mesh is required in the boundary layers for appropriate spatial resolution.

The electric potential distribution

$$\phi(x, y) = \frac{1}{2}\big(\phi(-1, \eta) + \phi(1, \eta)\big) + \frac{abP}{\sqrt{\mu\sigma}} \tilde{\varphi}(\xi, \eta). \tag{7}$$

To determine $\tilde{\varphi}$, we integrate the dimensionless *1st* order equation (2):

$$\frac{\partial \tilde{\varphi}}{\partial \xi} = -\frac{1}{H_a}\left(\frac{S}{1+S}\right) - \left(\frac{\partial h}{\partial \eta} + H_a u\right). \tag{8}$$

Without loss of generality, we set $\tilde{\varphi}(0,0) = 0$. $\tilde{\varphi}$ is odd in $\xi$: $\tilde{\varphi}(-\xi,\eta) = -\tilde{\varphi}(\xi,\eta)$ and even in $\eta$: $\tilde{\varphi}(\xi,-\eta) = \tilde{\varphi}(\xi,\eta)$. We use asymptotic methods to obtain explicit, approximate relations among the flow rate, pressure gradient, current, and potential difference across the electrodes.

## 3. Asymptotic Solution for Small Aspect Ratio *A*.

Previously, HS carried out an asymptotic analysis of MHD flow in conduits with rectangular cross-sections in the asymptotic limit of $H_a \to \infty$. We are, however, interested in systems wherein the magnetic fields are provided by permanent magnets (typically, $B_0 < 0.5$ *T*), and conduits' dimensions are on the *mm* scale. In other words, we focus on moderate values of the Hartmann number (*e.g., O(10)*). Furthermore, microfluidic systems are often shallow ($A \ll 1$). Hence, we examine $A \to 0$.

**Outer (core) solution**: We hypothesize that in the core (outer region), away from the electrodes ($\xi = \pm 1$), both $\frac{\partial^2 u}{\partial \xi^2}$ and $\frac{\partial^2 h}{\partial \xi^2}$ are *O(1)*. Thus, in the limit of $A \to 0$ (shallow conduit), we drop all terms with the pre-factor *A*. Equation (6) reduces to

$$\begin{pmatrix} \frac{\partial^2}{\partial \eta^2} & H_a \frac{\partial}{\partial \eta} \\ H_a \frac{\partial}{\partial \eta} & \frac{\partial^2}{\partial \eta^2} \end{pmatrix} \begin{pmatrix} u \\ h \end{pmatrix} = \begin{pmatrix} -1 \\ 0 \end{pmatrix}. \tag{9}$$

The solution of equation (9) is:

$$h_c(\eta) = H_a{}^{-1}\, \hat{h}_c(\eta) = -H_a{}^{-1} \left( \eta - \frac{Sinh(H_a \eta)}{Sinh(H_a)} \right) \tag{10}$$

and

$$u_c(\eta) = H_a{}^{-1}\, \hat{u}_c(\eta) = H_a{}^{-1} Coth(H_a) \left( 1 - \frac{Cosh(H_a \eta)}{Cosh(H_a)} \right). \tag{11}$$

In the above, we defined the scaled variables, $\hat{h}$ and $\hat{\eta}$, for later use.

When $H_a \gg 1$, $u_c$ includes the Hartman boundary layers of thickness $O(H_a^{-1})$ next to $\eta = \pm 1$. Away from $\eta = \pm 1$, the core flow is uniform $u_c \approx H_a^{-1}$. In the limit of small $H_a$, the velocity profile assumes a parabolic shape $\frac{1}{2}(1 - \eta^2)$, consistent with flow driven by uniform body force between two parallel plates. The vertical component of the current density in the core $\tilde{J}_{y,c} = 0$ for all Hartmann numbers. The dimensionless, horizontal component of the current density:

$$\tilde{J}_x = S + H_a \frac{\partial h}{\partial \eta}(S + 1) \tag{12}$$

We use $B_0 G^{-1}$ as the current density scale instead of the more obvious $a^{-1}\Delta H/2$ to accommodate open-circuit operation when $\Delta H = 0$. In the core,

$$\tilde{J}_{x,c} = (S+1)H_a \frac{Cosh(H_a \eta)}{Sinh(H_a)} - 1 \tag{13}$$

Integration of equation (8) yields the potential distribution in the core.

$$\tilde{\varphi}_c = \left(-\frac{Cosh(H_a)}{Sinh(H_a)} + \frac{1}{H_a}\left(\frac{G}{P}\right)\right)\xi = \left(-Coth(H_a) + \frac{1}{H_a(1+S)}\right)\xi \quad (-1+\delta < \xi < 1-\delta). \tag{14}$$

Witness that $\tilde{\varphi}_c$ is *independent* of $\eta$. $\delta$ represents the thickness of the lateral boundary layers.

As is common in singular perturbations, the core solution does not satisfy the boundary conditions at $\xi = \pm 1$. Next to $\xi = \pm 1$, our earlier hypothesis that $\frac{\partial^2 u}{\partial \xi^2}$ and $\frac{\partial^2 h}{\partial \xi^2}$ are *O(1)* cannot be true. We anticipate boundary layers next to $\xi = \pm 1$.

**Inner region:** To resolve the system's behavior next to $\xi = -1$, we introduce the stretched coordinate ($\zeta$) such that $\xi = -1 + \delta\,\zeta$, $(0 \le \zeta \le \infty, \delta \to 0)$. Dominant balance (equation 6) suggests, $\delta \sim A H_a^{-1/2}$. When $H_a \sim O(1)$, we have a Stokes layer of thickness *O(A)*, wherein the top and bottom surfaces screen the non-slip condition at $\xi = -1$. When $H_a \gg 1$, we have a Shercliff boundary layer of thickness $O\left(H_a^{-1/2}\right)$.

Rescaling the equations (6) in terms of the stretched coordinate $\zeta$ and introducing the scaled (hatted) dependent variables $h_i = H_a^{-1}\,\hat{h}_i$ and $u_i = H_a^{-1}\,\hat{u}_i$, we obtain the boundary layer equations $(0 \le \zeta \le \infty, -1 \le \eta \le 1)$:

$$\frac{\partial^2 \hat{u}_i}{\partial \zeta^2} + H_a{}^{-1}\frac{\partial^2 \hat{u}_i}{\partial \eta^2} + \frac{\partial \hat{h}_i}{\partial \eta} = -1 \tag{15}$$

$$\frac{\partial^2 \hat{h}_i}{\partial \zeta^2} + H_a{}^{-1}\frac{\partial^2 \hat{h}_i}{\partial \eta^2} + \frac{\partial \hat{u}_i}{\partial \eta} = 0 \tag{16}$$

with the boundary conditions:

$\hat{u}_i = \hat{u}_i(\zeta, \pm 1) = \hat{u}_i(\infty, \eta) - \hat{u}_c(\eta)$ and $\frac{\partial \hat{h}_i(0,\eta)}{\partial \zeta} = \hat{h}_i(\zeta, \pm 1) = \hat{h}_i(\infty, \eta) - \hat{h}_c(\eta) = 0$.

In the above, the subscripts *i* and *c* denote, respectively, inner (boundary layer) and (outer) core dependent variables. The conditions at $\zeta \to \infty$ are 1st order asymptotic matches. The solution of the boundary layer equations is independent of the aspect ratio $A$. For large $H_a$, to the 1st-order approximation, we could have dropped the $O(H_a^{-1})$ terms. However, we retain $O(H_a^{-1})$ terms because we wish to obtain approximations valid for moderate values of the Hartmann number. Serendipitously, $O(H_a^{-1})$ terms retention provides dissipation that stabilizes the numerical scheme (Supplement S1).

Once the inner solutions are available, we determine the deviation of the boundary layer potential from the linear core potential $\Delta\tilde{\varphi}_i = \tilde{\varphi}_i - \tilde{\varphi}_c$ by integrating the equation:

$$\frac{\partial \Delta\tilde{\varphi}_i}{\partial \zeta} = \frac{A}{\sqrt{H_a}}\left(\frac{\partial h_c}{\partial \eta} + H_a u_c - \frac{\partial h_i}{\partial \eta} - H_a u_i\right) \tag{17}$$

with $\Delta\tilde{\varphi}_i(\infty, \eta) = 0$, we have

$$\Delta\tilde{\varphi}_i(\zeta) = \frac{A}{H_a^{\frac{1}{2}}}\int_{\infty}^{\zeta}\left(\left(H_a^{-1}\frac{\partial \hat{h}_c}{\partial \eta} + \hat{u}_c\right) - \left(H_a^{-1}\frac{\partial \hat{h}_i}{\partial \eta} + \hat{u}_i\right)\right)dl, \tag{18}$$

Since the integrand approaches zero rapidly as $\zeta$ increases, it suffices to carry out the integration from a finite $\zeta_L > \zeta_{L,min}$ such that any increase in $\zeta_L$ beyond $\zeta_{L,min}$ does not significantly alter $\Delta\tilde{\varphi}_i(\zeta)$. Typically, $\zeta_L \sim 10$ suffices (Supplement S1). The difference at the left wall between the electrode potential extrapolated from the core expression and the actual electrode potential is of interest.

$$\Delta\tilde{\varphi}_{i,L} = \frac{A}{H_a^{\frac{1}{2}}}\widetilde{\Delta\varphi^*}_{i,L} = \frac{A}{H_a^{\frac{1}{2}}}\int_{\zeta_L}^{0}\left(\left(H_a^{-1}\frac{\partial \hat{h}_c}{\partial \eta} + \hat{u}_c\right) - \left(H_a^{-1}\frac{\partial \hat{h}_i}{\partial \eta} + \hat{u}_i\right)\right)dl\,. \tag{19}$$

$\Delta\tilde{\varphi}_{i,L}$ is independent of $\eta$. The result for the left boundary layer ($\xi = -1$) also applies to the right boundary layer ($\xi = 1$), with $\Delta\tilde{\varphi}_{i,R} = -\Delta\tilde{\varphi}_{i,L}$.

We numerically solved the boundary layer equations (15-16) (Supplement S1) for the velocity and current stream function.

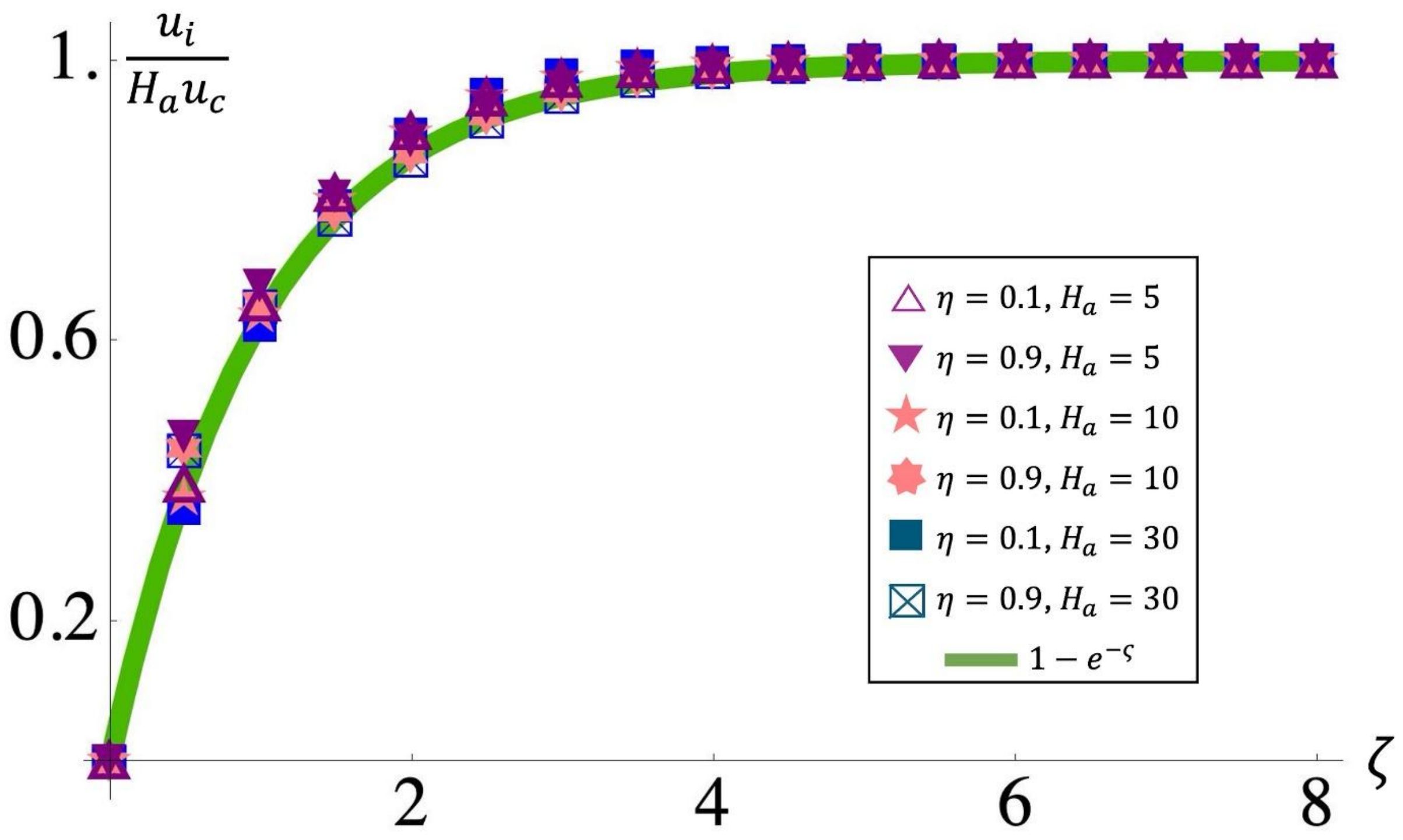

**Fig. 2:** The ratio of the computed boundary layer velocity and the core velocity $\frac{u_i(\zeta,\eta)}{H_a u_c(\eta)}$ as a function of the stretched boundary layer coordinate $\zeta$ when (**A**) $H_a$=5 (purple triangles) and $\eta$ =0.1 (hollow triangle), $\eta$ =0.9 (solid triangle), (**B**) $H_a$=10 (pink), and $\eta$ =0.1 (star), $\eta$ =0.9 (7 point star), and (C) $H_a$=30 (blue), and $\eta$ =0.1 (square), $\eta$ =0.9 (square with cross). $(1 - e^{-\zeta})$ is depicted with a solid (green) line.

The approximate velocity in the left lateral boundary layer

$$u(\xi, \eta) \sim u_c(\eta)\left(1 - e^{-\frac{\sqrt{H_a}}{A}(1+\xi)}\right) \quad (-1 \leq \xi \leq 1) \tag{20}$$

agrees well with the numerically computed velocities from the whole field equations (6) and boundary layer equations (15-16) (**Fig. 2).** The composite expression for the entire flow domain

$$u_{comp}(\xi, \eta) \sim u_c(\eta)\left(1 - 2e^{-A^{-1}H_a^{1/2}} Cosh\left(A^{-1}H_a^{1/2}\xi\right)\right) \ (-1 \leq \xi \leq 1) \tag{21}$$

satisfies the boundary conditions at $(\xi = \pm 1)$ only approximately (with an exponentially small error).

Using equation (20), we obtain the approximation for the flow rate $(Q_a)$ that accounts for the velocity deficits in the lateral boundary layers.

$$Q_a = \frac{4a^3b}{\mu} P H_a^{-2} \left(1 - \frac{A}{\sqrt{H_a}}\left(1 - e^{-\frac{\sqrt{H_a}}{A}}\right)\right)(H_a Coth(H_a) - 1) \ (m^3\ s^{-1}). \tag{22}$$

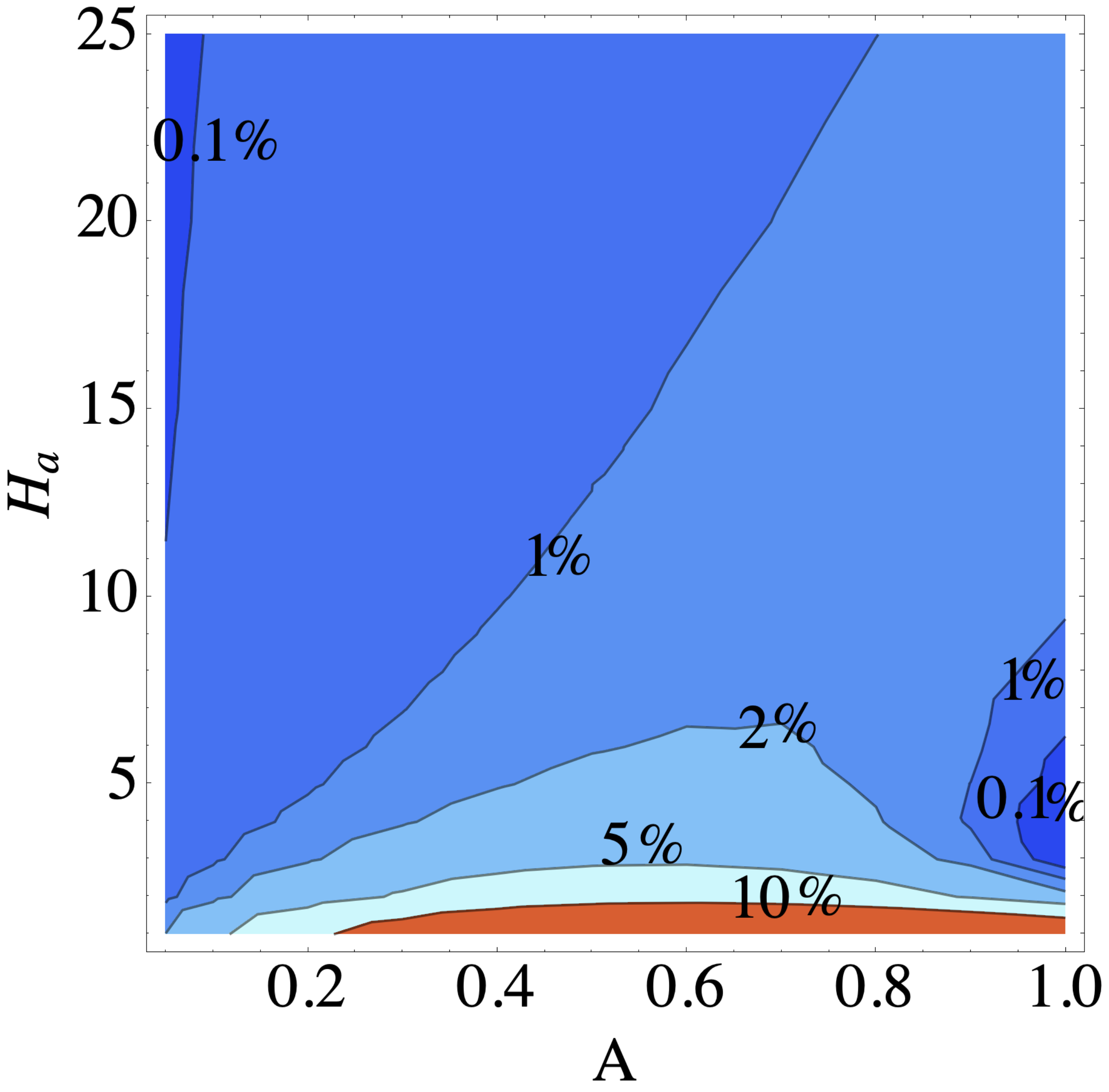

**Fig. 3:** The relative error (*%*) between the numerically computed flow rate and the approximate flow rate (equation 22) as a function of the Hartmann number $H_a$ and the aspect ratio $A$.

The approximate flow rate $Q_a$ agrees within *5%* with the numerical solution of equations (6) when $H_a$>5 and $0<A<1$. The error decreases as $H_a$ increases and/or $A$ decreases (**Fig. 3**). This error does not decrease monotonically as the Hartmann number ($H_a$) increases because our approximate velocity (equation 20) slightly underestimates (overestimates) the actual velocity when $H_a < 20$ ($H_a > 20$) and $A > 0.9$.

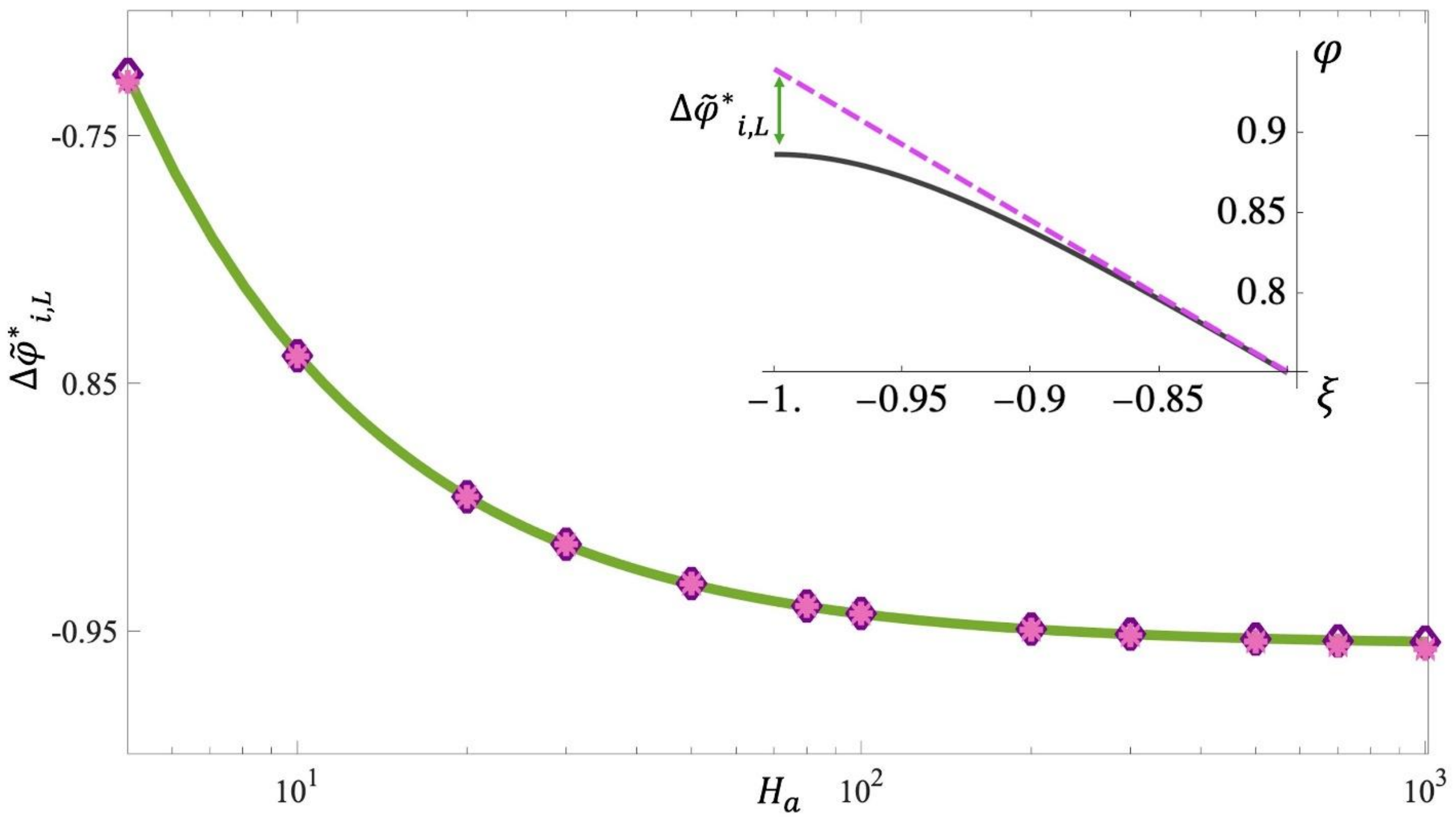

**Fig. 4:** The difference ($\Delta\tilde{\varphi}^*_{i,L}$) between the extrapolated core potential next to the left electrode ($\xi = -1$) and the actual numerically determined electrode potential as a function of the Hartmann number. The pink asterisks, purple open diamonds, and the solid line correspond, respectively, to $\Delta\tilde{\varphi}^*_{i,L}$obtained by numerical integration of the boundary layer equations (13-14 and 18), full-domain numerical simulations (with the core-potential subtracted), and the empirical correlation $\Delta\tilde{\varphi}^*_{i,L} = \alpha H_a^{\beta} + c$, where $\alpha = 1,\ \beta = -1, c = -0.96$ ($R^2 = 0.97$). The inset on the right depicts the potential in the left boundary layer (solid line) and the extrapolated core potential (dashed line) as functions of $\xi$ when $H_a = 5,\ A = 0.05,$ and $S = 1.5$.

The potential in the lateral boundary layer adjacent to the left electrode ($\xi$=−*1*) deviates from its core linear dependence on $\xi$, dipping below the extrapolated core potential (**Fig. 4**, inset). The deviation from the extrapolated core potential correlates with the Hartmann number. $\Delta\tilde{\varphi}^*_{i,L} = \alpha H_a^{\beta} + c$, where $\alpha \sim 1$, $\beta \sim -1, c \sim -0.96$ ($R^2 = 0.97$) (Fig. 4, solid line). The approximate difference ($\psi$) between the left and right electrodes' potential:

$$\psi = \Phi_L - \Phi_R = \frac{2\,a\,b\,P}{\sqrt{\mu\sigma}}\left(Coth(H_a) + \frac{1}{H_a}\left(-\frac{1}{1+S} + \frac{A}{\sqrt{Ha}}\left(\alpha H_a^{\beta} + c\right)\right)\right). \tag{23}$$

4. **Onsager Relations**

MHD systems, such as the one described here, can serve multiple functions, including electrical power generators (converting mechanical energy into electrical energy), actuators (converting electrical energy into mechanical motion, such as in pumps), as well as valves, brakes, and sensors. When designing, optimizing, controlling, and operating MHD-based systems, it is convenient to have simple relations among the system's inputs, such as the potential difference between the left and right electrodes $\psi$ and the pressure-gradient $G$, and the system's outputs, such as the current $\Delta H$ and the flow rate $Q$.

$$\begin{pmatrix} \Delta H \\ Q \end{pmatrix} = \begin{pmatrix} K_{EE} & K_{EH} \\ K_{HE} & K_{HH} \end{pmatrix}\begin{pmatrix} \psi \\ G \end{pmatrix}. \tag{24}$$

$K_{ij}$ are the Onsager conductivities, where the first subscript denotes the force ($E$ for electrical and $H$ for hydrodynamic), and the second subscript represents the flow ($E$ for electrical current and $H$ for fluid flow). $K_{EE}$ is the electric conductance – the inverse of the resistance. When $\psi > 0$ and $G > 0$, the flow-induced current $K_{EH}G < 0$ is in the opposite direction to the Ohmic current, and therefore $K_{EH} < 0$. Based on section 3, we have obtained the Onsager coefficients (Supplement S2):

$$K_{EE} = \frac{A\sigma}{\sqrt{H_a}\left(\sqrt{H_a}\mathrm{Coth}[H_a] + A\left(c + \alpha H_a^{\beta}\right)\right)} \qquad \left(\frac{1}{\Omega m}\right) \tag{25}$$

$$K_{EH} = -2a^2\sqrt{\frac{\sigma}{\mu}}\frac{\left(H_a\mathrm{Coth}[H_a] + A\sqrt{H_a}\left(c + \alpha H_a^{\beta}\right) - 1\right)}{H_a^{3/2}\left(\sqrt{H_a}\mathrm{Coth}[H_a] + A\left(c + \alpha H_a^{\beta}\right)\right)} \qquad \left(\frac{m^2 A}{N}\right) \tag{26}$$

$$K_{HE} = 2a^2\sqrt{\frac{\sigma}{\mu}}\frac{\left(\sqrt{H_a} - A\left(1 - e^{-\frac{\sqrt{H_a}}{A}}\right)\right)(H_a\mathrm{Coth}[H_a] - 1)}{H_a^2\left(A\left(c + \alpha H_a^{\beta}\right) + \mathrm{Coth}[H_a]\sqrt{H_a}\right)} \qquad \left(\frac{m^2 A}{N}\right) \tag{27}$$

$$K_{HH} = \frac{4a^4\left(\sqrt{H_a} - A\left(1 - e^{-\frac{\sqrt{H_a}}{A}}\right)\right)(H_a Coth[H_a] - 1)}{A\,\mu H_a^3\left(A\left(c + \alpha H_a^{\beta}\right) + \sqrt{H_a}Coth[H_a]\right)} \qquad \left(\frac{m^6}{N\,s}\right) \tag{28}$$

In the limit of $H_a >> 1$, these Onsager coefficients (25-28) reduce to

$$K_{EE}^{\infty} = \frac{A\sigma}{H_a}, \quad K_{HE}^{\infty} = -K_{EH}^{\infty} = \frac{2a^2}{H_a}\sqrt{\frac{\sigma}{\mu}}, \quad K_{HH}^{\infty} = \frac{4a^3 b}{\mu H_a^2} \tag{29}$$

In the limit of *A<<1*, our Onsager coefficients reduce to

$$K_{EE}^{0} = \frac{A\sigma}{H_a Coth[H_a]}, \quad K_{HE}^{0} = -K_{EH}^{0} = 2a^2\sqrt{\frac{\sigma}{\mu}}\frac{(H_a \mathrm{Coth}[H_a] - 1)}{H_a^2 \mathrm{Coth}[H_a]},$$

$$K_{HH}^{0} = \frac{4a^4(H_a Coth[H_a] - 1)}{\mu\, A\, H_a^3 Coth[H_a]} \tag{30}$$

In both limits (29) and (30), the coefficients satisfy the Onsager-Casimir reciprocity condition [23, page 90], $K_{EH} = -K_{HE}$. This *anti-symmetry* arises because the magnetic field breaks time-reversal symmetry, rendering the classical, symmetric Onsager reciprocity invalid. The Onsager coefficients (26) and (27) satisfy the Onsager-Casimir reciprocity only approximately, likely because of the empirical approximations we used for the velocity and the potential distributions in the lateral boundary layers near *(*$\xi = \pm 1$*)*. However, the difference between $|K_{EH}|$ and $|K_{HE}|$ is less than 4% when $H_a > 5$ and $A < 1$. This difference decreases rapidly as $H_a$ increases and/or $A$ decreases (**Fig. 5**).

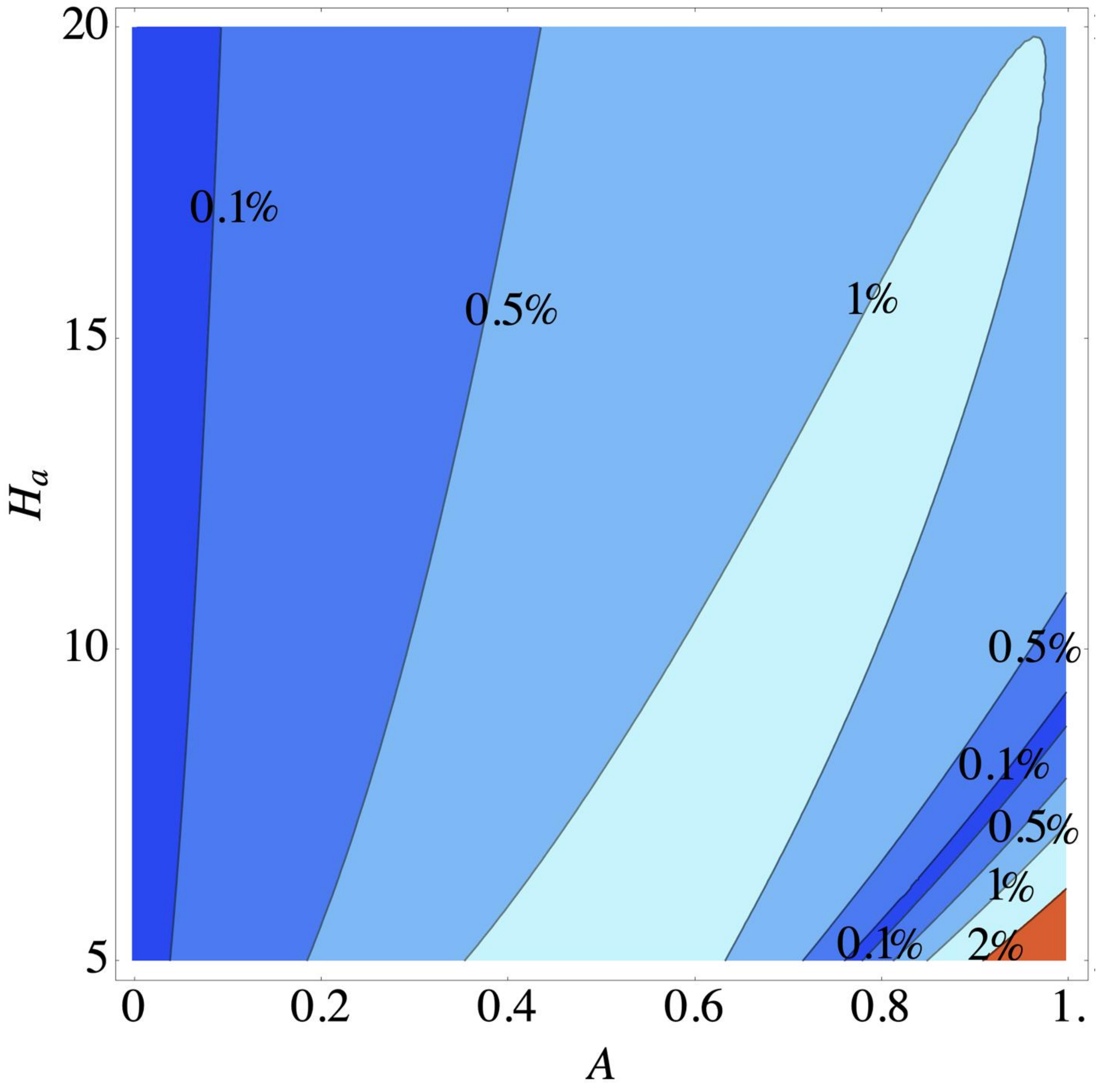

**Fig 5:** The relative difference (%) between the off-diagonal Onsager coefficients $2 \times 10^2 \left|\frac{K_{EH}+K_{HE}}{K_{EH}-K_{HE}}\right|$ as a function of $H_a$ and $A$.

Equations (24) apply to various MHD machines, including electrical power generators that convert mechanical power ($G > 0, Q > 0$) into electrical power ($\Delta H < 0, \psi > 0$); machines (e.g., pumps and actuators) that convert electrical power ($\Delta H$>0, $\psi$ >0) into mechanical power ($G < 0, Q > 0$) as well as hybrid systems combining these functions.

The net power dissipation is $\Delta H\Psi + GQ = K_{EE}\Psi^2 + K_{HH}G^2 \geq 0$. The off-diagonal terms cancel on account of Onsager-Casimir reciprocity. The term $K_{EE}\Psi^2$ represents ohmic losses, and the term $K_{HH}G^2$ viscous losses. When the Lorentz body force is equal in magnitude and opposed in direction to the pressure gradient $\frac{\Delta HB_0}{2a} + G = 0$, there is no flow $Q = 0$ at steady-state. This leads to $K_{HH}K_{EE} + K_{HE}^2 - \frac{2a}{B_0}K_{HE} = 0$.

## 5. Results and Discussion

### Velocity and Electrical Fields

We solved equations (6) for the velocity $u(\xi,\eta)$ and the current stream function $h(\xi,\eta)$ using finite elements in Mathematica (Supplement S1). Given symmetry, it suffices to solve the equations for the quarter domain $\{-1 < \xi < 0,\ 0 < \eta < 1\}$. The computational domain is divided into the coarse grid region $\xi^n + \eta^n < r_n$ and the boundary region, outside the coarse region, to facilitate using a fine mesh in the vicinity of the boundaries without incurring high computational costs. The whole domain numerical results were compared and agreed within 3% with the core (outer) results of Section 3 when $\frac{A}{\sqrt{H_a}} \leq 0.28$ and $-1 + 3\frac{A}{\sqrt{H_a}} \leq \xi \leq 1 - 3\frac{A}{\sqrt{H_a}}$.

Fig. 6 depicts a sample of our numerical results for the current stream function, velocity, and potential fields. Readers with access to Mathematica can readily solve the MHD equations for other values of $H_a$ and $A$ using the supplemental Mathematica Notebook (Supplement S1).

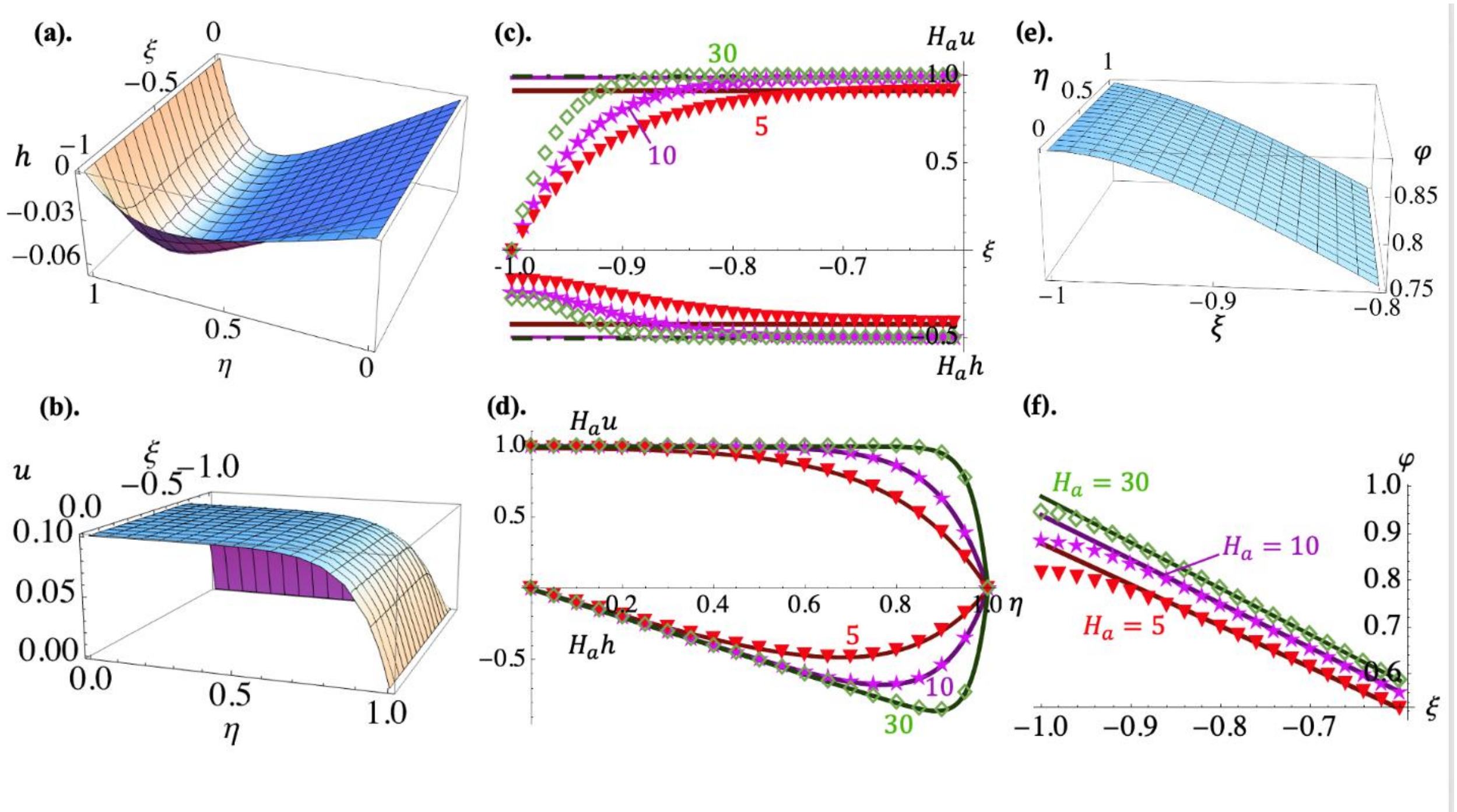

**Fig. 6:** (**a**) The current stream function $h(\xi,\eta)$ as a function of $\xi$ and $\eta$ ($H_a$=10, $A$=0.2). (**b**) The axial velocity $u(\xi, \eta)$ as a function of $\xi$ and $\eta$ ($H_a$=10, $A$=0.2). (**c**) The curves above and below the $\xi$ axis depict, respectively, the velocity $H_a u(\xi, 0.5)$ and the current stream function $H_a h(\xi, 0.5)$ as functions of $\xi$ when $H_a$=5 (red triangle), *10* (purple star), and *30* (green diamond). $A$=0.2. The solid and dashed lines show the corresponding core velocity and current stream function. (**d**) The curves above and below the $\eta$ axis depict, respectively, the velocity $H_a u(0, \eta)$ and the current stream function $H_a h(0, \eta)$ as functions of $\eta$ when $A$=0.2 and $H_a$=5 (red triangle), *10* (purple star), and *30* (green diamond). The solid and dashed lines show the corresponding core velocity and current stream function. (**e**) The potential $\varphi(\xi,\eta)$ as a function of $\xi$ and $\eta$ ($H_a$=10, $A$=0.2, $S$=2/3). (**f**) The potential $\varphi(\xi, 0.5)$ (symbol) and $\varphi_c(\xi)$ (solid line) as functions of $\xi$ when A=0.2, $S = 2/3$, and $H_a =$ 5 (red triangle), *10* (purple star), and 30 (green diamond).

With the exclusion of the $O(H_a^{-1/2})$ boundary layers adjacent to ($\xi = \pm 1$), the current stream function is independent of $\xi$, indicating that in most of the domain, the current flux density (***J***) is in the **$\xi$**-direction (Fig. 6a). Currents in the **$\eta$**-direction exist only next to the electrodes ($\xi = \pm 1$).

At moderate and high $H_a$, the velocity field away from the solid boundaries resembles plug-flow with a nearly uniform velocity $H_a^{-1}\left(\frac{B_0 \Delta H}{2a} - \frac{dp}{dz}\right)$ (Fig. 6b). The induced current produces damping body forces proportional to the velocity's magnitude (Hartmann break), leveling velocity variations away from the non-slip boundaries. As the Hartmann number increases, so does the $\xi$-component of the current density next to the insulating surfaces $\eta = \pm 1$, resulting in thin boundary layers. Although the physical mechanisms and length scales differ, high Hartmann number flow resembles electroosmotic flow in a conduit with a uniform zeta potential at its surface.

Both $u(\xi,\eta)$ and $h(\xi,\eta)$ attain their core values rapidly as the distance from the electrodes increases (Fig. 6c), even at moderate Hartmann numbers. The numerically computed $u(\xi,\eta)$ and $h(\xi,\eta)$ (symbols, Fig. 6d) favorably agree with the asymptotic core expressions $u_c(\xi,\eta)$ and $h_c(\xi,\eta)$.

The core region's electrical potential (Fig. 6e) is independent of $\eta$ and varies linearly with $\xi$ (Fig. 6f, solid line). However, the electrodes' potential cannot be obtained by linearly extrapolating the core potential. Witness the nonlinear dependence of $\varphi$ on $\xi$ next to the electrode (Figs. 4 inset (solid line), Figs. 6e and 6f ( symbols)).

The Onsager coefficients provide a simple means to determine the performance characteristics of various MHD machines, as illustrated next. Below, we applied Onsager-Casimir reciprocity in our analysis of various MHD machines.

**Conversion of Mechanical Power into Electrical Power (Electrical Generators)**

Consider pressure-driven flow $G = \left(-\frac{dp}{dz}\right) > 0$ in a conduit (**Fig. 1**). The electrodes are connected to an external load with resistance $R'_{ex}$ ($\Omega\ m$) per conduit's unit length. The generator's internal resistance is $R'_{in} = \left(\frac{b}{a\sigma}\right)$. The liquid metal's motion in the magnetic field induces an electromotive force (*emf*) $\Psi > 0$. $\Psi = -R'_{ex}\Delta H$. $\Psi$ varies from zero (short circuit, $R'_{ex}=0$) to open circuit potential $\Psi_{oc}$ ($R'_{ex} \to \infty$).

$$\Delta H = -\frac{K_{HE}}{1+K_{HH}R'_{ex}} G \text{ and } \Psi = \frac{K_{HE}R'_{ex}}{1+K_{EE}R'_{ex}} G \tag{31}$$

When the circuit is open ($R'_{ex} \to \infty$), $\Delta H = 0$, and $\Psi_{oc} = \frac{K_{HE}}{K_{EE}} G$. The current can be expressed as a function of the potential difference $\Delta H = K_{EE}(\Psi - \Psi_{oc})$. Maximum current intensity takes place when the electrodes are shortened ($R'_{ex}=0,\ \Psi=0$), $\Delta H_{max} = -K_{HE}G$. The corresponding $S$ value is $S_m = -\frac{K_{HE}}{2a} B_0$.

The flow rate $Q = K_{HH}G + K_{HE}\Psi = \left(K_{HH} + \frac{K_{HE}^2 R'_{ex}}{1+K_{EE}R'_{ex}}\right) G$ is lowest when the electrodes are short-circuited ($\Psi$=$R_{ex}$=0, $S$=$S_m$). The induced current generates Lorentz body force opposing $G$ throughout the conduit's cross-section (Fig. 7a). As the *emf* $\Psi$ increases (when the external load increases), it produces an Ohmic current opposing the induced current next to the insulating surfaces $\eta = \pm 1$, resulting in flow-assisting Lorentz force (Fig. 7b, $S = 0.5\, S_m$) and an increased flow rate. This effect is maximized under open circuit conditions (Fig. 7c, $S = 0$) when no current leaves the conduit. The electrical current forms a clockwise eddy when $\eta > 0$ and a counterclockwise eddy when $\eta < 0$ (Fig. 8).

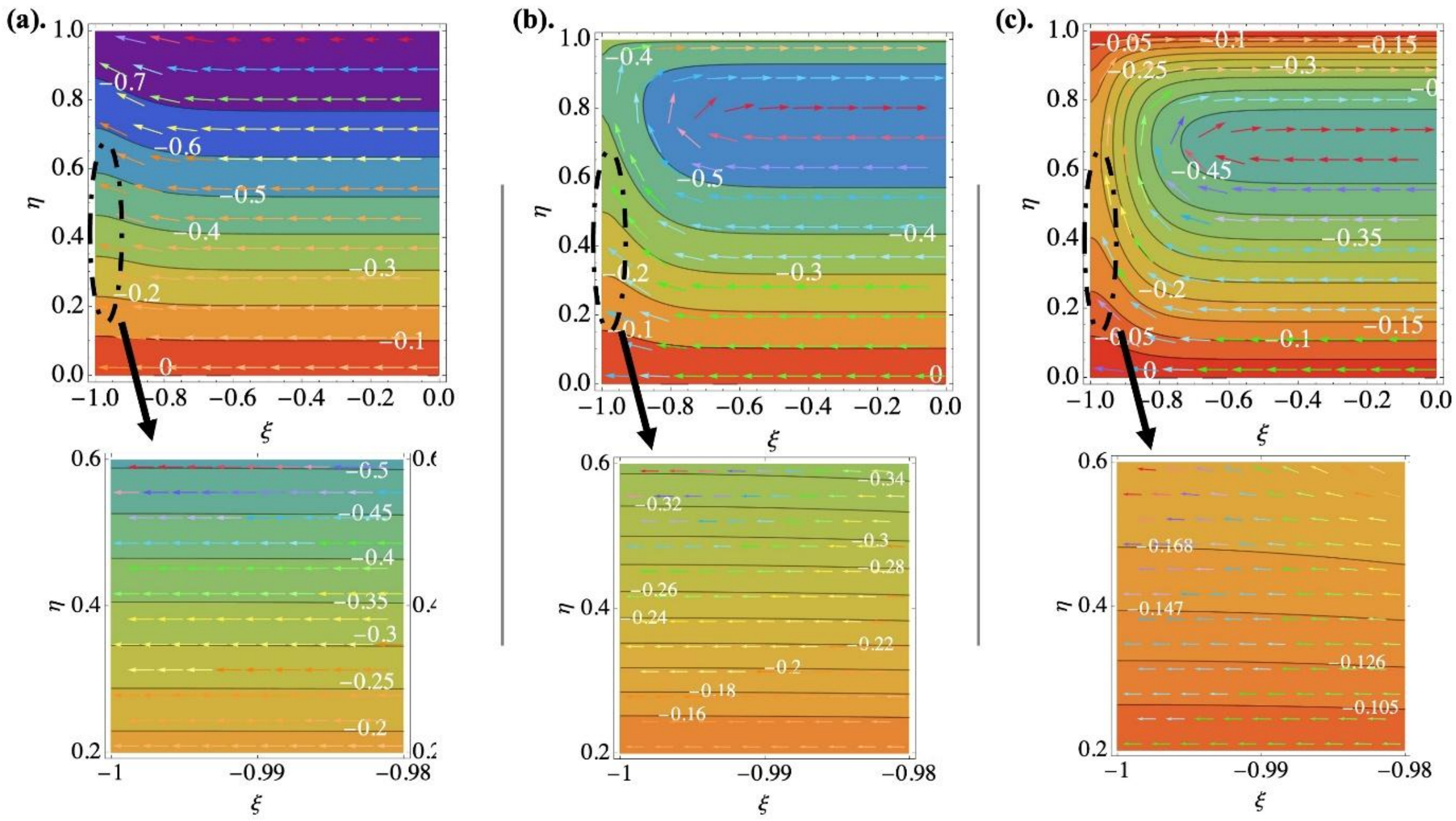

**Fig. 7:** Contour lines of the normalized current stream function $\widehat{H_z} = H_z \frac{B_0}{2aG} = S\eta + H_a(S+1)h(\xi,\eta)$ when $H_a = 5$ and $A = 0.2$. $S = S_m$ (**a**, $R'_{ex} = 0$), $S = \frac{S_m}{2}$ (**b**), and $S = 0$ (**c**). The bottom graphs provide a magnified view of the region ($-1 < \xi < -0.98$; $0.2 < \eta < 0.5$) adjacent to the electrode.

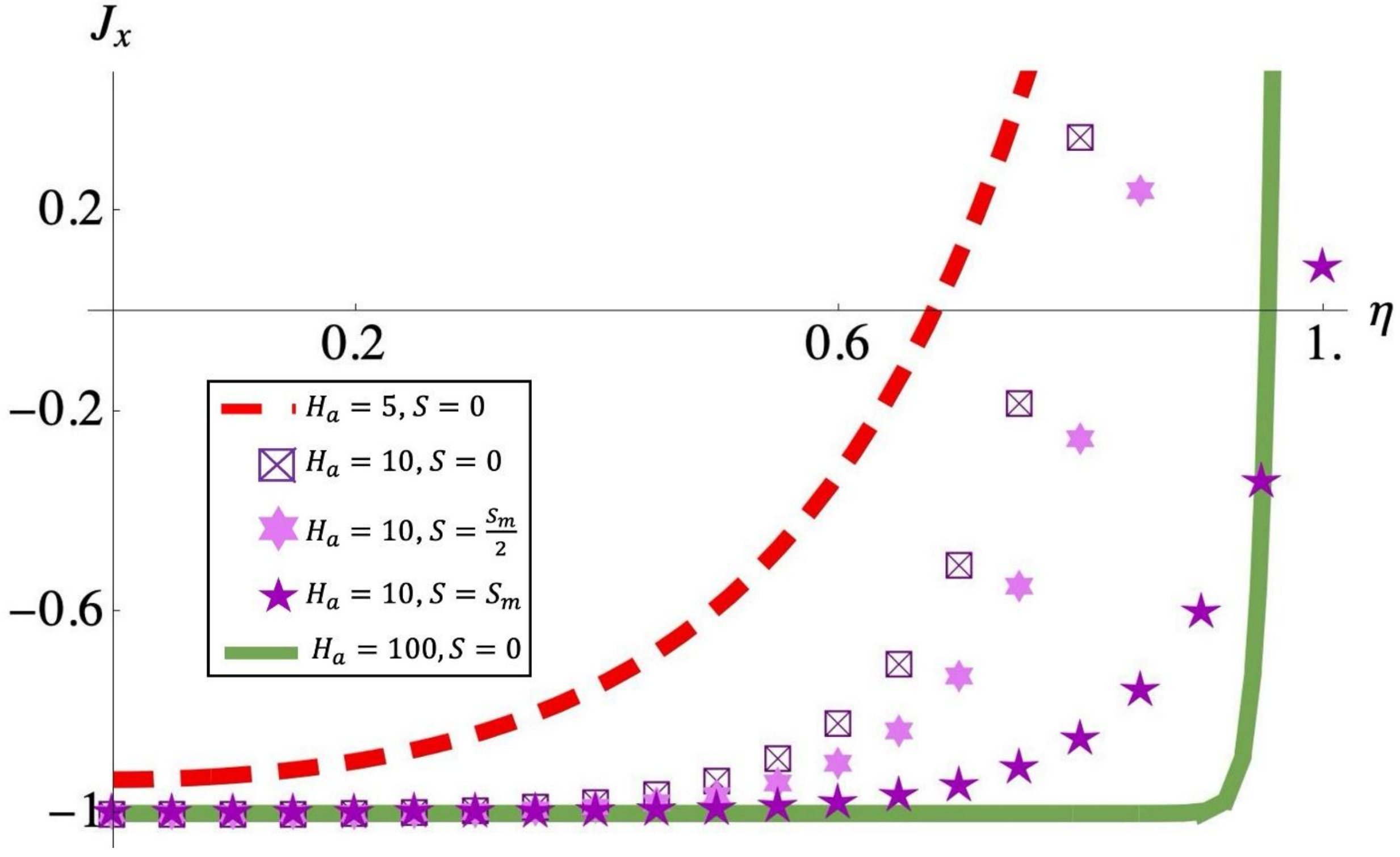

**Fig. 8:** The horizontal component of the normalized current $J_x(0, \eta)$ as a function of $\eta$ ($0 \leq \eta \leq 1$) at various Hartmann and $S$ numbers: $H_a = 5$ and $S = 0$ (dashed red line); $H_a = 10$ and $S = S_m$ (purple star), $S = S_m/2$ (six point star), and $S = 0$ (square with cross); and $H_a = 100$, $A = 0.2$, and $S = 0$ (green line).

The mechanical power consumed per unit length of the conduit is $QG$, and the electrical power produced per unit length of the conduit is $(-\Delta H\psi)$. The MHD generator conversion efficiency

$$\varepsilon_g = -\frac{\Delta H\Psi}{GQ} = \frac{K_{HE}{}^2 R'_{ex}}{(1+K_{EE}R'_{ex})\left(K_{HE}{}^2 R'_{ex} + K_{HH}(1+K_{EE}R'_{ex})\right)} \tag{32}$$

is independent of the driving pressure gradient *(G)* but it depends on the external load. As the resistance ratio $r = R'_{ex}/R'_{in}$ increases from zero, the generator's efficiency increases, attains a maximum, and then decreases (Fig. **9).** As the Hartmann number increases, the optimal resistance ratio $r$ and the maximum efficiency $\varepsilon_{g,m}$ increase and the efficiency's sensitivity to deviations from the optimal $r$ decreases, resulting in a broader peak. In the limit of $H_a \to \infty$,

$$\varepsilon_g^{\infty} = \frac{\frac{R'_{ex}}{R'_{in}}}{1 + \frac{R'_{ex}}{R'_{in}}} . \tag{33}$$

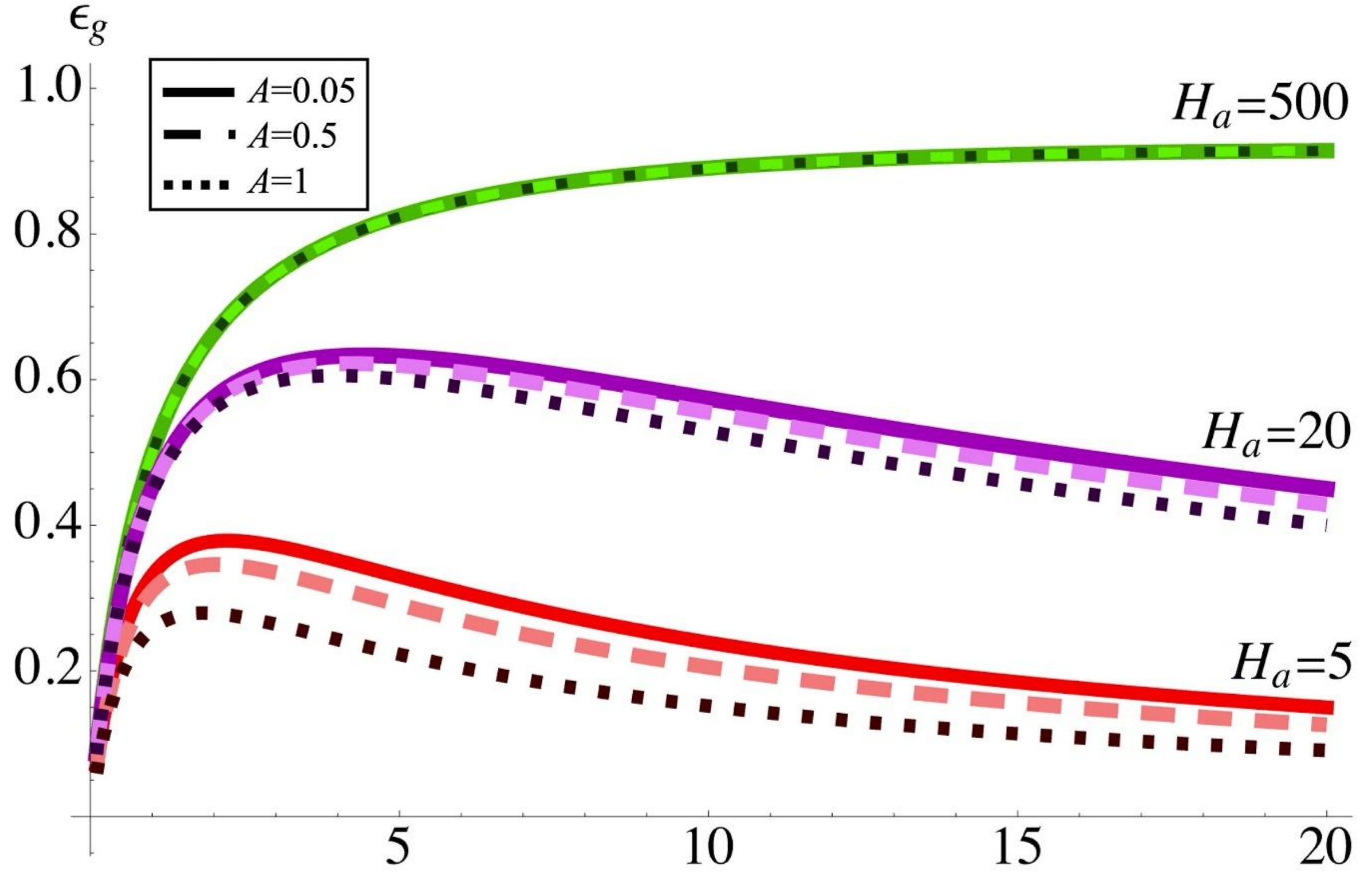

**Fig 9:** MHD generator's efficiency ($\varepsilon_g$) as a function of the external resistance and internal resistance ratio $r$ when : $H_a = 5,\ 20,\ 500$ and $A = 0.05$ (Solid line), 0.5 (dashed line), 1 (dotted line).

The maximum efficiency $\varepsilon_{g,m}$ is achieved when

$$r = \frac{R'_{ex}}{R'_{in}} = A\,\sigma \frac{\sqrt{K_{HH}}}{\sqrt{K_{EE}(K_{HE}^2 + K_{EE}K_{HH})}} \tag{34}$$

The efficiency at the optimal load

$$\varepsilon_{g.m} = \frac{K_{HE}^2 + 2K_{EE}K_{HH} - 2\sqrt{K_{HH}}\sqrt{K_{EE}\left(K_{HE}^2 + K_{EE}K_{HH}\right)}}{K_{HE}^2} \tag{35}$$

increases as the Hartmann number increases (Fig. 10). When $H_a \to \infty$, $\epsilon_{g,m} \to 1$.

The max efficiency occurs when $\left(\frac{R'_{ex}}{R'_{in}}\right)_{H_a \gg 1} \sim \sqrt{H_a}$. In the range of aspect ratios $A$ considered. The maximum efficiency does not vary much with variations in $A$.

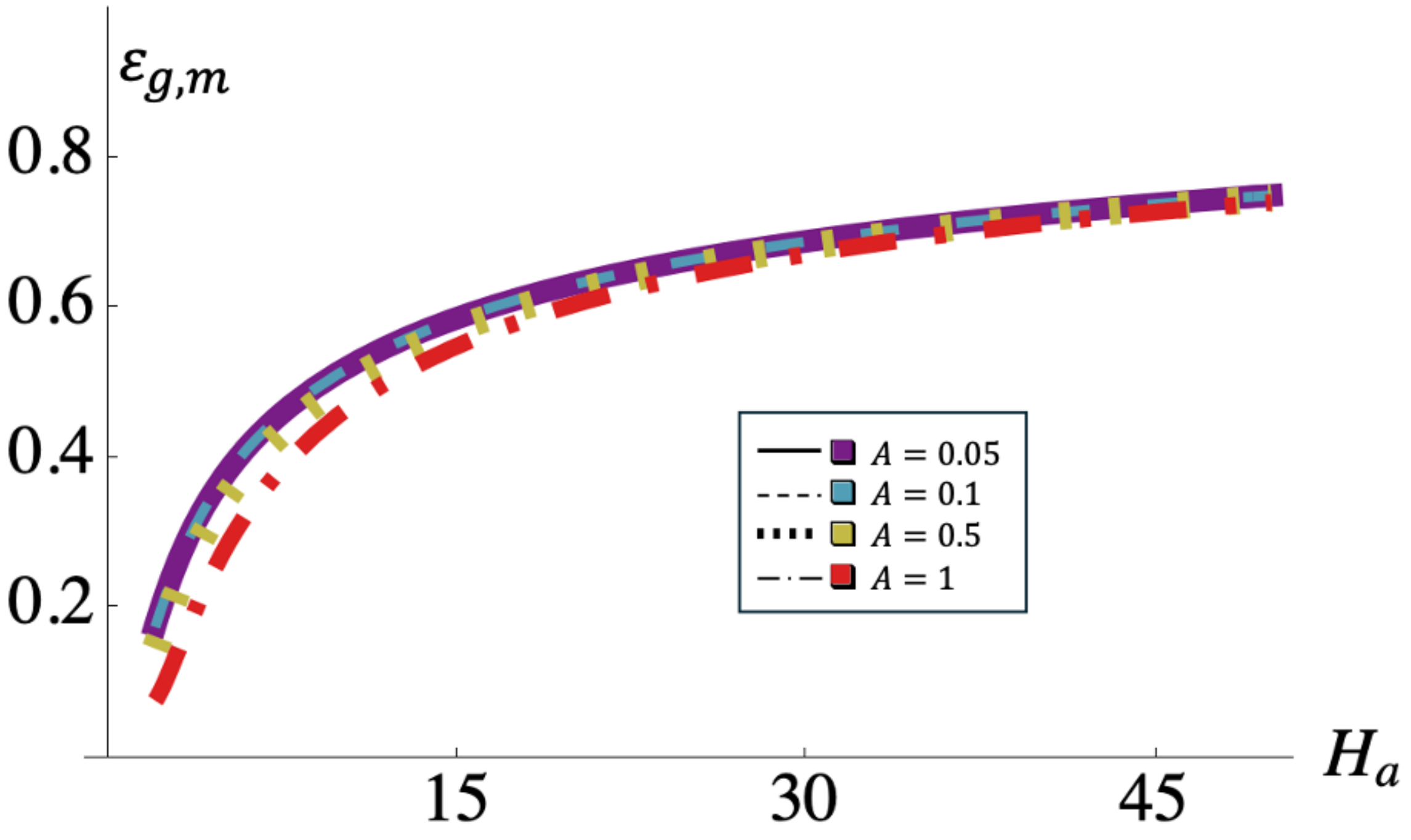

**Fig 10:** MHD generator's efficiency ($\varepsilon_{g,m}$) at optimal load as a function of $H_a$ when $A = 0.05$ (solid, purple line), 0.1 (horizontal dashed, blue line), 0.5 (vertical dashed, yellow line), and 1 (dash-dot, red line).

**Conversion of Electrical Power into Mechanical Power (Actuator, Pump)**

Next, we consider the case when the MHD machine is connected to an external power supply, and current is injected into the liquid metal through the electrodes. The resulting Lorentz body force can propel the liquid metal against an adverse pressure gradient ($G < 0$) or adverse force. Here, $G$ is the pressure gradient induced by the external load. This MHD system can circulate liquid metal for thermal control, displace immiscible fluids, or displace a piston or elastic members for actuation, as well as provide flow control without mechanical valves in fluidic networks [8]. In the absence of flow, the stagnation pressure gradient $G_s = -K_{HH}^{-1} K_{HE} \psi$ balances the Lorentz forces generated by the injected current.

The pump's mechanical efficiency $\varepsilon_p = -\frac{QG}{\Delta H \psi}$ varies with the adverse pressure head ($G$)

$$\varepsilon_p = \frac{G}{G_s}\left(1 - \frac{G}{G_s}\right)\left(\frac{K_{EE}K_{HH}}{K_{HE}^2} + \frac{G}{G_s}\right)^{-1} \tag{35}$$

As $G/G_S$ increases, the efficiency ($\varepsilon_p$) increases, attains a maximum, and decreases (**Fig. 11**). $\varepsilon_p$ attains its maximum when

$$\frac{G_{op}}{G_s} = \frac{-K_{EE}K_{HH} + \sqrt{K_{EE}K_{HH}(K_{HE}^2 + K_{EE}K_{HH})}}{K_{HE}^2} \tag{36}$$

The optimal conversion efficiency is:

$$\varepsilon_{p,m} = \left(\frac{G_{op}}{G_s}\right)^2 \frac{K_{HE}^2}{K_{EE}K_{HH}} \tag{37}$$

As the Hartmann number increases, so does the maximum efficiency $\varepsilon_{p,m}$ (Fig. 11). When $H_a \to \infty$, $\epsilon_{p,m} \to 1$ (**Fig. 12**).

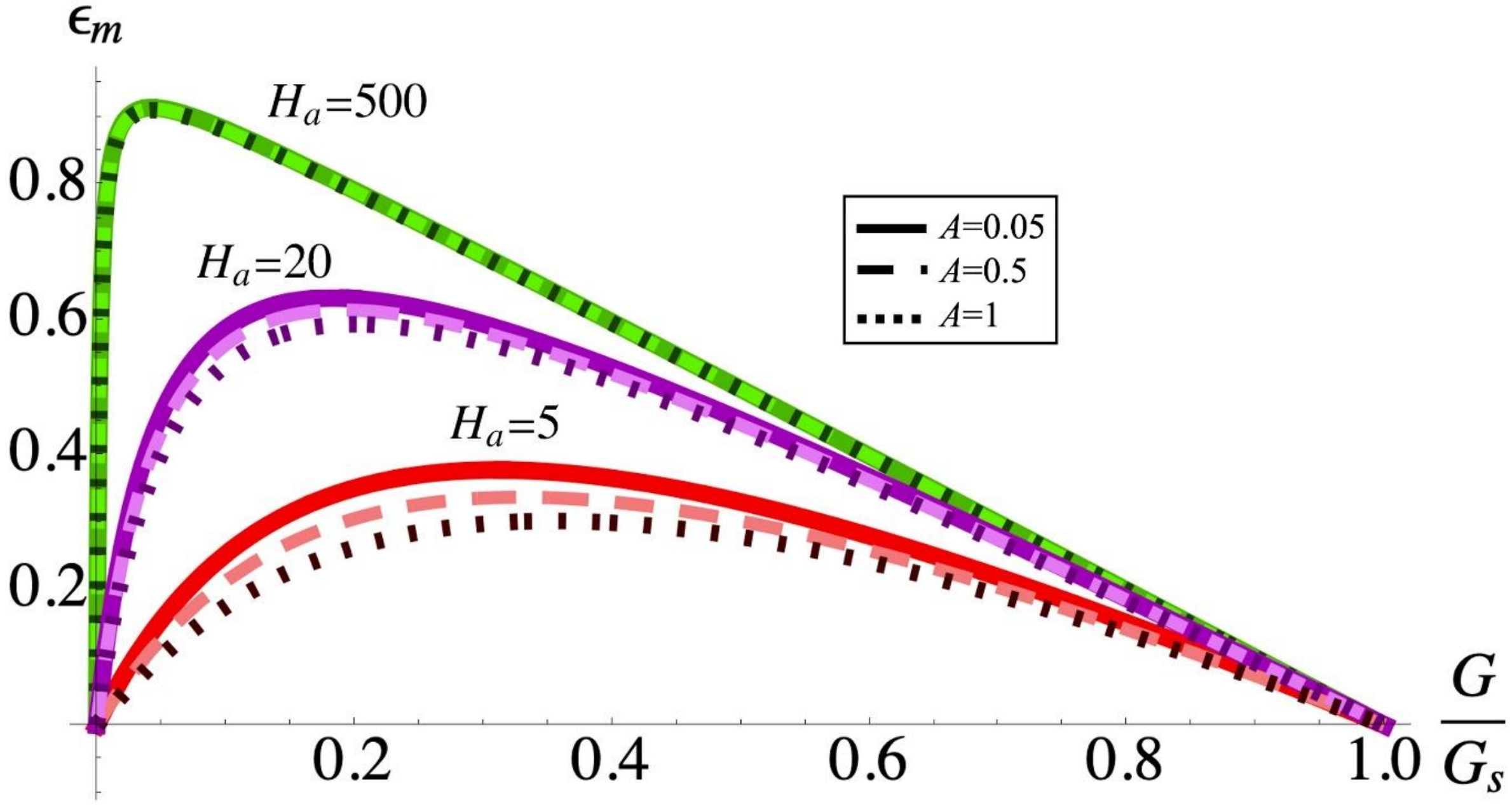

**Fig. 11**: MHD pump efficiency ($\varepsilon_p$) as a function of $\frac{G}{G_s}$. $H_a = 5$ (red), 20 (purple), and 500 (green). $A = 0.05$ (solid line), 0.5 (dashed line), and 1 (dotted line).

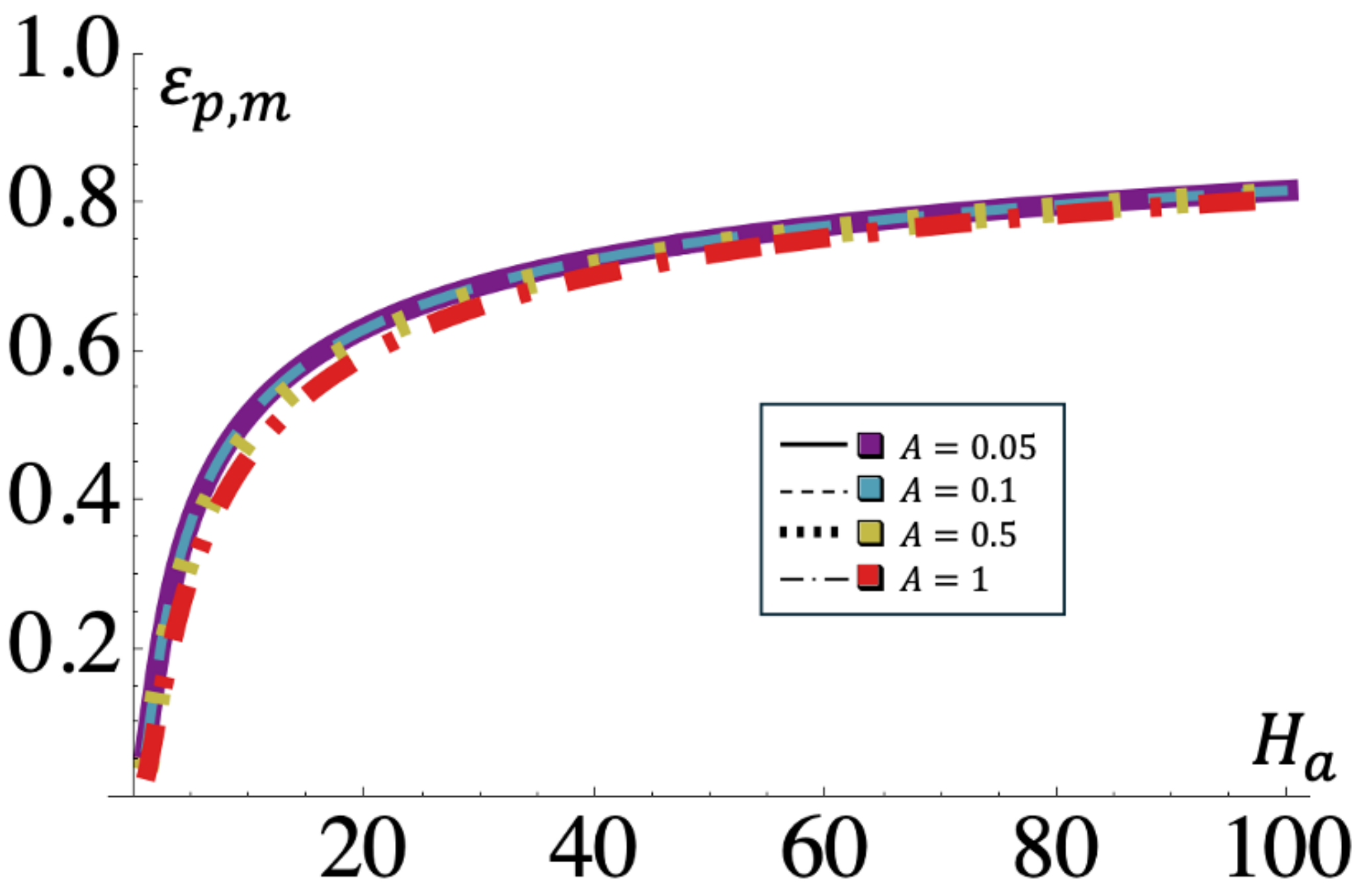

**Fig 12:** Optimal pump efficiency ($\varepsilon_{p,m}$) as a function of the Hartmann number $Ha$. $A =$ 0.05 (solid, purple line), 0.1 (horizontal dashed, blue line), 0.5 (vertical yellow, solid line), and 1 (orange, dash-dot line).

**MHD sensors**

Measurement of electrical signals, such as the potential difference across or total current passing through the electrodes, informs the flow rate and the pressure drop in the conduit. Consider, for example, open circuit conditions ($\Delta H = 0$). $G = \frac{K_{EE}}{K_{HE}} \Psi_{oc}$ and $Q = K_{HE}\left(1 + \frac{K_{HH}K_{EE}}{K_{HE}^2}\right)\Psi_{oc}$.

**MHD Valve and Brake**

The induced current in the moving liquid metal generates a Lorentz body force that opposes the flow, providing a (Hartmann) brake. The Lorentz body force can be modulated to further slow the flow by adjusting the magnetic field intensity and/or supplying current from a power supply ($\psi < 0$ when $G > 0$). Thus, the MHD machine provides a means to control the flow rate without a mechanical valve [8].

When $|\psi|$ is sufficiently large to stop motion, we have a MHD brake. MHD brakes provide several advantages over conventional friction-based brakes. Instead of dissipating the braking energy as heat, the MHD brake provides the opportunity to recover and harvest most of the braking energy. The MHD brake provides contactless, soft braking, avoiding the wear and performance degradation of friction-based brakes.

**Conclusions**

Recently, interest has grown in applying room- and near-room-temperature liquid metals to miniature systems for power harvesting, actuation, flow control, and sensing. These systems typically operate with permanent magnets at moderate Hartmann numbers. Analyzing such systems often requires repeated numerical solutions of the MHD equations during design, optimization, and control processes. This paper reformulates the relationships between control inputs, such as the electrodes' potential difference ($\Psi$) and the pressure gradient ($G$), and outputs, such as flow rate ($Q$) and current ($\Delta H$), in the form of Onsager relations. These simplified algebraic relations facilitate the control and optimization of various MHD machines for pumping, flow control, and actuation when the pressure gradient and electrodes' potentials are independently controlled (**Fig. 13** left) and for electrical power generation and sensing when an external load controls electrode potential (**Fig. 13** right).

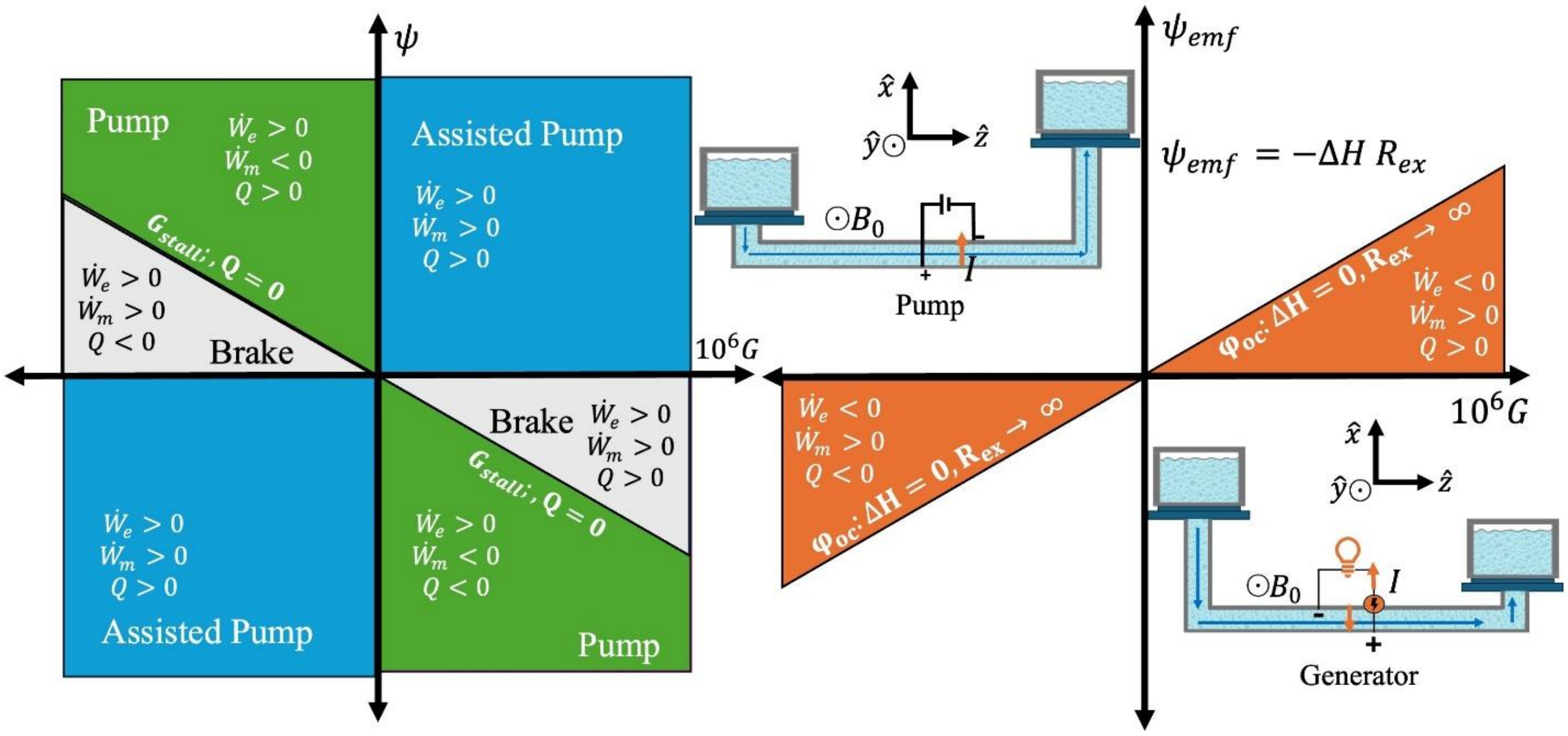

**Fig 13:** Characterization of the operating regimes of the MHD machine as a function of the pressure gradient $G$ and the potential difference across the electrodes $\Psi$ (left). $\dot{W}_m = GQ$ ($Wm^{-1}$) and $\dot{W}_e = \Psi\Delta H$ ($Wm^{-1}$) are, respectively, the mechanical and electrical powers per unit length. $Q$ is the flow rate ($m^3 s^{-1}$),

taken as positive when in the positive **z**-direction. $\Delta H$ is taken as positive when in the positive x direction. When operating as a generator (right), the external load resistance $R'_{ex}$ ($\Omega m$) controls the electrode potential difference $\psi$.

We explicitly demonstrated that the cross Onsager coefficients satisfy the anti-symmetric Onsager–Casimir reciprocity $K_{HE} = -K_{EH}$. This *anti-symmetry* arises because the magnetic field breaks time-reversal symmetry, and the classical, symmetric Onsager's reciprocal relations are no longer valid.

To determine the Onsager coefficients, we employed multiple methods, including asymptotic analysis in the limits of small aspect ratio ($A \to 0$), small Hartmann number ($H_a \to 0$), and large Hartmann number ($H_a \to \infty$)), as well as full domain, finite element simulations. All approaches yielded consistent results. This favorable agreement serves as a mutual verification of our findings.

The Onsager conductivities $K_{ij}$ can be determined from experimental data, enabling validation of this paper's results. The flow rate ($Q$), current ($\Delta H$), pressure gradient ($G$), and potential difference across the electrodes ( $\Psi$ ) are all measurable quantities. For example, a simple experimental setup could resemble the MHD machines sketched in Fig. 13 (right). In this setup, a duct with a small rectangular cross-sectional area, like the one in Fig. 1, connects two relatively large reservoirs containing liquid metal at different levels. The difference in liquid levels controls the pressure gradient. By varying the pressure gradient ($G$) and the potential difference ( $\Psi$ ) and measuring the flow rate ($Q$) and the electrical current ($\Delta H$) repeatedly, one can extract the Onsager conductivities $K_{ij}$ reliably by minimizing an objective (cost) function such as $\sum_{\ell}\left(\left(Q_{\ell}^{predicted} - Q_{\ell}^{measured}\right)^2 + \left(\Delta H_{\ell}^{predicted} - \Delta H_{\ell}^{measured}\right)^2\right)$.

**Supplementary Material**

S1. Finite Element Calculations of the Velocity and Electrical Fields. A *Mathematica* Notebook containing the scripts used to solve and visualize the velocity, potential, and current distributions.

S2. Onsager Coefficients and Performance Characteristics. A *Mathematica* Notebook presenting Onsager Coefficients and performance calculations of various magnetohydrodynamic (MHD devices.

**Funding statement**

SES acknowledges support from the USA National Science Foundation Graduate Research Fellowship (DGE-2236662). Any opinions, findings, conclusions, or recommendations expressed in this paper are those of the authors and do not necessarily reflect the views of the National Science Foundation.

# Supplement S1: Finite Element Simulation of MHD Flow in a Long Conduit with a Rectangular Cross-Section

Supplement to: Sindu Shanmugadas & Haim H. Bau*, 2025, “Onsager Coefficients for Liquid Metal Flow in a Conduit under a Magnetic Field,”
Dept. Mechanical Engineering and Applied Mechanics, University of Pennsylvania, Philadelphia, PA 1904-6315
*Corresponding author: *bau@seas.upenn.edu*

Mathematica bersion 14.2

We solve the two coupled, dimensionless *pdes* for the steady, fully-developed axial velocity *u(ξ,η)* and the current stream function *h(ξ,η)*.

$$\begin{pmatrix} A^2 \frac{\partial^2}{\partial\xi^2} + \frac{\partial^2}{\partial\eta^2} & H_a \frac{\partial}{\partial\eta} \\ H_a \frac{\partial}{\partial\eta} & A^2 \frac{\partial^2}{\partial\xi^2} + \frac{\partial^2}{\partial\eta^2} \end{pmatrix} \begin{pmatrix} u \\ h \end{pmatrix} = \begin{pmatrix} -1 \\ 0 \end{pmatrix}$$

$$u(\pm 1, \eta) = u(\xi, \pm 1) = \frac{\partial h(\pm 1, \eta)}{\partial \xi} = h(\xi, \pm 1) = 0.$$

Hence to forth, we refer, respectively, to the normalized coordinates $-1<\eta<1$ and $-1<\xi<1$ as “y” and “x” and the *dimensionless* axial velocity as v(x,y).

## 1. Finite Element Solution for a Quarter Domain

```
In[ ]:= Needs["DifferentialEquations`InterpolatingFunctionAnatomy`"]
```

Exploiting symmetries, we solve the MHD equations for the quarter domain {0<y<1} {-1<x<0}, The velocity u(x,y) is even with respect to x=0 and y=0. $\frac{\partial u(0,y)}{\partial x}=\frac{\partial u(x,0)}{\partial y}=0$. h(x,y) is odd with respect to y=0 and even with respect to x=0. $h(x,0)=\frac{\partial h(0,y)}{\partial x}=0$. The computational domain is divided into a core (outer) region $x^n + y^n < 0.9$ with relatively coarse mesh and an inner region $x^n + y^n > 0.9$ with a finer grid.

```
In[1]:= L1 = ImplicitRegion[-1 ≤ x ≤ 0 && 0 ≤ y ≤ 1, {x, y}];
refinementRegion := ImplicitRegion[x^2 + y^2 > 0.9, {{x, -1, 0}, {y, 0, 1}}];
mrfl = With[{rmf = RegionMember[refinementRegion]},
   Function[{vertices, area}, Block[{x, y}, {x, y} = Mean[vertices];
     If[rmf[{x, y}], area > 0.000005, area > 0.0025]]]];
```

In[4]:=
```
Show[RegionPlot[L1], RegionPlot[refinementRegion], AspectRatio → Automatic]
```

Out[4]=
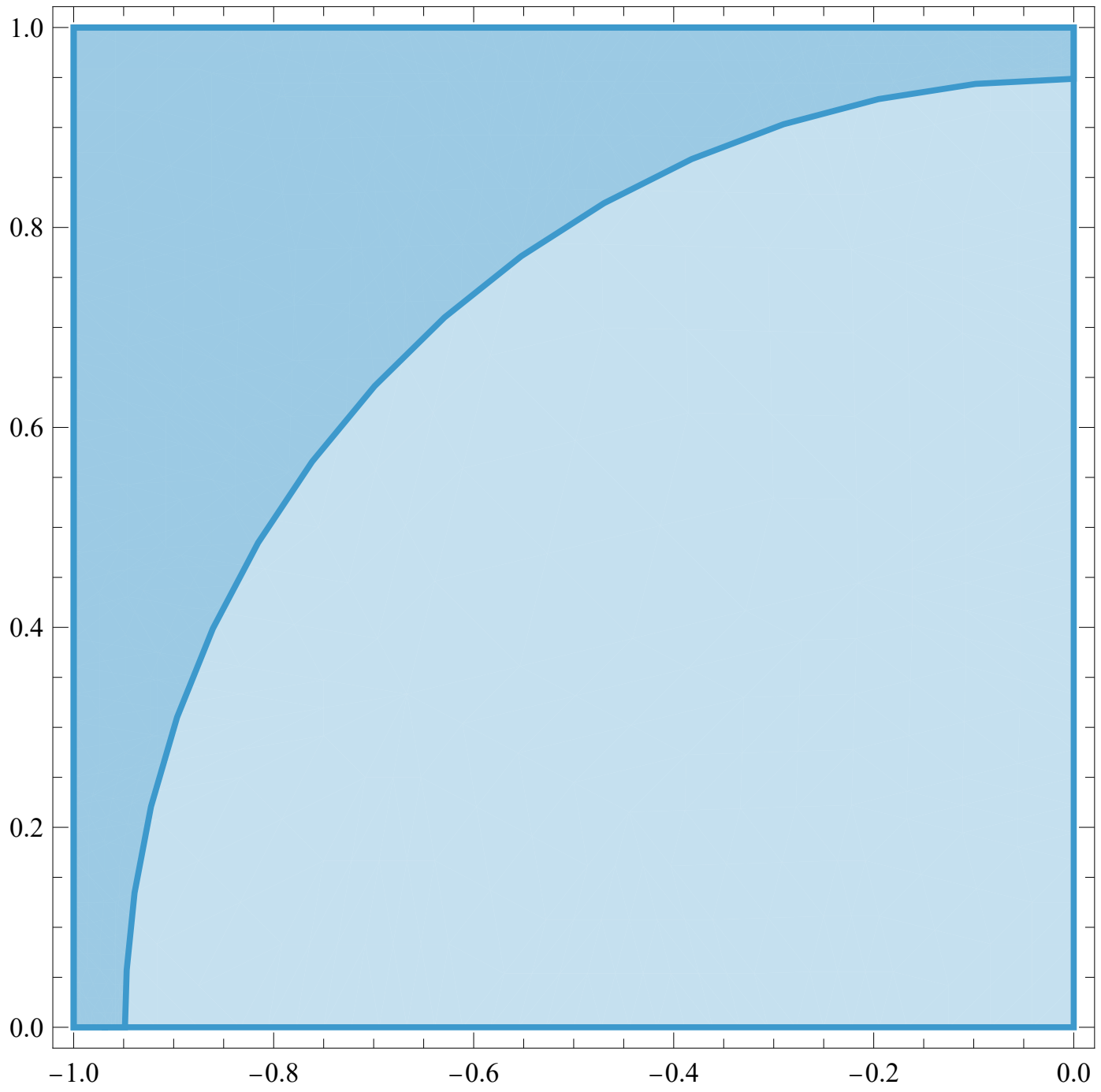

Below, we specify the coupled PDEs and boundary conditions.

In[5]:=
```
op1[A_, Ha_] :=
  Inactive[Div][{{A^2 , 0}, {0, 1}}.Inactive[Grad][v[x, y], {x, y}], {x, y}] + Ha D[h[x, y], y] + 1;
op2[A_, Ha_] :=
  Inactive[Div][{{A^2 , 0}, {0, 1}}.Inactive[Grad][h[x, y], {x, y}], {x, y}] + Ha D[v[x, y], y];
BC$Dirichlethq = {v[x, 1] == 0, v[-1, y] == 0, h[x, 0] == 0, h[x, 1] == 0};
```

In[8]:=
```
Ha = 10; A = 0.2;
{nh, nv} =
 NDSolveValue[{Activate[op2[A, Ha]] == 0, Activate[op1[A, Ha]] == 0, BC$Dirichlethq}, {h, v},
  Element[{x, y}, L1], Method → {"FiniteElement", "MeshOptions" → {MeshRefinementFunction → mrfl}}]
```

Out[9]=
```
{InterpolatingFunction[ Domain: {{-1., 0.}, {0., 1.}}
                        Output: scalar
                        Data not saved. Save now ],
 InterpolatingFunction[ Domain: {{-1., 0.}, {0., 1.}}
                        Output: scalar
                        Data not saved. Save now ]}
```

Apparently, listing the second PDE first in NDSolve leads to a better conditioned stiffness matrix than otherwise.

# 2. Data on the FE grid

In[10]:=
```
Show[nv["ElementMesh"]["Wireframe"]] (*Show the finite element mesh *)
```

Out[10]=

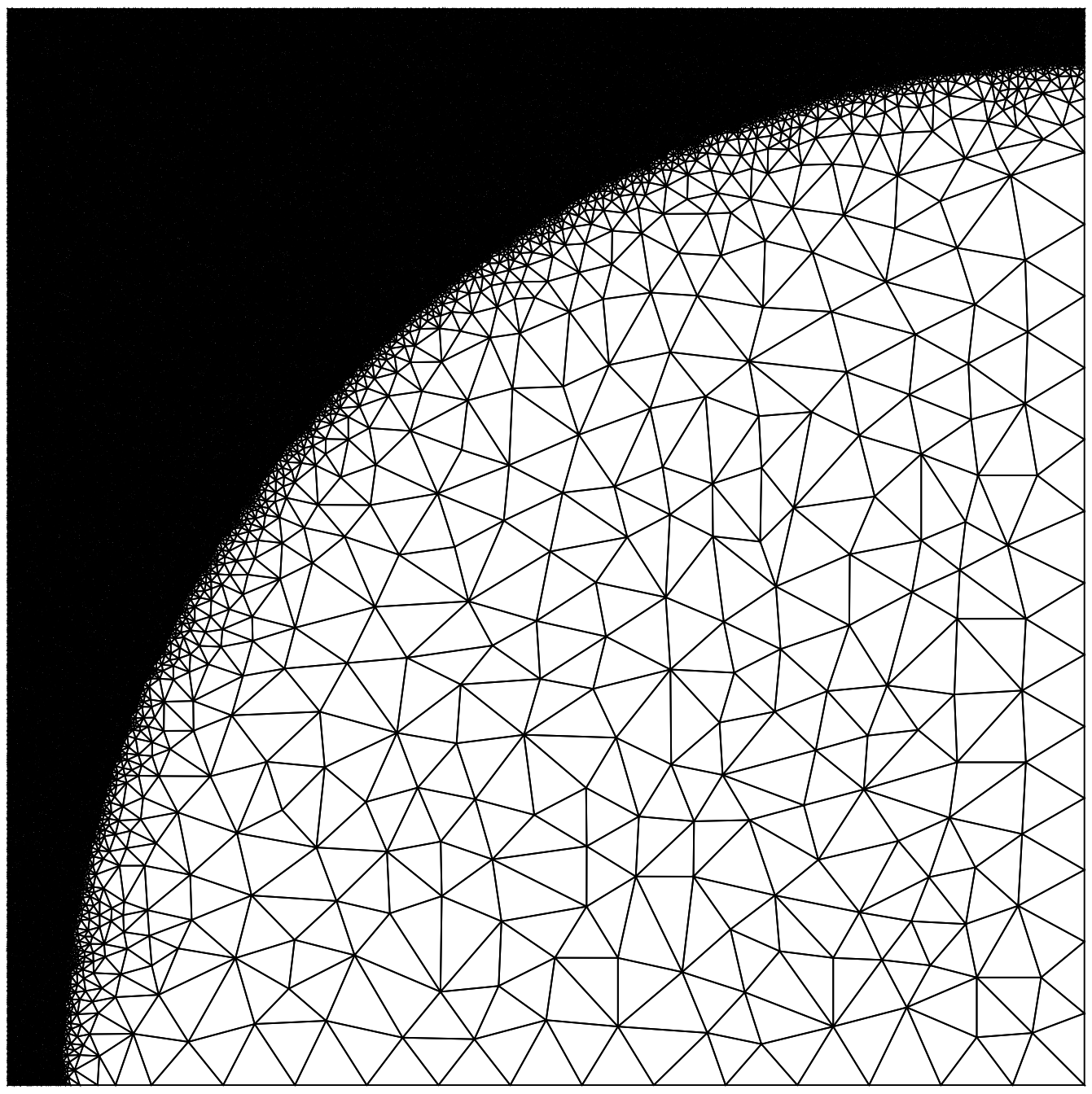

In[11]:=
```
nv["ElementMesh"] (* number of elements *)
```

Out[11]=
```
NDSolve`FEM`ElementMesh[{{-1., 0.}, {0., 1.}}, {NDSolve`FEM`TriangleElement[<95256>]}]
```

In[12]:=
```
nv["ElementMesh"]["MeshOrder"] (* verify that Mathematica uses quadratic elements*)
```

Out[12]=
```
2
```

# 3. Core (outer) Solution :

In[13]:=
```
vc[Ha_, y_] := 1 / Ha * Cosh[Ha] / Sinh[Ha] (1 - Cosh[Ha * y] / Cosh[Ha])
hc[Ha_, y_] := -1 / Ha (y - Sinh[Ha * y] / Sinh[Ha])
jxc[Ha_, y_, S_] := (S + 1) Ha Cosh[Ha y] / Sinh[Ha] - 1 (* current component in the x-direction*)
fc[Ha_, x_, S_] := (-Coth[Ha] + 1 / Ha / (1 + S)) x   (*potential*)
```

In[17]:=
```
Plot[{Ha vc[Ha, y], Ha hc[Ha, y], Ha^(-1/2) jxc[10, y, 0]}, {y, -1, 1}, AxesLabel → {"y"}, Epilog →
  {Text["Ha vc", {0.3, 1.2}, Background → White], Text["Ha hc", {-0.5, 0.7}, Background → White],
   Text["Ha^(-1/2) Jx c", {-0.8, 2}, Background → White]}, PlotTheme → "Classic", PlotStyle → Thick]
```

Out[17]=

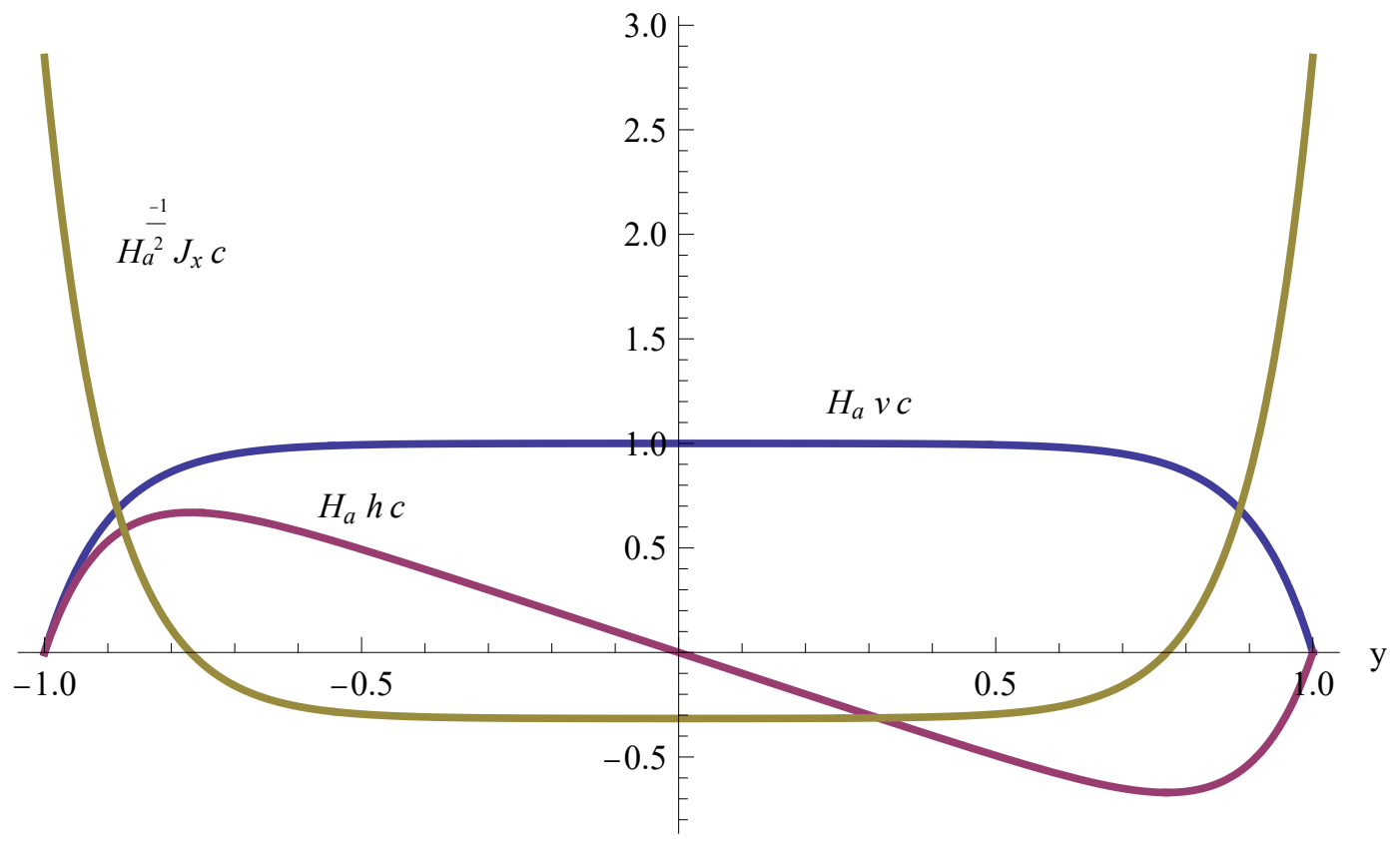

# 4. The velocity v(x,y) and the current stream function h(x,y)

In[18]:=
```
gv = Plot3D[nv[x, y], {x, -1, 0}, {y, 0, 1},
  LabelStyle → Directive[Medium], AxesLabel → {"ξ", "η", "v"}, PlotTheme → "Classic"]
```

Out[18]=

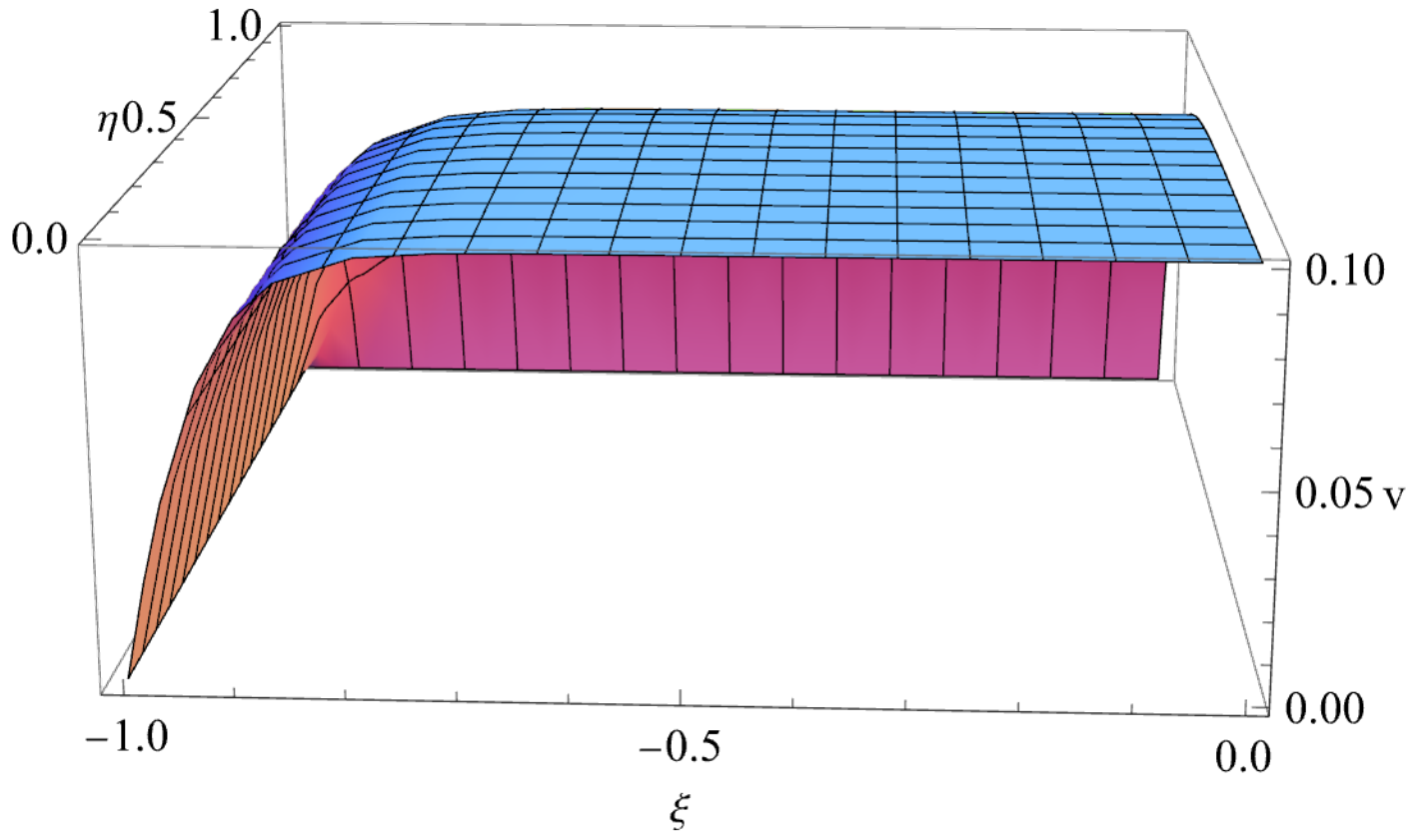

```
In[19]:= gh = Plot3D[Ha nh[x, y], {x, -1, 0}, {y, 0, 1},
   AxesLabel → {"ξ", "η", "Ha h"}, LabelStyle → Directive[Medium], PlotTheme → "Classic"]
```

Out[19]=

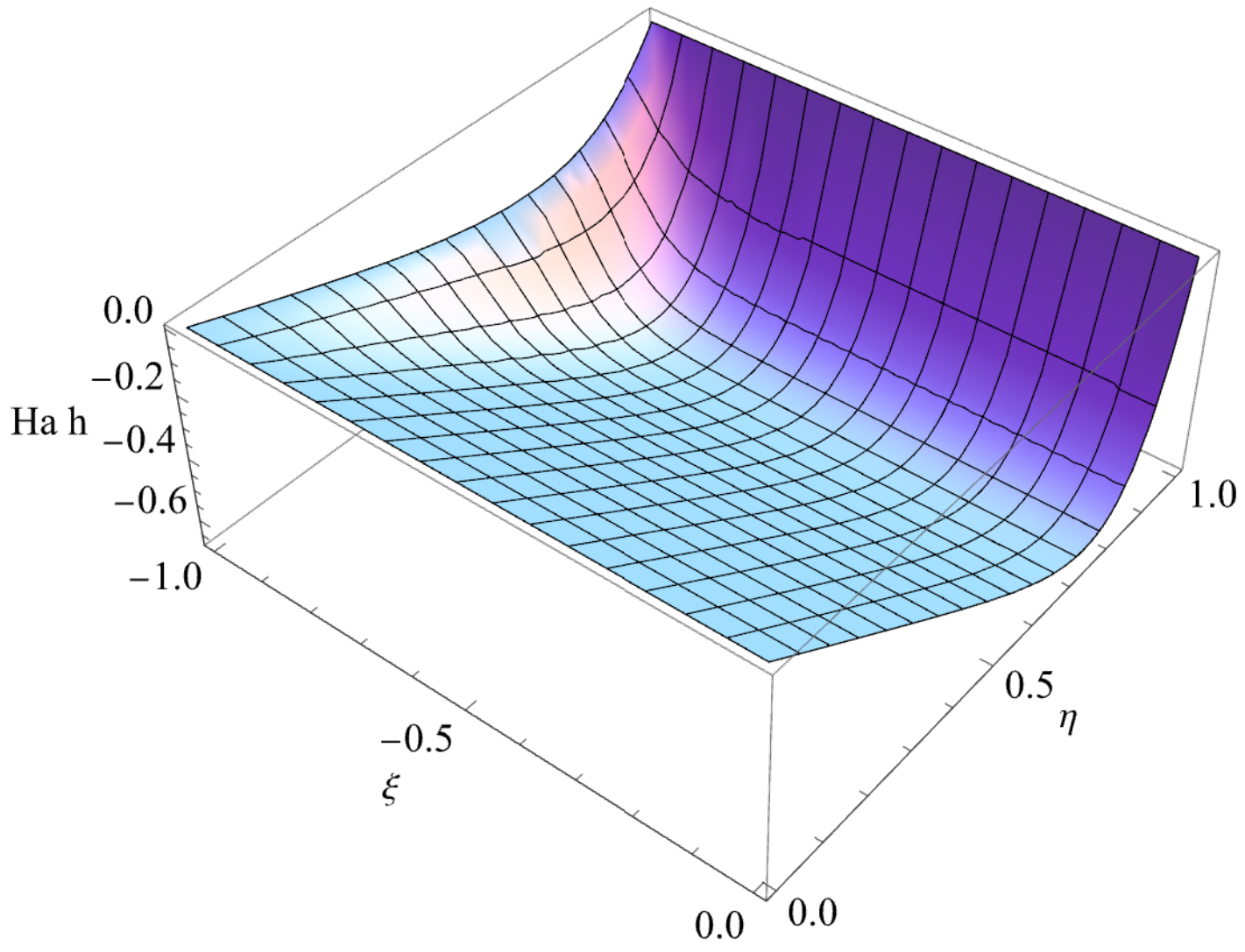

In[20]:=
```
Plot[{Ha nv[0, y], Ha nh[0, y]}, {y, 0, 1}, AxesLabel → {"η"},
 LabelStyle → Directive[Medium], AxesOrigin → {0, 0}, PlotStyle → Thickness[0.01],
 Epilog → {Table[{Red, Circle[{y, Ha vc[Ha, y]}, 0.02]}, {y, 0, 1, 0.1}],
   Table[{Green, Circle[{y, Ha hc[Ha, y]}, 0.02]}, {y, 0, 1, 0.1}], Text["Ha v(0,η)",
    {0.3, 0.9}, Background → White], Text["Ha h(0,η)", {0.6, -0.5}, Background → White]},
 AspectRatio → Automatic, ScalingFunctions → None, PlotStyle → "Classic"]
(*FE solution (solid line); analytical core solution (symbols) *)
```

Out[20]=

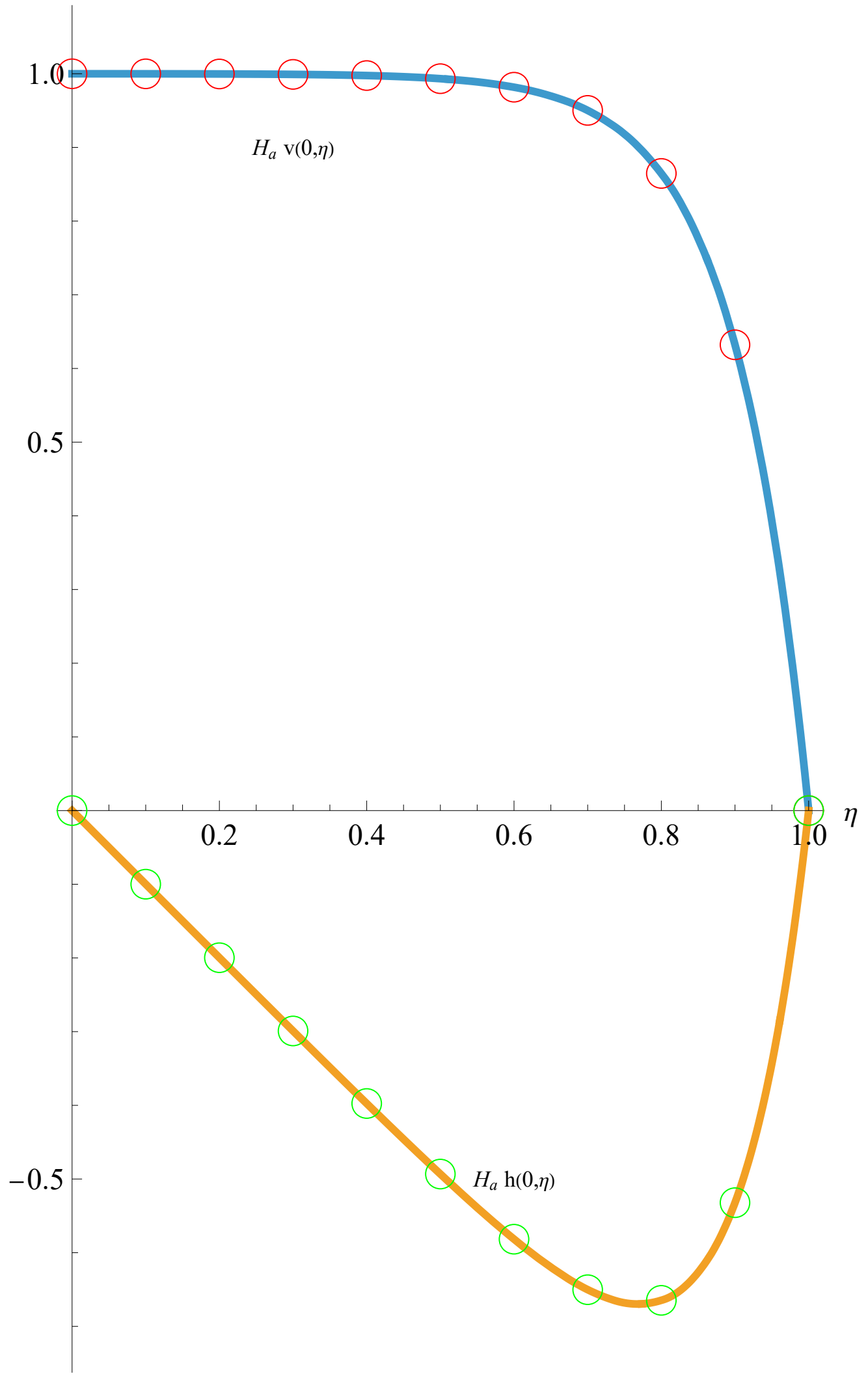

In[21]:=
```
Plot[{Ha nv[x, 0], Ha nh[x, 0.3]}, {x, -1, 0}, AxesLabel → {"ξ"},
 LabelStyle → Directive[Medium], AxesOrigin → {0, 0}, PlotStyle → Thickness[0.01],
 Epilog → {Table[{Red, Circle[{x, Ha vc[Ha, 0]}, 0.02]}, {x, -1, 0, 0.1}],
   Table[{Green, Circle[{x, Ha hc[Ha, 0.3]}, 0.02]}, {x, -1, 0, 0.1}], Text["Ha v(ξ,0)",
    {-0.4, 0.9}, Background → White], Text["Ha h(ξ,0.3)", {-0.5, -0.2}, Background → White]},
 AspectRatio → Automatic, ScalingFunctions → None] (*FE solution (solid line);
analytical core solution (symbols) *)
```

Out[21]=

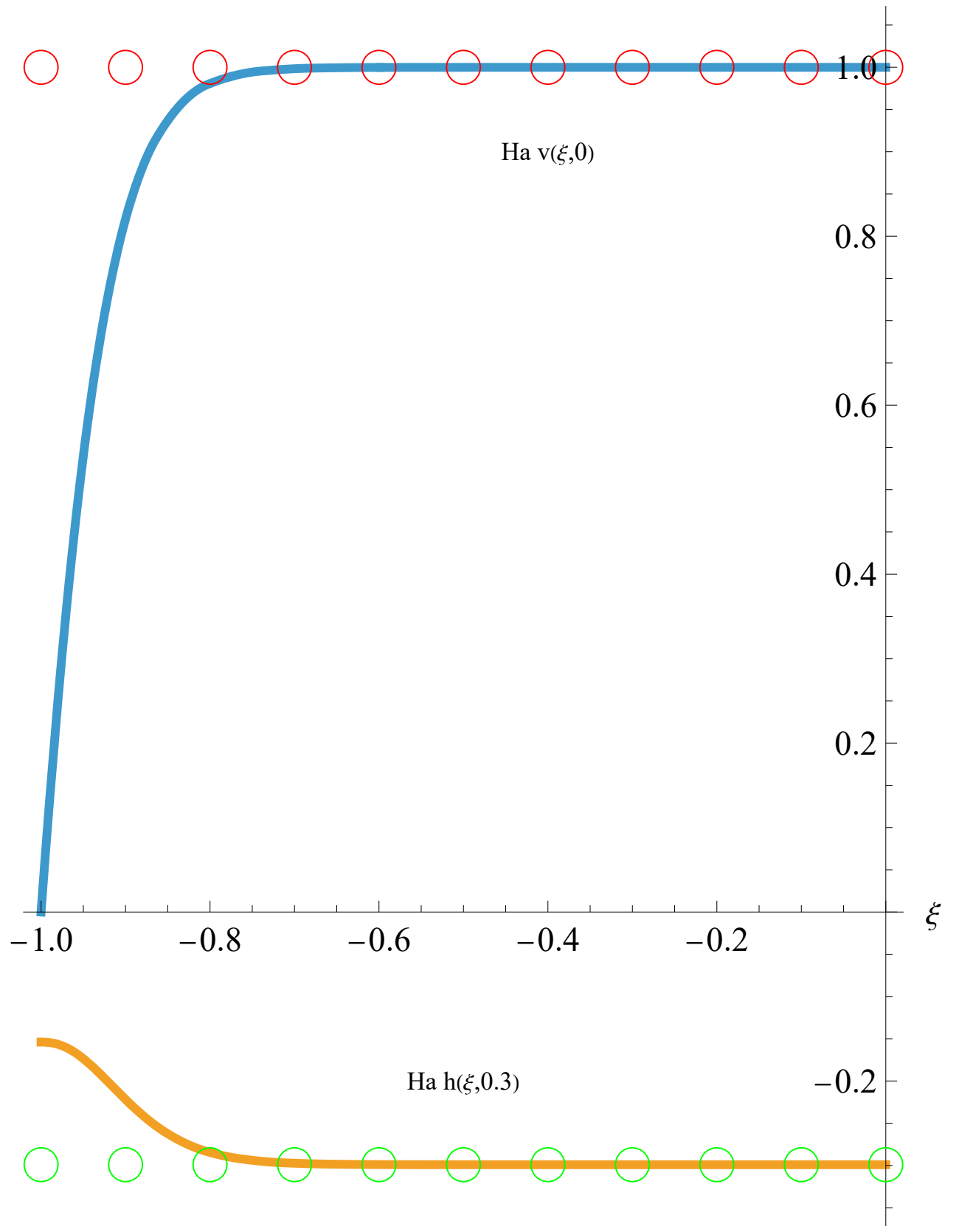

# 5. Potential

$$\frac{\partial\tilde{\varphi}}{\partial\xi} = -\frac{1}{H_a}\left(1+\frac{1}{P}\frac{d\,p}{d\,z}\right)-\left(\frac{\partial h}{\partial\eta}+H_a\,u\right) = -\frac{1}{H_a}\left(\frac{S}{1+S}\right)-\left(\frac{\partial h}{\partial\eta}+H_a\,u\right)$$

$$S = \frac{B_0\,\Delta\,H}{2\,a\,G}$$

In[ ]:=
```
S = 2 / 3;
```

In[ ]:=
```
dphidx[x_, y_, S_] := - (Derivative[0, 1][nh][x, y] + 10 * nv[x, y]) - 1 / 10 (S / (1 + S))
phi10[x_, y_] := NIntegrate[dphidx[xx, y, S], {xx, 0, x}, PrecisionGoal → 6, MaxRecursion → 50]
```

In[•]:=
```
nvphi0 = Plot[phi10[x, 0], {x, -1, 0}, PlotTheme → "Classic",
  AxesLabel → {"ξ", "φ"}, LabelStyle → Directive[Medium]]
```

NIntegrate: Numerical integration converging too slowly; suspect one of the following: singularity, value of the integration is 0, highly oscillatory integrand, or WorkingPrecision too small.

Out[•]=

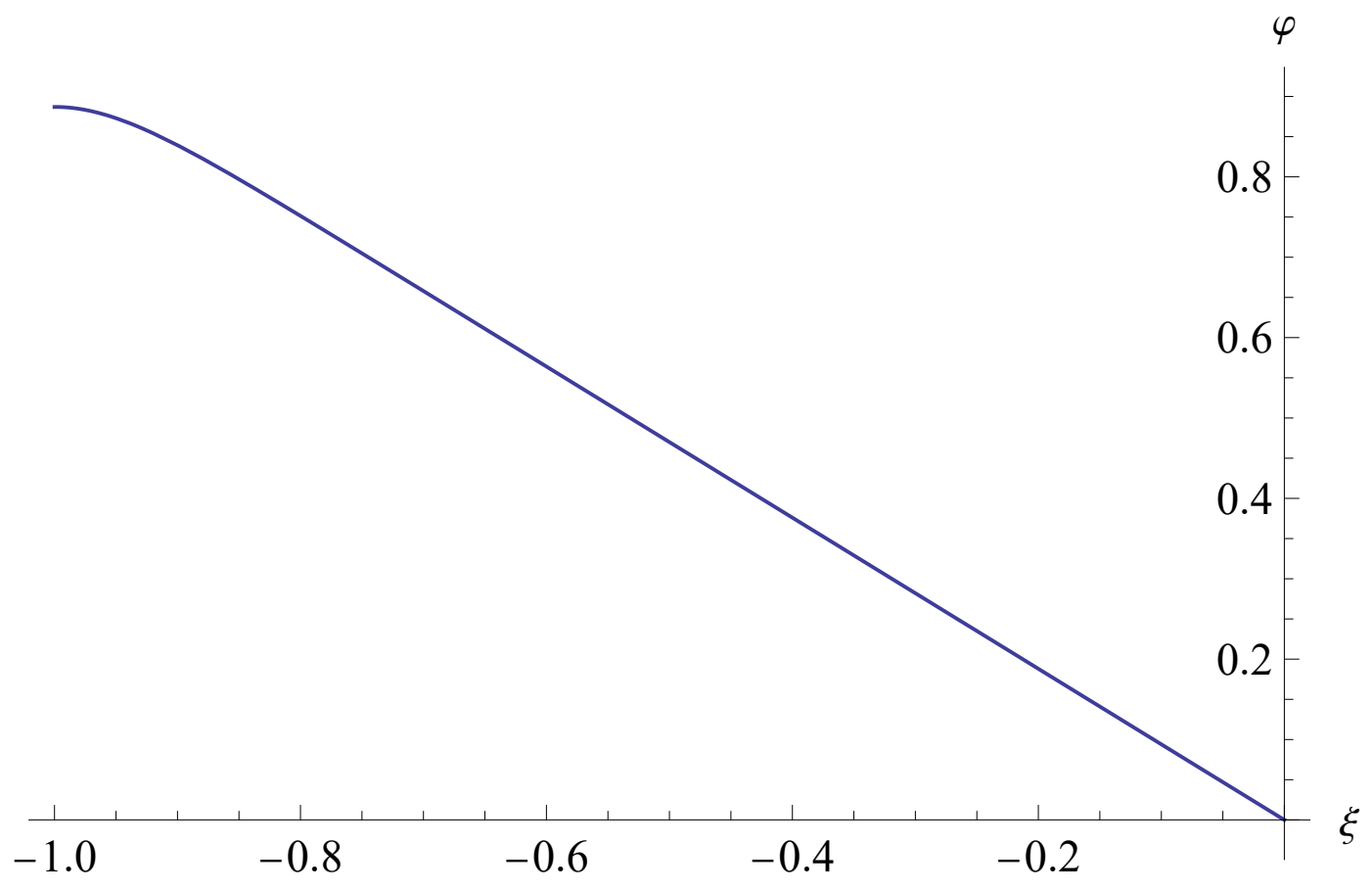

```
nvphi = Plot3D[phi10[x, y], {x, -1, -0.8}, {y, 0, 1},
  PlotTheme → "Classic", AxesLabel → {"ξ", "η", "φ"}, LabelStyle → Directive[Medium]]
```

NIntegrate: Numerical integration converging too slowly; suspect one of the following: singularity, value of the integration is 0, highly oscillatory integrand, or WorkingPrecision too small.

Out[•]=

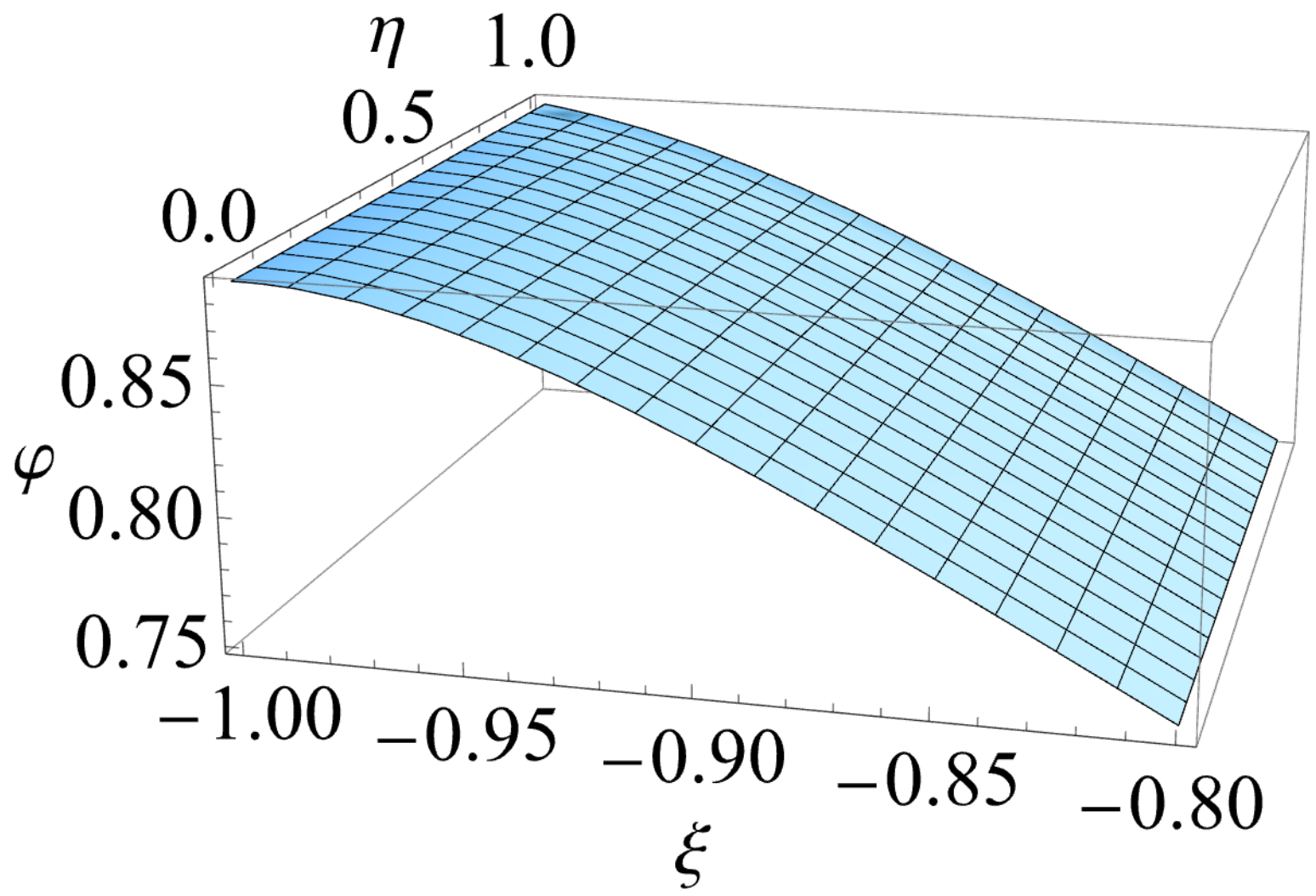

```
Plot comparing core phi (dashed lines) to FEA solutions for Ha=10 (purple)
```

In[ ]:=
```
Plot[{phi10[x, 0], fc[Ha, x, 2/3]}, {x, -1, -0.8},
 PlotStyle → "DarkRainbow", PlotTheme → "Monochrome",
 PlotRange → {{-1, -0.8}, {0.8, 1.0}}, AxesOrigin → {-1, 0.8}, PlotStyle → Thick]
```

NIntegrate: Numerical integration converging too slowly; suspect one of the following: singularity, value of the integration is 0, highly oscillatory integrand, or WorkingPrecision too small.

Out[ ]=

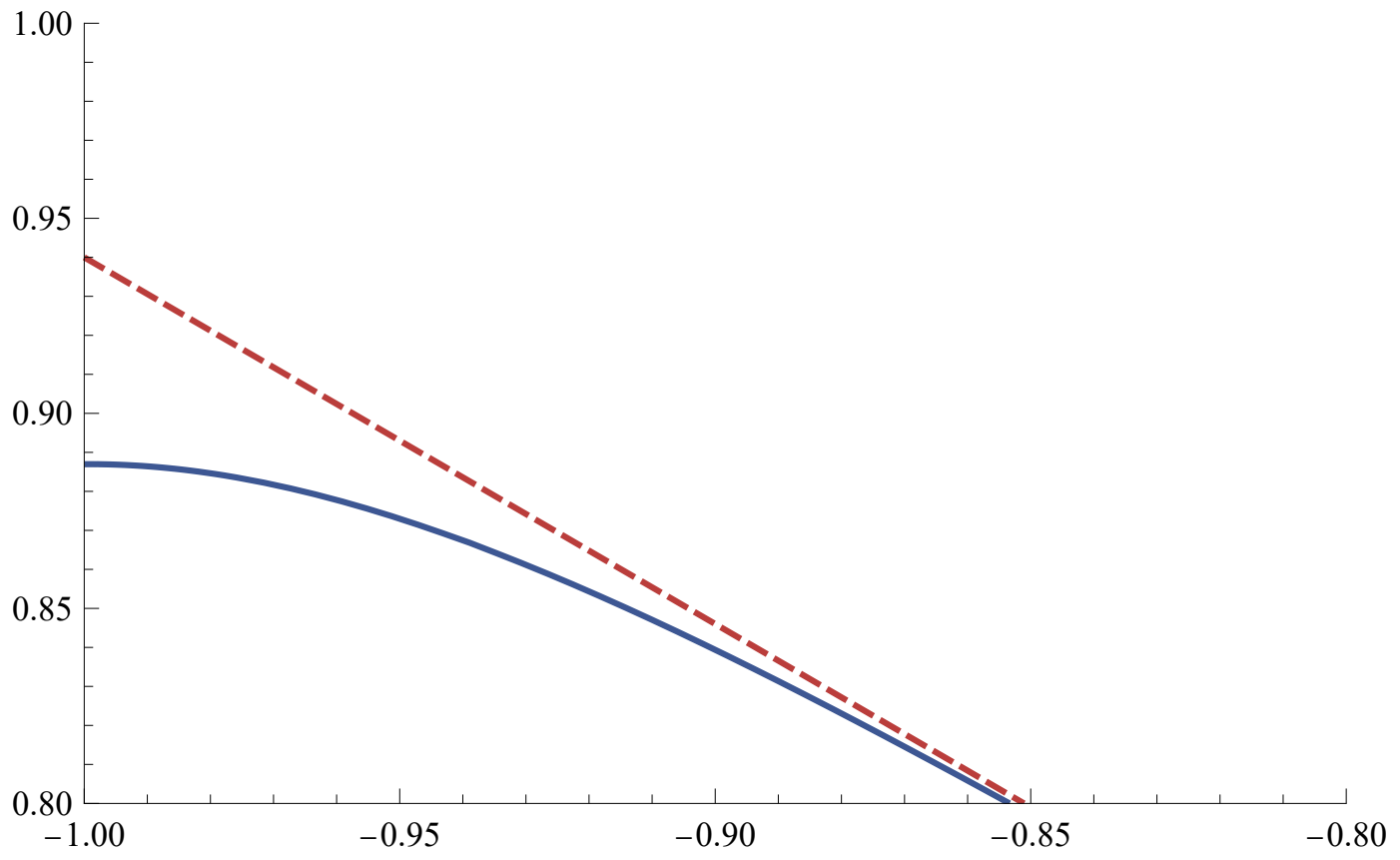

# 6. Boundary Layer Problem

To compute the system’s behavior in the left boundary layer next to *ξ=-1*, we introduce the stretched coordinate $\xi = -1 + \frac{A}{\sqrt{H_a}} \zeta$.

$$\frac{\partial^2 \hat{u}_i}{\partial \varsigma^2} + H_a^{-1} \frac{\partial^2 \hat{u}_i}{\partial \eta^2} + \frac{\partial \hat{h}_i}{\partial \eta} = -1$$

$$\frac{\partial^2 \hat{h}_i}{\partial \varsigma^2} + H_a^{-1} \frac{\partial^2 \hat{h}_i}{\partial \eta^2} + \frac{\partial \hat{u}_i}{\partial \eta} = 0$$

$$\hat{u}_i(0, \eta) = \hat{u}_i(\zeta, 1) = \frac{\partial \hat{u}_i(\zeta,0)}{\partial \eta} = \hat{u}_i(\infty, \eta) - H_a\, \hat{u}_c(\eta) = \hat{h}_i(\zeta, 0) = \hat{h}_i(\zeta, 1) = \frac{\partial \hat{h}_i(0,\eta)}{\partial \zeta} = \hat{h}_i(\infty, \eta) - H_a\, \hat{h}_c(\eta) = 0$$

In[ ]:=
```
lbl = 12 (* the BL thickness *); Ha = 5.;
BLdomain = Rectangle[{0, 0}, {lbl, 1}]; (* define domain and equations*)
refinementRegionBL := ImplicitRegion[y > 0.8, {{xx, 0, lbl}, {y, 0.8, 1}}];
mrfb = With[{rmf = RegionMember[refinementRegionBL]},
   Function[{vertices, area}, Block[{xx, y}, {xx, y} = Mean[vertices];
     If[rmf[{xx, y}], area > 0.00005, area > 0.0025]]]];
be1[Ha_] := D[vb[xx, y], {xx, 2}] + Ha^(-1) D[vb[xx, y], {y, 2}] + D[hb[xx, y], y] + 1

be2[Ha_] := D[hb[xx, y], {xx, 2}] + Ha^(-1) D[hb[xx, y], {y, 2}] + D[vb[xx, y], y]

bbBC$Dirichlet[Ha_] := {vb[xx, 1] == 0, vb[0, y] == 0,
   vb[lbl, y] == Ha * vc[Ha, y], hb[xx, 1] == 0, hb[xx, 0] == 0, hb[lbl, y] == Ha * hc[Ha, y]};
```

In[ ]:=
```
Show[RegionPlot[BLdomain], RegionPlot[refinementRegionBL], AspectRatio → Automatic]
```

Out[ ]=

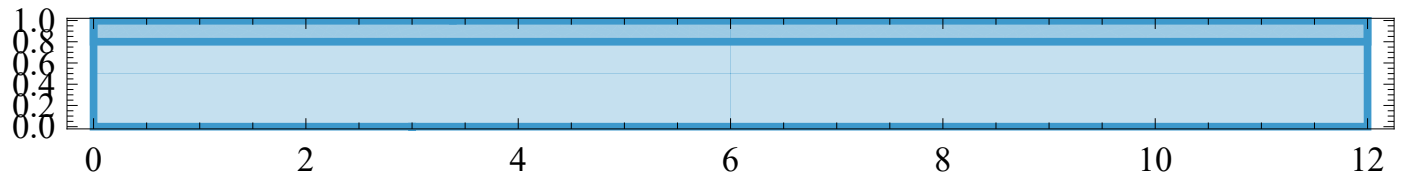

In[∘]:=
```
{bv, bh} = NDSolveValue[
  {be1[Ha] == 0, be2[Ha] == 0, bbBC$Dirichlet[Ha]}, {vb, hb}, Element[{xx, y}, BLdomain],
  Method → {"FiniteElement", "MeshOptions" → {MeshRefinementFunction → mrfb}}]
```

Out[∘]=
```
{InterpolatingFunction[ Domain: {{0., 12.}, {0., 1.}}
                        Output: scalar
                        Data not saved. Save now ],

 InterpolatingFunction[ Domain: {{0., 12.}, {0., 1.}}
                        Output: scalar
                        Data not saved. Save now ]}
```

In[∘]:=
```
Plot[{bv[xx, 0.] / Ha / vc[Ha, 0.], bh[xx, 0.5] / Ha / hc[Ha, 0.5]},
 {xx, 0, lbl}, PlotRange → {{0, lbl}, {0, 1.1}}, PlotStyle → Thickness[0.01]]
```

Out[∘]=

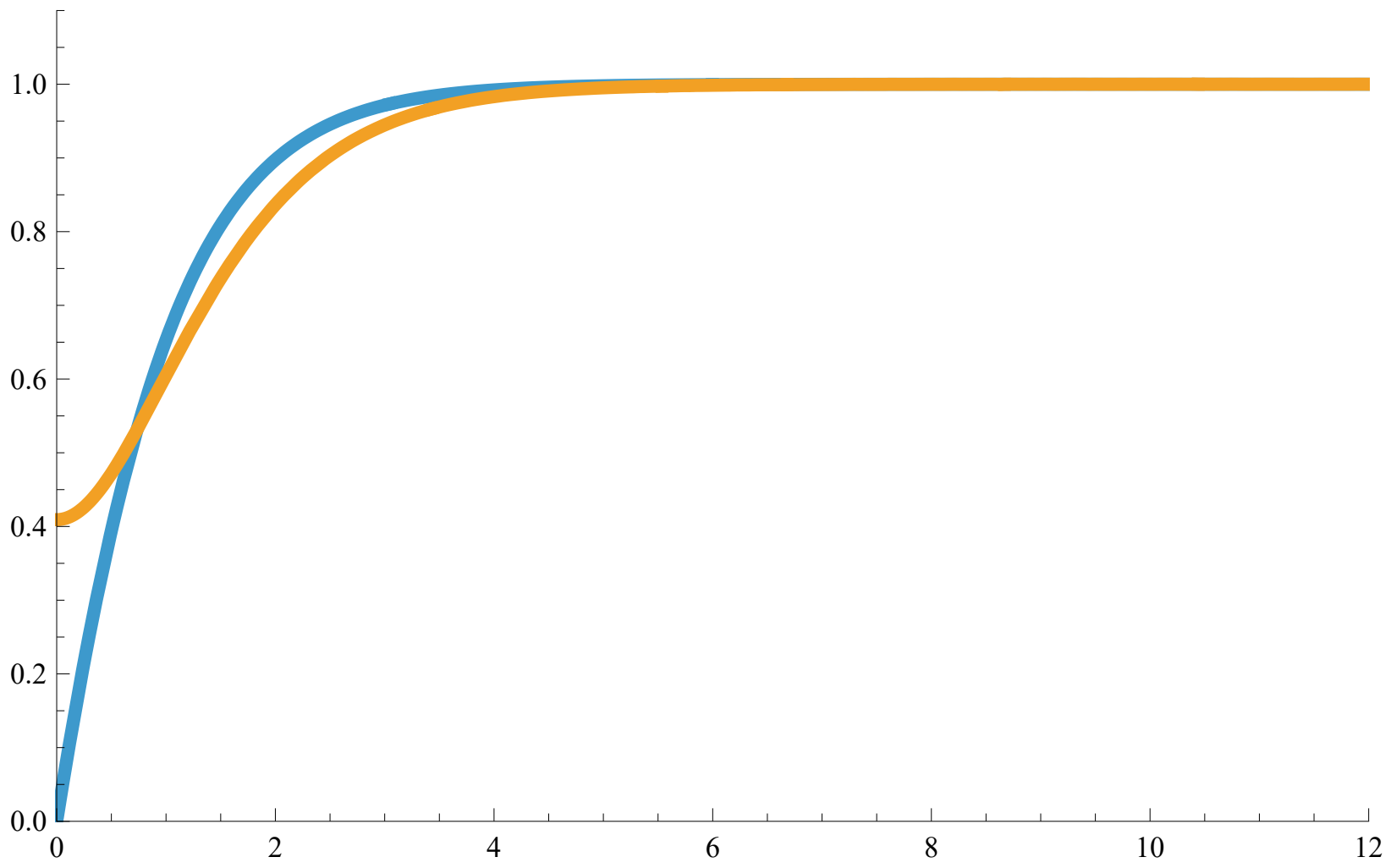

When Ha=5, the velocity and stream function achieve their asymptotic values within 1% at xx~5.

In[∘]:=
```
Plot[{bv[1, y], bh[1, y]}, {y, 0, 1},  PlotStyle → Thickness[0.01]]
```

Out[∘]=

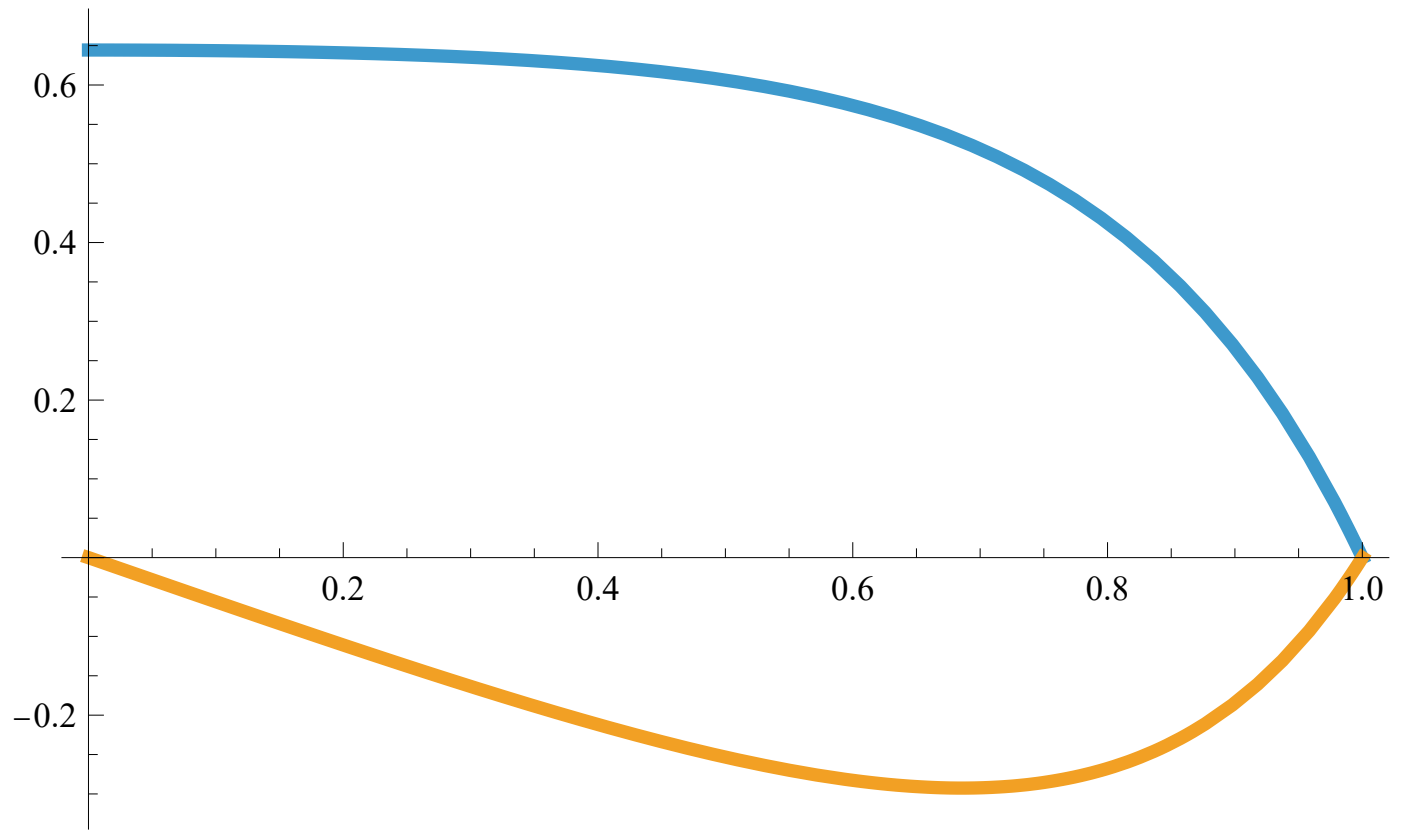

Witness the smooth dependence on y.

# 7. Data on the FE grid in the Boundary Layer

In[∘]:=
```
Show[bv["ElementMesh"]["Wireframe"]] (*Show the finite element mesh *)
```
Out[∘]=

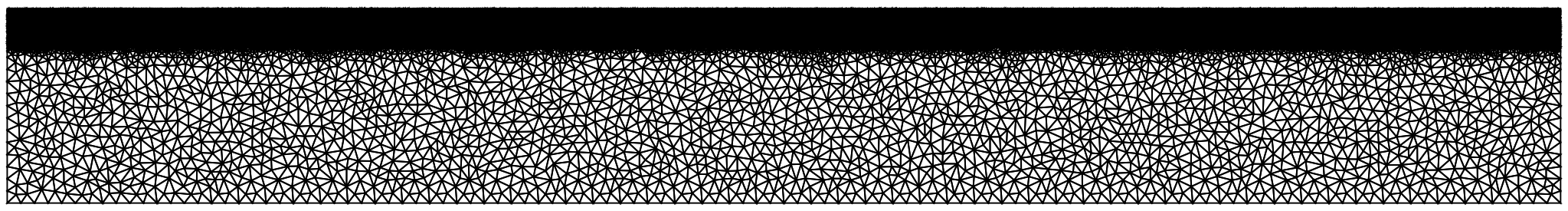

In[∘]:=
```
bv["ElementMesh"] (* number of elements *)
```
Out[∘]=
```
NDSolve`FEM`ElementMesh[{{0., 8.}, {0., 1.}}, {NDSolve`FEM`TriangleElement[<57483>]}]
```
In[∘]:=
```
bv["ElementMesh"]["MeshOrder"] (* verify that Mathematica uses quadratic elements*)
```
Out[∘]=
```
2
```

# 8. How far is far enough?

Since we solve the BL equations numerically, we must define a finite computational domain. We hypothesize that the inner (BL) variables approach their asymptotic core values once lbl exceeds a certain value. The results of the previous section suggest that when Ha>5, the boundary layer variables reach their asymptote value at xx~5. Hence, in most of our calculations,when Ha>5, we use lbl=8.

# 9. The Perils of Neglecting ($H_a^{-1}$) terms in the BL equations

For low and moderate Hartmann numbers, the terms proportional to $H_a^{-1}$ cannot be ignored. Although in the limit $H_a$--> infinity, the .$H_a^{-1}$ terms could be formally neglected, they provide dissipation to dampen spurious numerical (Gibbs) oscillations. Below, we compare results obtained from integrating the BL equations in the presence and the absence of $H_a^{-1}$ terms.

In the following BL equations the $H_a^{-1}$ terms are absent. The solutions still depend on the $H_a$ number through the far field matching conditions.

Without the $H_a^{-1}$ term, the boundary layer equations become:

In[∘]:=
```
be1h[Ha_] := D[vb[xx, y], {xx, 2}] + D[hb[xx, y], y] + 1
be2h[Ha_] := D[hb[xx, y], {xx, 2}] + D[vb[xx, y], y]
```
In[∘]:=
```
{bvh, bhh} =
  NDSolveValue[{be1h[Ha] == 0, {be2h[Ha] == 0}, bbBC$Dirichlet[Ha]}, {vb, hb}, Element[{xx, y},
    BLdomain],
   Method → {"FiniteElement", "MeshOptions" → {MeshRefinementFunction → mrfb}}]
```
Out[∘]=
```
{InterpolatingFunction[ Domain: {{0., 12.}, {0., 1.}} Output: scalar ],
 InterpolatingFunction[ Domain: {{0., 12.}, {0., 1.}} Output: scalar ]}
```
Data not saved. Save now

Data not saved. Save now

In[•]:=
```
Plot[{bvh[0.5, y], bhh[0.5, y]}, {y, 0.9, 1}, PlotStyle → Thickness[0.01]]
```

Out[•]=

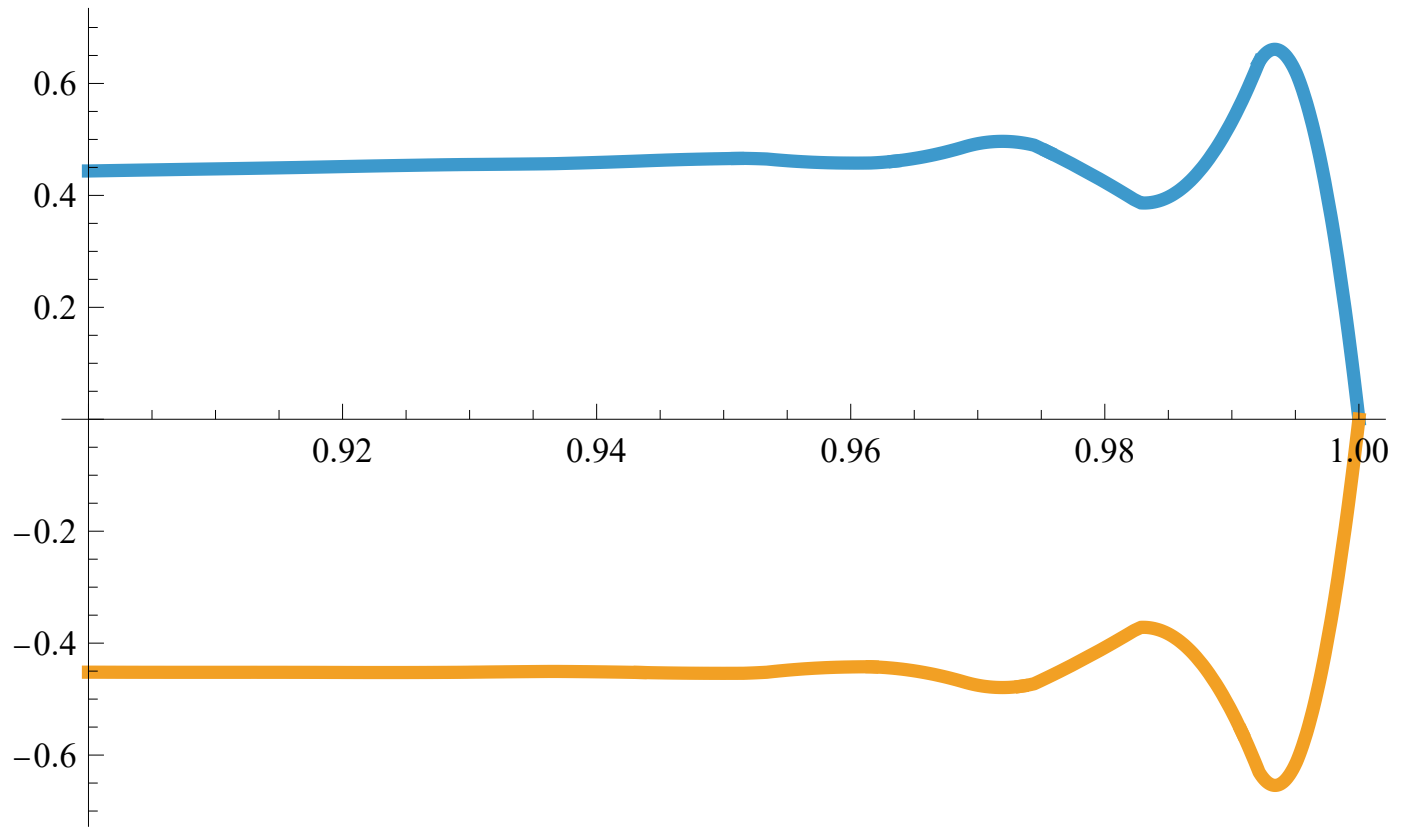

In the absence of the dissipate terms $H_a^{-1}\frac{\partial^2 \hat{u}_i}{\partial \eta^2}$ and $H_a^{-1}\frac{\partial^2 \hat{h}_i}{\partial \eta^2}$, the solutions exhibit numerical, non-physical oscillations.

## 11. How well does exponential behavior $v_C$[Ha, y] $(1 - e^{-xx})$ approximates the numerical BL solution for the velocity

The relative difference between the exponential approximation $v_c(y)(1-e^{-x\,x})$ and the boundary layer finite element solution as a function of xx when Ha=5, 10, and 30. The error is smaller than 3%.

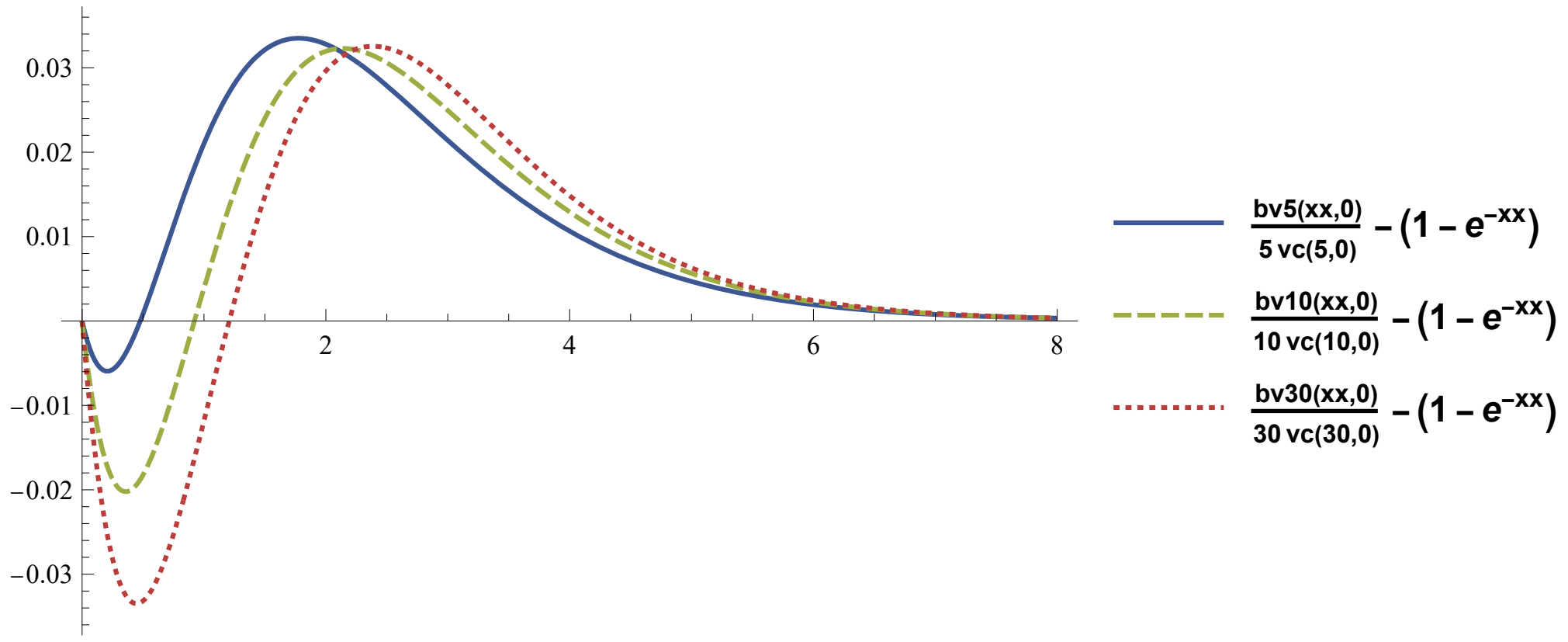

## 12. The deviation of the electrode potential $\Delta\,\tilde{\varphi}^*$ from extrapolated core value

$$\Delta\,\tilde{\varphi}_{i,L} = \frac{A}{H_a^{\frac{1}{2}}}\,\Delta\,\tilde{\varphi}^*{}_{i,L} = \frac{A}{H_a^{\frac{1}{2}}}\int_{\zeta_L}^{0}\left(\left(H_a^{-1}\frac{\partial \hat{h}_c}{\partial \eta} + \hat{u}_c\right) - \left(H_a^{-1}\frac{\partial \hat{h}_i}{\partial \eta} + \hat{u}_i\right)\right) d\,l\,.$$

When $|\zeta_L| >$ lbl, the BL (inner solution denoted with subscript i) matches the core solution (denoted with subscript c), The integrand equals to zero. $\hat{h}_c = H_a\,h_c$ and $\hat{u}_c = H_a\,u_c$

In[ ]:=
```
D[hc[Ha, nu], nu] + Ha * vc[Ha, nu] // FullSimplify
```
Out[ ]=

$$-\frac{1}{Ha} + \mathrm{Coth}[Ha]$$

We carry out our calculations without the prefactor $\frac{A}{\sqrt{H_a}}$ and find $\triangle \tilde{\varphi}^*_{i,L}$ at the electrode's surface

In[ ]:=
```
dphidx[xx_, y_, Ha_] := (-1/Ha + Coth[Ha] - (Derivative[0, 1][bh][xx, y] / Ha + bv[xx, y]))
```
In[ ]:=
```
deltaf[l_, y_, Ha_] :=
  NIntegrate[dphidx[xx, y, Ha], {xx, lbl, l}, PrecisionGoal → 6, MaxRecursion → 50]
deltaf[0, 0, 10] (*difference in potential between
   core and bl solution at the left electrode Δφ̃*_{i,L} at y=0 and Ha=10*)
```
Out[ ]=

-0.83899

# 13. Whole region FE-computed and Approximate Flow Rates

In[ ]:=
```
Integrate[2 * vc[Ha, y] (1 - Exp[-((x + 1) Sqrt[Ha]) / A]), {y, -1, 1}, {x, -1, 0}] //
  Simplify(*approximate flow rate*)
```
Out[ ]=

$$\frac{4\, e^{-\frac{\sqrt{Ha}}{A}} \left(A + e^{\frac{\sqrt{Ha}}{A}} \left(-A + \sqrt{Ha}\right)\right) (-1 + Ha\, \mathrm{Coth}[Ha])}{Ha^{5/2}}$$

In[ ]:=
```
Qapprox[A_, Ha_] := 4 (1 - A/Sqrt[Ha] (1 - E^(-Sqrt[Ha]/A))) (-1 + Ha Coth[Ha]) / Ha^2
```
$$\mathrm{Qapprox}[A_, Ha_] := \frac{4 \left(1 - \frac{A}{\sqrt{Ha}} \left(1 - e^{-\frac{\sqrt{Ha}}{A}}\right)\right) (-1 + Ha\, \mathrm{Coth}[Ha])}{Ha^2}$$

In[ ]:=
```
Qapprox[0.2, 10]
```
Out[ ]=

0.337232

FE whole domain solution for the flow rate. Below, Ha=10 and A= 0.2

In[ ]:=
```
NIntegrate[4 * nv[x, y], {y, 0, 1}, {x, -1, 0}] (*FE flow rate*)
```
Out[ ]=

0.338776

# 14. Current Density

The current stream function $H_z = \frac{H_1 + H_2}{2} + \frac{\Delta H}{2} \eta + a^2 \left(\frac{B_0 \Delta H}{2 a} + G\right) \sqrt{\frac{\sigma}{\mu}}\, h(\xi, \eta)$

Let $H_1 = \frac{\Delta H}{2}$, and $H_2 = -\frac{\Delta H}{2}$

$H_z = \frac{\triangle H}{2} \eta + a^2 \left(\frac{B_0 \triangle H}{2 a} + G\right) \sqrt{\frac{\sigma}{\mu}}\, h(\xi, \eta)$ = S \eta + H_a h[\xi, \eta] (S + 1)

The electric current is maximized when the device is short circuited ($\varphi = 0$). Based on our Onsager relations, the max current is: $\Delta H_{max} = K_{EH}\, G = -K_{HE}\, G$.

$K_{HE}$ (Supplemental S2) :

In[ ]:=
```
KHE[A_, Ha_] :=
 (2 a^2 (1 + A (-1 + ⅇ^(-√Ha/A)) / √Ha) √σ (-1 + Ha Coth[Ha])) / (Ha^(3/2) √μ (A (c + Ha^β α) + √Ha Coth[Ha]))
```

For concreteness, we use thermochemical properties of liquid Gallium and specific conduit dimensions. [Sindu: please add units. You did not specify B0]

```
μ := 2.4 * 10^(-3)
σ := +2.3 * 10^6
a := .001
α := 1
β := -1
c := -0.956
B0 := Ha / a * Sqrt[μ / σ]  (*T*)
```

At maximum current, $S_m = -\frac{K_{HE} * G B_0}{2 a G}$,

```
Sm[A_, Ha_] := -KHE[A, Ha] * B0 / (2 a)
```

The x-component of current density $J_x = S + H_a \frac{\partial h}{\partial \eta} (S + 1)$

$J_y = -H_a \frac{\partial h}{\partial \xi} (S + 1)$

When Ha=10 and A=0.2:

In[ ]:=
```
Hz[x_, y_, S_] := S * y + S 10 * nh[x, y] + 10 * nh[x, y]
```

In[ ]:=
```
Jx[x_, y_, S_] := S + 10 * (S + 1) Derivative[0, 1][nh][x, y]
Jy[x_, y_, S_] := -10 * (S + 1) Derivative[1, 0][nh][x, y]
```

In[ ]:=
```
contourplot = ContourPlot[Hz[x, y, Sm[0.2, 10]], {x, -1, 0}, {y, 0, 1}, PlotLegends → Automatic,
    ContourLabels → True, FrameLabel → {"ξ", "η"}, BaseStyle → {FontSize → 21, FontColor → White},
    FrameStyle → Directive[21], ColorFunction → (ColorData["Rainbow"][Rescale[#, {-0.8, 0}]] &)
, PlotRange → All, ColorFunctionScaling → False];
vectorfieldsmax = VectorPlot[{Jx[x, y, Sm[0.2, 10]], Jy[x, y, Sm[0.2, 10]]}, {x, -1, 0},
   {y, 0, 1}, VectorPoints → Coarse, VectorSizes → 1.3, VectorStyle → Arrowheads[0.05],
   VectorColorFunction → "BrightBands", PlotLegends → Automatic, FrameLabel → {"ξ", "η"}]
Show[{contourplot, vectorfieldsmax}]
```

Out[ ]=

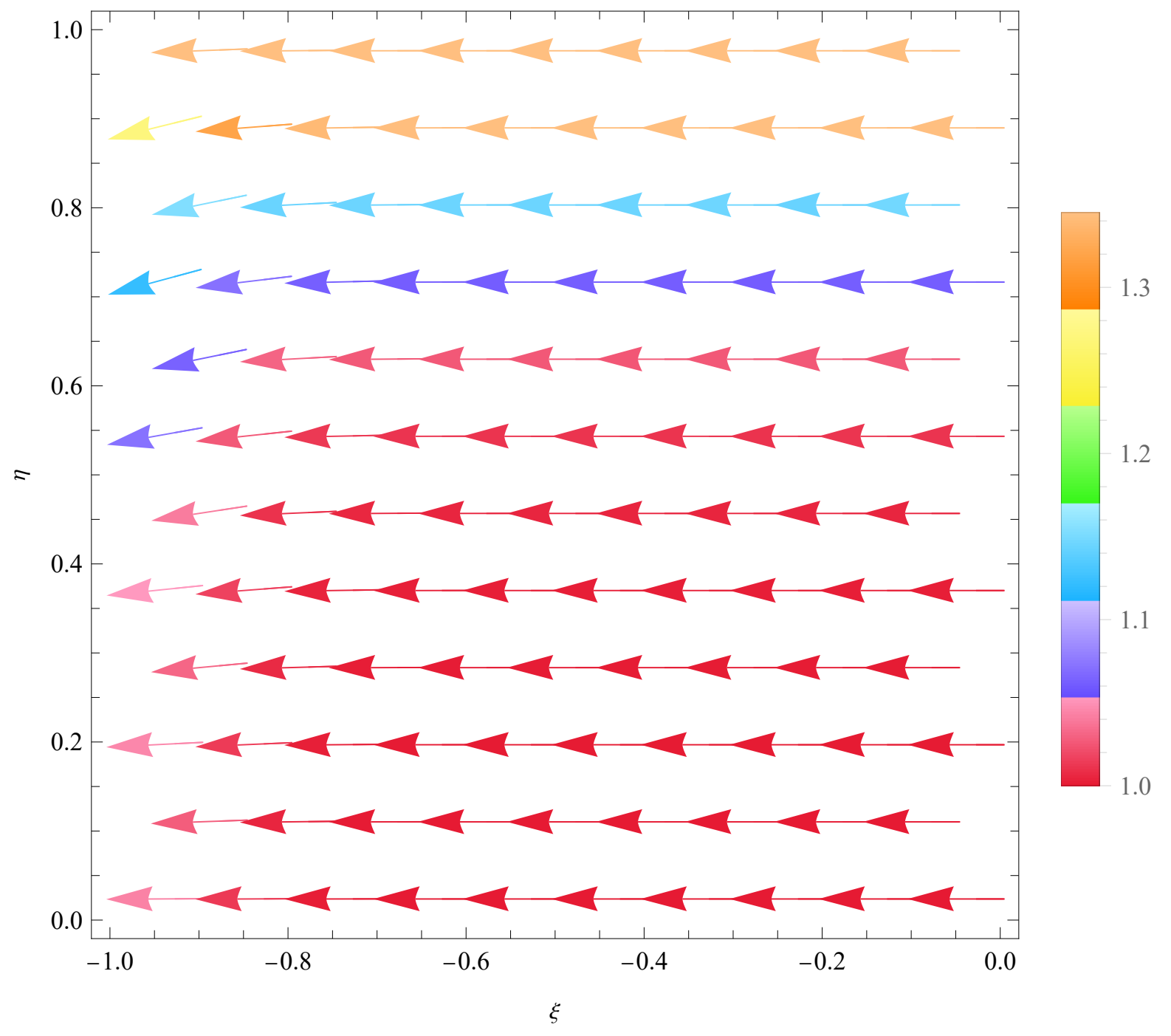

Out[ ]=

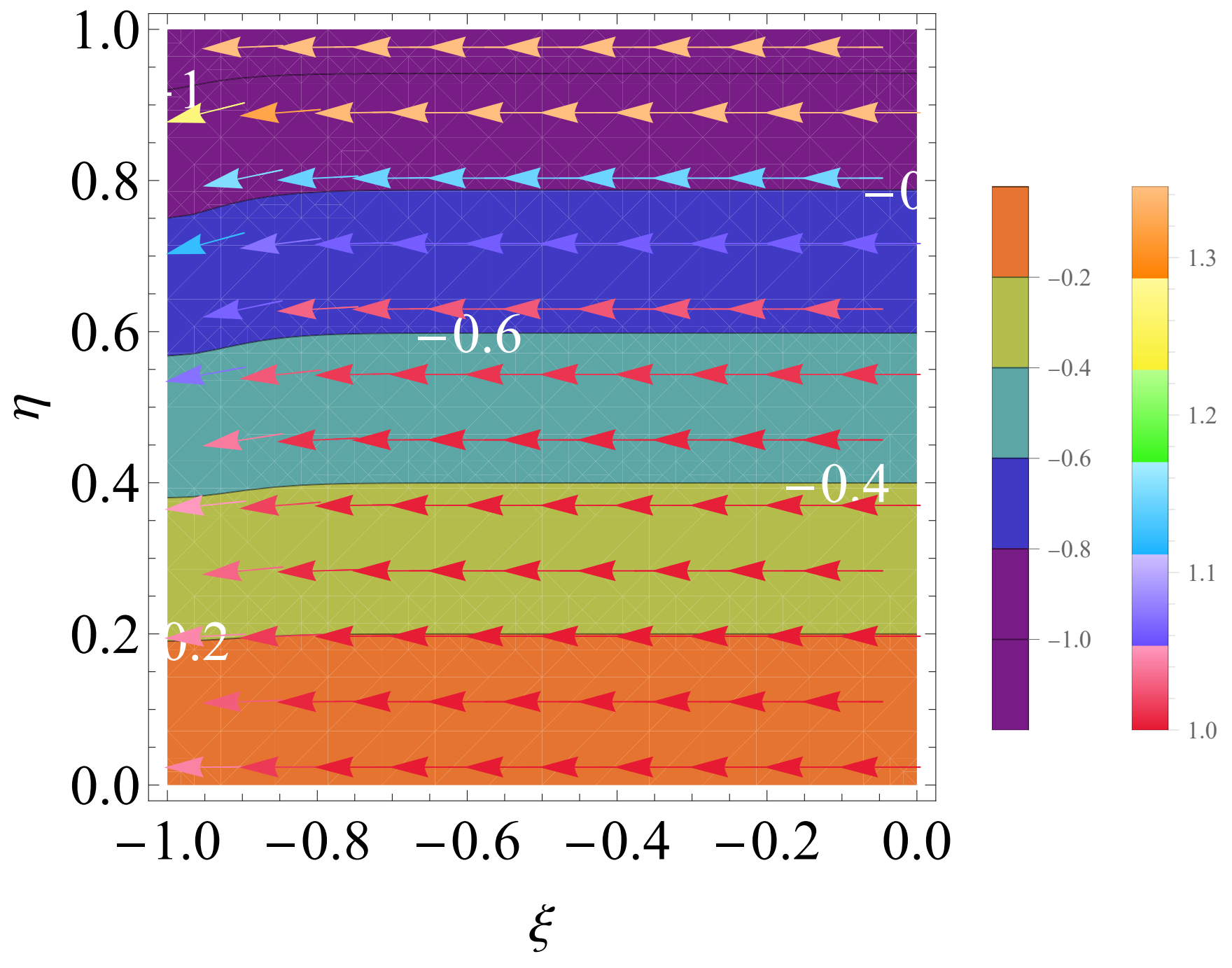

# Supplement S2: Finite Element Simulation of MHD Flow in a Long Conduit with a Rectangular Cross-Section

Supplement to: Sindu Shanmugadas & Haim H. Bau*, 2025, "Onsager Coefficients for Liquid Metal Flow in a Conduit under a Magnetic Field,"

Dept. Mechanical Engineering and Applied Mechanics, University of Pennsylvania, Philadelphia, PA 1904-6315
*Corresponding author: *bau@seas.upenn.edu*

## Onsager Coefficients & Efficiency Calculations

*Onsager coefficients are provided in a format suitable for calculations with Mathematica. These coefficients are then used to estimate performance characteristics of various MHD machines.*

$\Psi$ - the difference between the (computed) left electrode potential and the electrode potential extrapolated from the core solution.

$\Delta\tilde{\Phi}_L = \frac{A}{\sqrt{H_a}}(\alpha H_a{}^{\beta} + c)$ - empirical correlation

$\tilde{\Phi}_c(\xi) = \left(-Coth(H a) - \frac{1}{H_a}\left(\frac{1}{P}\frac{dp}{dz}\right)\right)\xi$ - core potential as a function of position

$\Psi = \frac{2abP}{\sqrt{\sigma\mu}}\left(\frac{A}{\sqrt{H_a}}(\alpha H_a{}^{\beta} + c) + Coth(H a) + \frac{1}{H_a}\left(\frac{1}{P}\frac{dp}{dz}\right)\right)$ - the potential difference between the left (x=-1) and right (x=1) electrodes

$DH = \Delta H = H_1 - H_2$ - total current between the electrodes per conduit's unit length.

The pressure gradient $G = \frac{-dp}{dz}$ = -dpdz.

For convenience, in Mathematica expressions, we replaced $\xi$ and $\eta$ with "x" and "y," respectively.

## A. Onsager Coefficients

### A1. Current ($\Delta H$) Expression

We start with the $\Psi$ expression:

In[110]:=

```
P[dpdz_] := -dpdz + DH*B0/(2 a)
Psi[Ha_, dpdz_, A_, DH_] :=
 (2*a*b*P[dpdz]/(√(σ*μ))) ((A/√Ha) (α*Ha^β + c) + Coth[Ha] + 1/Ha (dpdz/P[dpdz]))
Psi[Ha, dpdz, A, DH]
```

Out[112]=

$$\frac{2ab\left(\frac{B0\,DH}{2a} - dpdz\right)\left(\frac{dpdz}{\left(\frac{B0\,DH}{2a} - dpdz\right)Ha} + \frac{A\left(-0.956 + \frac{1}{Ha}\right)}{\sqrt{Ha}} + Coth[Ha]\right)}{\sqrt{\mu\,\sigma}}$$

The above expression for Psi is simplified and the coefficients of $G$ and $\Psi$ are identified;

$$\Psi = a_{11}\left(-\frac{dP}{dz}\right) + a_{12}\, DH$$

In[113]:=

$$a11[A_, Ha_] := \frac{2\, a\, b\left(-\frac{1}{Ha} + \frac{A\,(c + Ha^{\beta}\alpha)}{\sqrt{Ha}} + \text{Coth}[Ha]\right)}{\sqrt{\mu\,\sigma}}$$

$$a12[A_, Ha_] := \frac{b\, B0\, \left(A\left(c + Ha^{\beta}\alpha\right) + \sqrt{Ha}\, \text{Coth}[Ha]\right)}{\sqrt{Ha}\,\sqrt{\mu\,\sigma}}$$

In[115]:=

```
Psi[Ha, dpdz, A, DH] - (DH * a12[A, Ha] + (-dpdz) * a11[A, Ha]) // FullSimplify
 (*verify that the coefficient form is equal to original expression*)
```

Out[115]=

```
0.
```

The expressions are rearranged to isolate DH.

$$DH = \frac{1}{a_{12}}\psi + \frac{-a_{11}}{a_{12}}\left(-\frac{dP}{dz}\right)$$

$$DH = K_{EE}\,\psi + K_{EH}\left(-\frac{dP}{dz}\right)$$

In[116]:=

$$KEE[A_, Ha_] := \frac{A\,\sigma}{\sqrt{Ha}\,\left(A\left(c + Ha^{\beta}\alpha\right) + \sqrt{Ha}\,\text{Coth}[Ha]\right)}$$

In[117]:=

$$KEH[A_, Ha_] := -\left(\left(2\, a\verb|^|2\,\sqrt{\sigma/\mu}\,\left(-1 + A\, c\,\sqrt{Ha} + A\, Ha^{\frac{1}{2}+\beta}\alpha + Ha\,\text{Coth}[Ha]\right)\right)\Big/\left(Ha\verb|^|(3/2)\,\left(A\left(c + Ha^{\beta}\alpha\right) + \sqrt{Ha}\,\text{Coth}[Ha]\right)\right)\right)$$

In[118]:=

```
DH[Ha_, A_, G_, ψ_] := KEE[A, Ha] * ψ + KEH[A, Ha] * (G) (*DH expressed with Onsager Coefficients*)
```

## A2. Flow Rate ($Q$) expression

The approximate velocity field that accounts for the velocity deficit in the boundary layer next to x=-1:

$$u(\xi, \eta) \approx u_c(\eta)\left(1 - e^{\left(-\frac{\sqrt{Ha}}{A}(1+\xi)\right)}\right)$$

In[119]:=

```
uc[y_] := 1 / Ha * Coth[Ha] (1 - Cosh[Ha * y] / Cosh[Ha]);

Integrate[2 uc[y] (1 - Exp[-(x + 1) Sqrt[Ha] / A]), {x, -1, 0}, {y, -1, 1}] // FullSimplify
 (* integral of the velocity field over the dimensionless cross-section *)
```

Out[120]=

$$\frac{4\left(1 + \frac{A\left(-1 + e^{-\frac{\sqrt{Ha}}{A}}\right)}{\sqrt{Ha}}\right)(-1 + Ha\,\text{Coth}[Ha])}{Ha^2}$$

In[121]:=

$$\text{Q[Ha, dpdz, A, DH]} := \frac{4\left(1+\frac{A\left(-1+e^{-\frac{\sqrt{Ha}}{A}}\right)}{\sqrt{Ha}}\right)(-1+\text{Ha Coth[Ha]})}{Ha^2} * a^{\wedge}3 * b * P[dpdz]/\mu$$

The above expression for the dimensional flow-rate is rewritten as

$Q = a_{21}\,\Delta H + a_{22}\left(-\frac{d\,p}{d\,z}\right)$. Below, we replace $\Delta H$ with the expression from section A1.

In[122]:=

$$\text{a21[A_, Ha_]} := \frac{1}{Ha^2\,\mu}\,2\,a^2\,b\,B0\left(1+\frac{A\left(-1+e^{-\frac{\sqrt{Ha}}{A}}\right)}{\sqrt{Ha}}\right)(-1+\text{Ha Coth[Ha]})$$

$$\text{a22[A_, Ha_]} := \frac{1}{Ha^2\,\mu}\,4\,a^3\,b\left(1+\frac{A\left(-1+e^{-\frac{\sqrt{Ha}}{A}}\right)}{\sqrt{Ha}}\right)(-1+\text{Ha Coth[Ha]})$$

In[124]:=

```
Q[Ha, dpdz, A, DH] - (DH * a21[A, Ha] + (-dpdz) * a22[A, Ha]) // FullSimplify
 (*verify that the coefficient form is equal to original expression*)
```

Out[124]=

```
0
```

$Q = K_{H\,E}\,\Psi + K_{H\,H}\left(-\frac{d\,p}{d\,z}\right)$

The simplified coefficients are below:

In[125]:=

$$\text{KHE[A_, Ha_]} := \left(2\,a^{\wedge}2\left(1+\frac{A\left(-1+e^{-\frac{\sqrt{Ha}}{A}}\right)}{\sqrt{Ha}}\right)\sqrt{\sigma}\,(-1+\text{Ha Coth[Ha]})\right)\Big/\left(Ha^{3/2}\,\sqrt{\mu}\,\left(A\left(c+Ha^{\beta}\,\alpha\right)+\sqrt{Ha}\,\text{Coth[Ha]}\right)\right)$$

$$\text{KHH[A_, Ha_]} := \left(4\,a^4\left(1+\frac{A\left(-1+e^{-\frac{\sqrt{Ha}}{A}}\right)}{\sqrt{Ha}}\right)(-1+\text{Ha Coth[Ha]})\right)\Big/\left(Ha^{5/2}\,A\,\mu\,\left(A\left(c+Ha^{\beta}\,\alpha\right)+\sqrt{Ha}\,\text{Coth[Ha]}\right)\right)$$

In[127]:=

```
Q[Ha_, A_, G_, ψ_] :=
 KHE[A, Ha] * ψ + KHH[A, Ha] * G (*The flow rate Q expressed with Onsager Coefficients*)
```

## A3. Onsager Coefficient Matrix

$$\begin{pmatrix} K_{EE} & K_{EH} \\ K_{HE} & K_{HH} \end{pmatrix}\begin{pmatrix} \Psi \\ G \end{pmatrix}$$

In[128]:=

```
onsmat := {{KEE[A, Ha], KEH[A, Ha]}, {KHE[A, Ha], KHH[A, Ha]}}
```

In[20]:= `onsmat // MatrixForm`

Out[20]//MatrixForm=

$$\begin{pmatrix} \frac{A\,\sigma}{\sqrt{Ha}\,\left(A\left(c+Ha^{\beta}\alpha\right)+\sqrt{Ha}\,Coth[Ha]\right)} & -\frac{2\,a^2\sqrt{\frac{\sigma}{\mu}}\left(-1+A\,c\,\sqrt{Ha}+A\,Ha^{\frac{1}{2}+\beta}\alpha+Ha\,Coth[Ha]\right)}{Ha^{3/2}\left(A\left(c+Ha^{\beta}\alpha\right)+\sqrt{Ha}\,Coth[Ha]\right)} \\ \frac{2\,a^2\left(1+\frac{A\left(-1+e^{-\frac{\sqrt{Ha}}{A}}\right)}{\sqrt{Ha}}\right)\sqrt{\sigma}\,(-1+Ha\,Coth[Ha])}{Ha^{3/2}\sqrt{\mu}\left(A\left(c+Ha^{\beta}\alpha\right)+\sqrt{Ha}\,Coth[Ha]\right)} & \frac{4\,a^4\left(1+\frac{A\left(-1+e^{-\frac{\sqrt{Ha}}{A}}\right)}{\sqrt{Ha}}\right)(-1+Ha\,Coth[Ha])}{A\,Ha^{5/2}\,\mu\left(A\left(c+Ha^{\beta}\alpha\right)+\sqrt{Ha}\,Coth[Ha]\right)} \end{pmatrix}$$

## A4. Onsager-Casimir Reciprocity

Since we use approximate expressions for the velocity and potential in the boundary layers next to the electrodes, we cannot demonstrate reciprocity symbolically except in limiting cases (see manuscript). Instead, we show numerically that $K_{EH} + K_{HE} \approx 0$.

To verify that the O-C reciprocity is approximately satisfied, we compute (KEH+KHE)/(KEH-KHE) as a function of $H_a$ and $A$.

In[21]:= `(KEH[A, Ha] + KHE[A, Ha]) / (KEH[A, Ha] - KHE[A, Ha]) // FullSimplify`

Out[21]=

$$-\frac{\frac{\left(1+\frac{A\left(-1+e^{-\frac{\sqrt{Ha}}{A}}\right)}{\sqrt{Ha}}\right)\sqrt{\sigma}\,(-1+Ha\,Coth[Ha])}{\sqrt{\mu}}-\sqrt{\frac{\sigma}{\mu}}\left(-1+A\,c\,\sqrt{Ha}+A\,Ha^{\frac{1}{2}+\beta}\alpha+Ha\,Coth[Ha]\right)}{\frac{\left(1+\frac{A\left(-1+e^{-\frac{\sqrt{Ha}}{A}}\right)}{\sqrt{Ha}}\right)\sqrt{\sigma}\,(-1+Ha\,Coth[Ha])}{\sqrt{\mu}}+\sqrt{\frac{\sigma}{\mu}}\left(-1+A\,c\,\sqrt{Ha}+A\,Ha^{\frac{1}{2}+\beta}\alpha+Ha\,Coth[Ha]\right)}$$

The above expression simplifies to

In[129]:=

$$\text{diffons} := -\left(\left(A\left(1-e^{-\frac{\sqrt{Ha}}{A}}-Ha\left(c+Ha^{\beta}\alpha\right)+\left(-1+e^{-\frac{\sqrt{Ha}}{A}}\right)Ha\,Coth[Ha]\right)\right)\Big/\left(\sqrt{Ha}\left(-1+A\,c\,\sqrt{Ha}+A\,Ha^{\frac{1}{2}+\beta}\alpha+Ha\,Coth[Ha]+\left(1+\frac{A\left(-1+e^{-\frac{\sqrt{Ha}}{A}}\right)}{\sqrt{Ha}}\right)(-1+Ha\,Coth[Ha])\right)\right)\right)$$

The following contour plot shows the relative difference (%) between the off-diagonal Onsager coefficients: $2*10^2\left|\frac{K_{EH}+K_{HE}}{K_{EH}-K_{HE}}\right|$

In[173]:=

```
α := 1
β := -1
c := -0.956
```

In[130]:=

```
ContourPlot[2*10^2 Abs[diffons], {A, 0, 1}, {Ha, 5, 20}, PlotRange → All,
 ColorFunction → "DarkBands", Contours → {0.1, 0.5, 1, 2}, FrameLabel → {"A", "Ha"},
 BaseStyle → {FontSize → 15, Black}, PlotLegends → Automatic, ContourLabels → True]
```

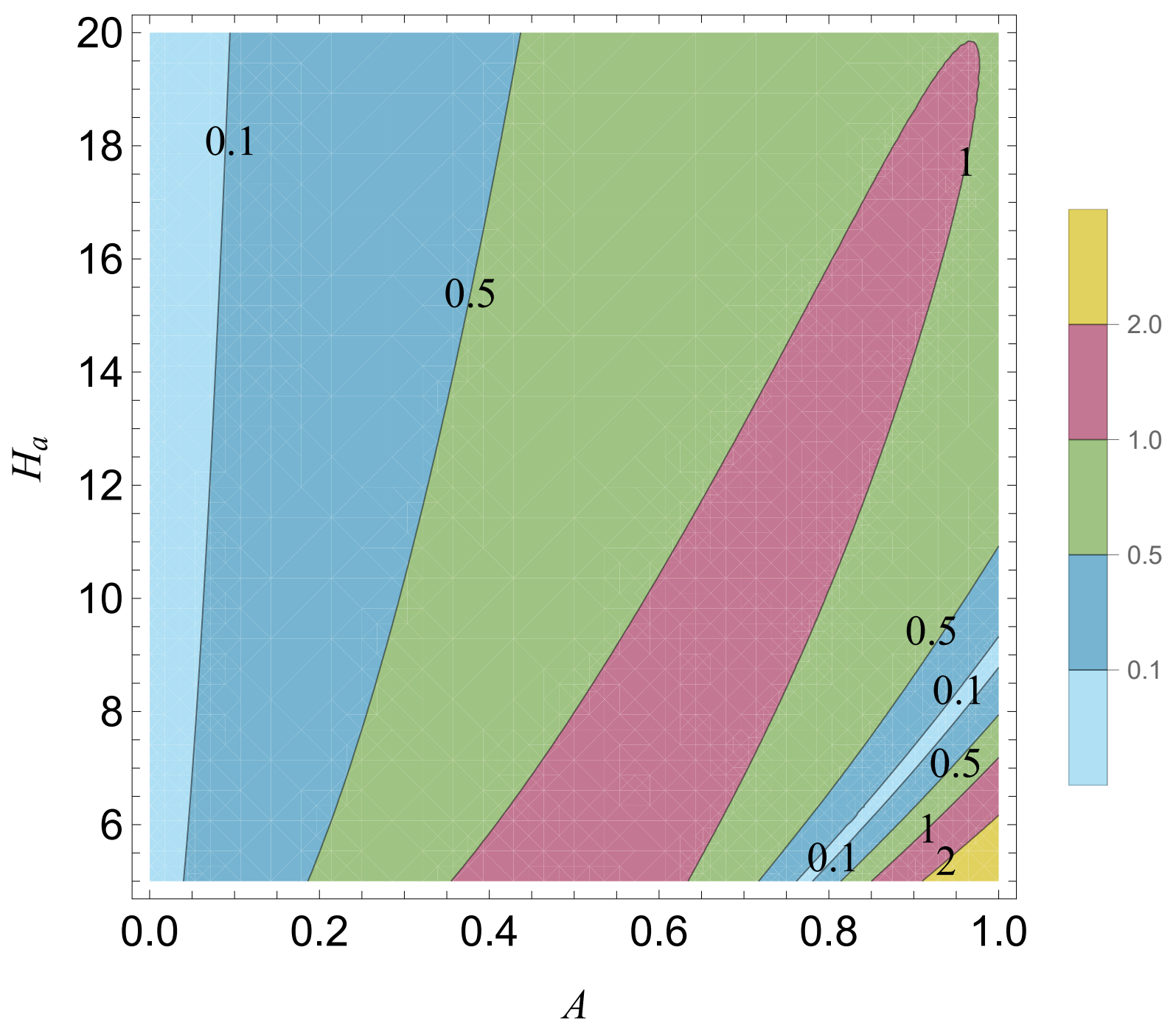

When A<0.8 and Ha>5, the O-C reciprocity is satisfied with less than 2% difference.

# B Efficiency Calculations

## B1. MHD Generator

Consider pressure-driven flow in a conduit (**Fig. 1**). The electrodes are connected to an external load with resistance $R_x$ ($\Omega$ $m$) per conduit's unit length. The internal resistance $R_{in} = \frac{1}{\sigma A}$ ($\Omega$ $m$). The conductor's motion in the magnetic field induces an electromotive force (*EMF*) $\Psi$>0.

The generator efficiency is the ratio of the electrical power produced and the mechanical power consumed. $\psi = -R x * D H$ .

$$\epsilon_g = \frac{-\Delta H \Psi}{Q G} = \frac{\Delta H^2 R x}{Q G}$$

We rewrite DH and Q in terms of Rx instead of $\psi$. We rename the expressions for the flow rate and current Qg and DHg;

In[131]:=

```
Qg[Ha_, A_, G_, Rx_] := (KHH[A, Ha] - (KHE[A, Ha] * KEH[A, Ha] * Rx) / (1 + KEE[A, Ha] * Rx)) (G)
```

In[132]:=

```
DHg[Ha_, A_, G_, Rx_] := (KEH[A, Ha] * G) / (1 + KEE[A, Ha] * Rx)
```

The generator efficiency is:

In[30]:= (DHg[Ha, A, G, Rx] ^2 Rx) / (Qg[Ha, A, G, Rx] *G) // Simplify

Out[30]=

$$\left(\mathrm{Rx}\,\sigma\left(A\left(-0.956+\frac{1}{\mathrm{Ha}}\right)+\sqrt{\mathrm{Ha}}\,\mathrm{Coth}[\mathrm{Ha}]\right)^2\left(-1+\frac{A\,(1.-0.956\,\mathrm{Ha})}{\sqrt{\mathrm{Ha}}}+\mathrm{Ha}\,\mathrm{Coth}[\mathrm{Ha}]\right)^2\right)\Big/$$
$$\left(\left(1+\frac{A\left(-1+e^{-\frac{\sqrt{\mathrm{Ha}}}{A}}\right)}{\sqrt{\mathrm{Ha}}}\right)\left(\frac{A\,(1.-0.956\,\mathrm{Ha}+1.\sqrt{\mathrm{Ha}}\,\mathrm{Rx}\,\sigma)}{\mathrm{Ha}}+1.\sqrt{\mathrm{Ha}}\,\mathrm{Coth}[\mathrm{Ha}]\right)^2(-1+\mathrm{Ha}\,\mathrm{Coth}[\mathrm{Ha}])\right.$$
$$\left.\left(\left(-0.956+\frac{1}{\mathrm{Ha}}\right)\sqrt{\mathrm{Ha}}+\frac{\mathrm{Ha}\,\mathrm{Coth}[\mathrm{Ha}]}{A}+\frac{\mathrm{Rx}\sqrt{\mu}\sqrt{\sigma}\sqrt{\frac{\sigma}{\mu}}\left(-1+\frac{A\,(1.-0.956\,\mathrm{Ha})}{\sqrt{\mathrm{Ha}}}+\mathrm{Ha}\,\mathrm{Coth}[\mathrm{Ha}]\right)}{1+\frac{A\,\mathrm{Rx}\,\sigma}{\sqrt{\mathrm{Ha}}\left(A\left(-0.956+\frac{1}{\mathrm{Ha}}\right)+\sqrt{\mathrm{Ha}}\,\mathrm{Coth}[\mathrm{Ha}]\right)}}\right)\right)$$

Introducing $R_i$ and the resistance ratio RxRi = Rex/Rin, the expression for efficiency becomes:

In[31]:=
$$-\left(\left(e^{\frac{\sqrt{\mathrm{Ha}}}{A}}\sqrt{\mathrm{Ha}}\,\mathrm{RxRi}\left(-1+A\,c\sqrt{\mathrm{Ha}}+A\,\mathrm{Ha}^{\frac{1}{2}+\beta}\alpha+\mathrm{Ha}\,\mathrm{Coth}[\mathrm{Ha}]\right)^2\right)\Big/\left(\left(-A+e^{\frac{\sqrt{\mathrm{Ha}}}{A}}\left(A-\sqrt{\mathrm{Ha}}\right)\right)(1+\mathrm{RxRi})\right.\right.$$
$$\left.\left.(-1+\mathrm{Ha}\,\mathrm{Coth}[\mathrm{Ha}])\left(A\left(c\sqrt{\mathrm{Ha}}+\mathrm{Ha}^{\frac{1}{2}+\beta}\alpha\right)+\mathrm{RxRi}+\mathrm{Ha}\,\mathrm{Coth}[\mathrm{Ha}]\right)\right)\right)\ //\ \mathrm{FullSimplify}$$

Out[31]=

$$-\frac{e^{\frac{\sqrt{\mathrm{Ha}}}{A}}\sqrt{\mathrm{Ha}}\,\mathrm{RxRi}\left(-1+\frac{A\,(1.-0.956\,\mathrm{Ha})}{\sqrt{\mathrm{Ha}}}+\mathrm{Ha}\,\mathrm{Coth}[\mathrm{Ha}]\right)^2}{\left(-A+e^{\frac{\sqrt{\mathrm{Ha}}}{A}}\left(A-\sqrt{\mathrm{Ha}}\right)\right)(1+\mathrm{RxRi})\,(-1+\mathrm{Ha}\,\mathrm{Coth}[\mathrm{Ha}])\left(\frac{A\,(1.-0.956\,\mathrm{Ha})}{\sqrt{\mathrm{Ha}}}+\mathrm{RxRi}+\mathrm{Ha}\,\mathrm{Coth}[\mathrm{Ha}]\right)}$$

The above is equivalent to:

In[133]:=

$$\mathrm{Effg}[\mathrm{Ha_},A_,\mathrm{RxRi_}]:=\left(\mathrm{RxRi}\left(-1+A\,c\sqrt{\mathrm{Ha}}+A\,\mathrm{Ha}^{\frac{1}{2}+\beta}\alpha+\mathrm{Ha}\,\mathrm{Coth}[\mathrm{Ha}]\right)^2\right)\Big/$$
$$\left(\left(1+\frac{A\left(-1+e^{-\frac{\sqrt{\mathrm{Ha}}}{A}}\right)}{\sqrt{\mathrm{Ha}}}\right)(-1+\mathrm{Ha}\,\mathrm{Coth}[\mathrm{Ha}])\left(A\left(c\sqrt{\mathrm{Ha}}+\mathrm{Ha}^{\frac{1}{2}+\beta}\alpha\right)+\mathrm{RxRi}+\mathrm{Ha}\,\mathrm{Coth}[\mathrm{Ha}]\right)(1+\mathrm{RxRi})\right)$$

Below, the MHD generator’s efficiency is depicted as a function of the resistances ratio RxRi for various Ha and A:

```
In[38]:= Plot[{Effg[5, 0.05, RxRi], Effg[5, 0.5, RxRi], Effg[5, 1, RxRi], Effg[20, 0.05, RxRi],
   Effg[20, 0.5, RxRi], Effg[20, 1, RxRi], Effg[500, 0.05, RxRi], Effg[500, 0.5, RxRi],
   Effg[500, 1, RxRi]}, {RxRi, 0.1, 20}, PlotStyle → {Thick, Dashed, Dotted},
  AxesLabel → {"Rex/Rin", "ϵg"}, PlotLegends → Automatic, PlotTheme → "Classic"]
```

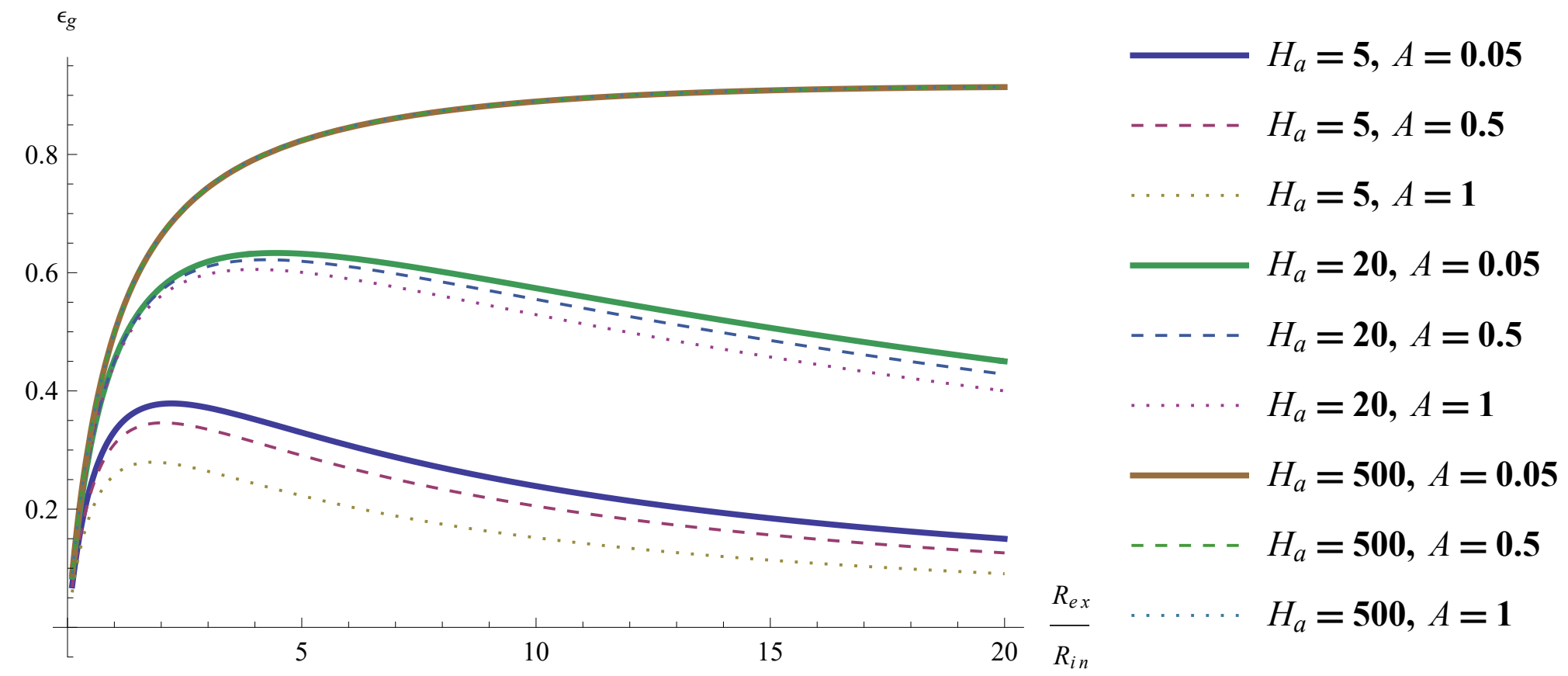

In[43]:=
```
Limit[Effg[Ha, A, RxRi], {Ha → Infinity}]
```

Out[43]=

$$\frac{\text{RxRi}}{1. + \text{RxRi}} \quad \text{if } \text{RxRi} \in \mathbb{R} \,\&\&\, A > 0$$

Next, we find the resistance ratio that maximizes MHD generator's efficiency:

In[44]:=
```
Solve[D[Effg[Ha, A, RxRi], RxRi] == 0, RxRi]
```

Out[44]=

$$\left\{\left\{\text{RxRi} \to -1.\sqrt{\frac{1.\,A}{\sqrt{\text{Ha}}} - 0.956\,A\sqrt{\text{Ha}} + 1.\,\text{Ha}\,\text{Coth}[\text{Ha}]}\right\}, \left\{\text{RxRi} \to \sqrt{\frac{1.\,A}{\sqrt{\text{Ha}}} - 0.956\,A\sqrt{\text{Ha}} + 1.\,\text{Ha}\,\text{Coth}[\text{Ha}]}\right\}\right\}$$

In[45]:=

$$\text{Effg}\left[\text{Ha}, A, \sqrt{A\,c\,\sqrt{\text{Ha}} + A\,\text{Ha}^{\frac{1}{2}+\beta}\,\alpha + \text{Ha}\,\text{Coth}[\text{Ha}]}\right] \text{ // FullSimplify}$$

Out[45]=

$$\left(\left(-1 + \frac{A\,(1. - 0.956\,\text{Ha})}{\sqrt{\text{Ha}}} + \text{Ha}\,\text{Coth}[\text{Ha}]\right)^2 \sqrt{\frac{A\,(1. - 0.956\,\text{Ha})}{\sqrt{\text{Ha}}} + \text{Ha}\,\text{Coth}[\text{Ha}]}\right) \Big/ \left(\left(1 + \frac{A\left(-1 + e^{-\frac{\sqrt{\text{Ha}}}{A}}\right)}{\sqrt{\text{Ha}}}\right)(-1 + \text{Ha}\,\text{Coth}[\text{Ha}])\left(1 + \sqrt{\frac{A\,(1. - 0.956\,\text{Ha})}{\sqrt{\text{Ha}}} + \text{Ha}\,\text{Coth}[\text{Ha}]}\right)\left(\frac{A\,(1. - 0.956\,\text{Ha})}{\sqrt{\text{Ha}}} + \text{Ha}\,\text{Coth}[\text{Ha}] + \sqrt{\frac{A\,(1. - 0.956\,\text{Ha})}{\sqrt{\text{Ha}}} + \text{Ha}\,\text{Coth}[\text{Ha}]}\right)\right)$$

Further simplification leads to the max generator efficiency Effgmax:

In[52]:=

$$\text{Effgmax}[\text{Ha_}, A_] := \left(1 + A\,c\,\sqrt{\text{Ha}} + A\,\text{Ha}^{\frac{1}{2}+\beta}\,\alpha + \text{Ha}\,\text{Coth}[\text{Ha}] - 2\sqrt{A\,c\,\sqrt{\text{Ha}} + A\,\text{Ha}^{\frac{1}{2}+\beta}\,\alpha + \text{Ha}\,\text{Coth}[\text{Ha}]}\right) \Big/ \left(\left(1 + \frac{A\left(-1 + e^{-\frac{\sqrt{\text{Ha}}}{A}}\right)}{\sqrt{\text{Ha}}}\right)(-1 + \text{Ha}\,\text{Coth}[\text{Ha}])\right)$$

The MHD generator's  efficiency at optimal resistance ratio as a function of the Hartman number for various aspect ratios A.

In[56]:=
```
Plot[{Effgmax[Ha, 0.05], Effgmax[Ha, 0.1], Effgmax[Ha, 0.5], Effgmax[Ha, 0.9]},
 {Ha, 2, 50}, PlotRange → {0, 1}, AxesLabel → {"Ha", "ϵg,m"},
 LabelStyle → {18, GrayLevel[0]}, Ticks → {{0, 15, 30, 45}, {0.2, 0.4, 0.6, 0.8}},
 PlotLegends → {"A=0.05", "A=0.1", "A=0.5", "A=0.9"},
 PlotStyle → "Rainbow", PlotTheme → "ThickLines"]
```

Out[56]=

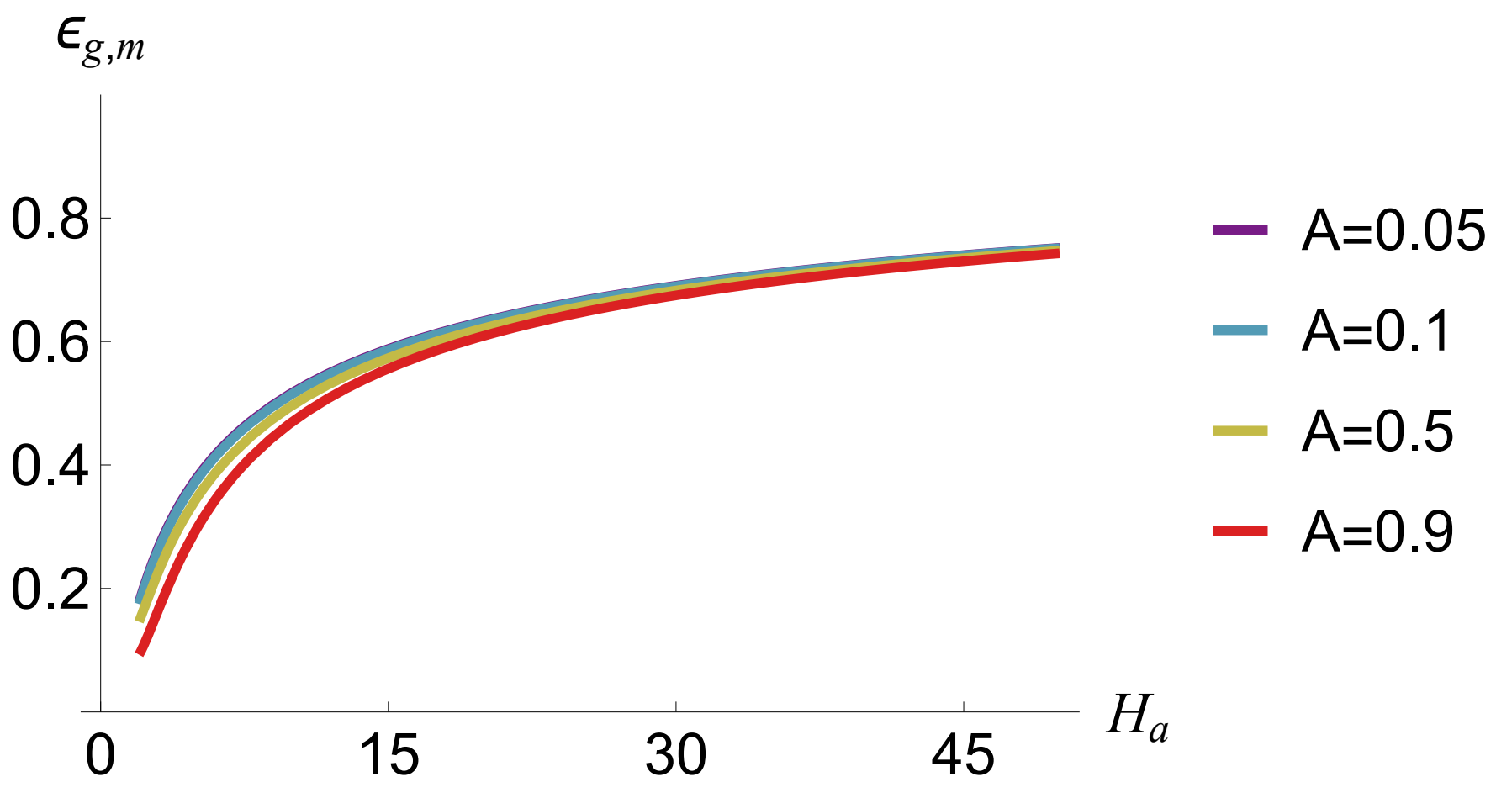

### B2. MHD Pump

Next, consider the case when current is injected into the electrodes. As the result of the Lorentz body force, the liquid metal is propelled against an adverse pressure gradient $G<0$. The maximum flow rate is obtained in the absence of adverse pressure (G=0)

Efficiency $= \frac{-Q\,G}{\Delta H\,\psi}$. The stagnation pressure gradient $G_s = \frac{-K_{H\,E}\,\Psi}{K_{H\,H}}$ balances the Lorentz forces generated by the injected current. We rewrite the efficiency expression in terns of $\frac{G}{G_s}$:

In[141]:=
```
dh := kee * Ψ + keh (G)
q := khe * Ψ + khh * G
Gs := - khe Ψ / khh
```

In[60]:=
```
(-q G) / (dh Ψ) // FullSimplify
```

Out[60]=

$$-\frac{G\,(G\,\text{khh} + \text{khe}\,\Psi)}{\Psi\,(G\,\text{keh} + \text{kee}\,\Psi)}$$

In[145]:=
```
G / Gs (1 - G / Gs) (khh (kee) / khe^2 + G / Gs (1 - (keh + khe) / (khe)))^(-1) // FullSimplify
 (*Verify that we get the same pump efficiency expression using the G/Gs term*)
```

Out[145]=

$$-\frac{G\,(G\,\text{khh} + \text{khe}\,\Psi)}{\Psi\,(G\,\text{keh} + \text{kee}\,\Psi)}$$

This pump efficiency:

$$\epsilon_p = \frac{G}{G_s}\left(1 - \frac{G}{G_s}\right)\left(b_1 + \frac{G}{G_s}(1 - b_2)\right)^{-1}$$

In the above, b1 & b2 are:

```
In[62]:= b1[A_, Ha_] := KEE[A, Ha] * KHH[A, Ha] / (KHE[A, Ha] ^2) // FullSimplify
In[79]:= FullSimplify[b2[A, Ha] := (KEH[A, Ha] + KHE[A, Ha]) / KHE[A, Ha]]
In[80]:= FullSimplify[b2[A, Ha], Assumptions → μ > 0] // Chop
Out[80]=
```

$$\frac{A\,(1. - 1.\,\mathrm{Ha}\,\mathrm{Coth}[\mathrm{Ha}]) + e^{\frac{\sqrt{\mathrm{Ha}}}{A}}\,(-0.956\,A\,\mathrm{Ha} + 1.\,A\,\mathrm{Ha}\,\mathrm{Coth}[\mathrm{Ha}])}{\left(-1.\,A + e^{\frac{\sqrt{\mathrm{Ha}}}{A}}\,(1.\,A - 1.\,\sqrt{\mathrm{Ha}})\right)(-1. + \mathrm{Ha}\,\mathrm{Coth}[\mathrm{Ha}])}$$

In[81]:= $$\mathrm{b2}[A_, \mathrm{Ha}] := 1 - \frac{-1 + A\,c\,\sqrt{\mathrm{Ha}} + A\,\mathrm{Ha}^{\frac{1}{2}+\beta}\,\alpha + \mathrm{Ha}\,\mathrm{Coth}[\mathrm{Ha}]}{\left(1 + \frac{A\left(-1+e^{-\frac{\sqrt{\mathrm{Ha}}}{A}}\right)}{\sqrt{\mathrm{Ha}}}\right)(-1 + \mathrm{Ha}\,\mathrm{Coth}[\mathrm{Ha}])}$$

```
In[82]:= effpggs[Ha_, A_, GGS_, Ψ_] := GGS (1 - GGS) (b1[A, Ha] + GGS (1 - b2[A, Ha])) ^ (-1)
In[73]:= Plot[{effpggs[5, 0.05, GGS, 0.3], effpggs[5, 0.5, GGS, 0.3], effpggs[5, 1, GGS, 0.3],
   effpggs[20, 0.05, GGS, 0.3], effpggs[20, 0.5, GGS, 0.3], effpggs[20, 1, GGS, 0.3],
   effpggs[500, 0.05, GGS, 0.3], effpggs[500, 0.5, GGS, 0.3], effpggs[500, 1, GGS, 0.3]}, {GGS, 0, 1},
  PlotStyle → {Thick, Dashed, Dotted}, PlotTheme → "ThickLines", AxesLabel → {"G/G_stall", "ϵ_m"},
  LabelStyle → Directive[Medium], PlotRange → All, PlotLegends → Automatic]
```

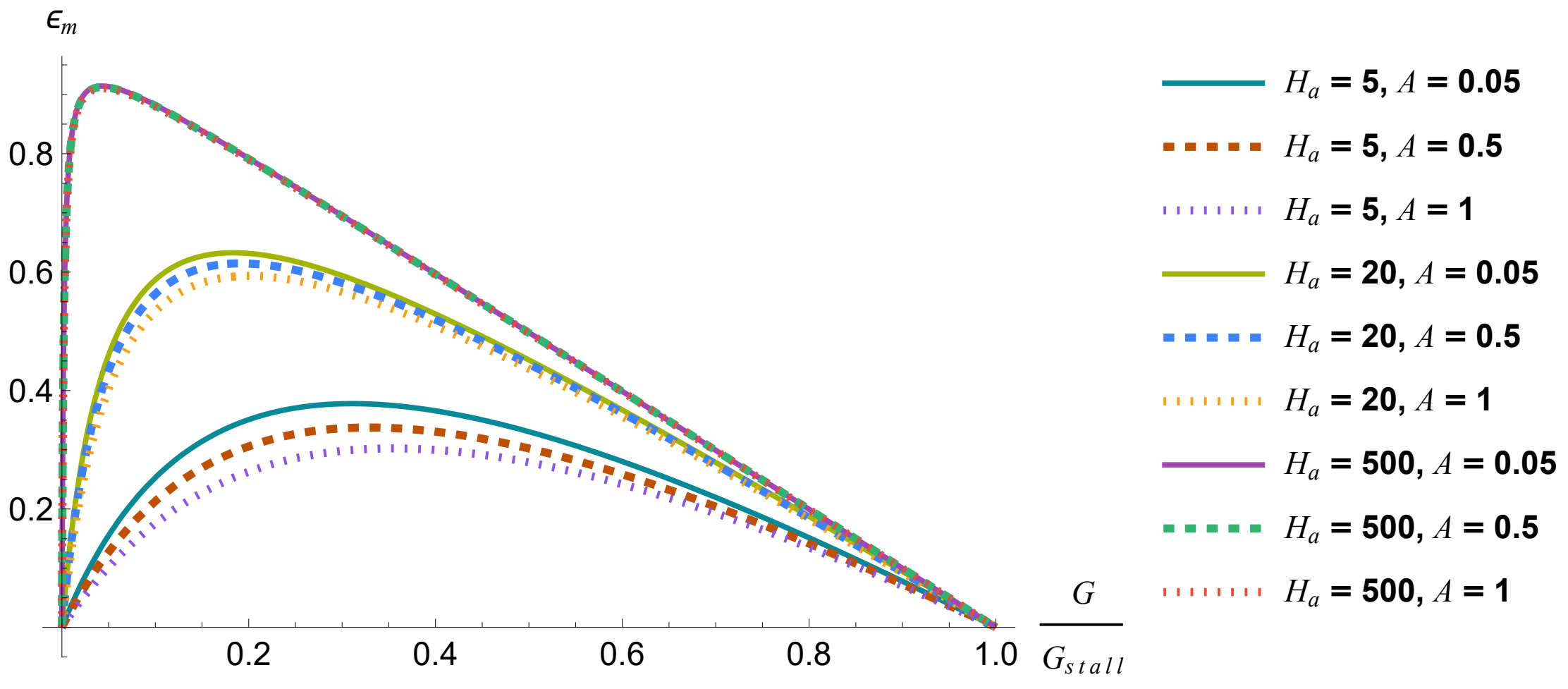

Below, we calculate the pressure gradient ratio that optimizes the pump/actuator efficiency as well as the optimal efficiency.

```
In[93]:= Solve[D[effpggs[Ha, A, GGS, Ψ], GGS] == 0, GGS, Assumptions → μ > 0] // FullSimplify
```

$$\left\{\left\{\mathrm{GGS} \to \frac{1}{1 - \sqrt{\left(A\,c\,\sqrt{\mathrm{Ha}} + A\,\mathrm{Ha}^{\frac{1}{2}+\beta}\,\alpha + \mathrm{Ha}\,\mathrm{Coth}[\mathrm{Ha}]\right)}}\right\}, \left\{\mathrm{GGS} \to \frac{1}{1 + \sqrt{\left(A\,c\,\sqrt{\mathrm{Ha}} + A\,\mathrm{Ha}^{\frac{1}{2}+\beta}\,\alpha + \mathrm{Ha}\,\mathrm{Coth}[\mathrm{Ha}]\right)}}\right\}\right\}$$

Select the positive GGS and plug into the efficiency expression:

In[94]:= $$\mathrm{GGSmax} := \frac{1}{1 + \sqrt{A\,c\,\sqrt{\mathrm{Ha}} + A\,\mathrm{Ha}^{\frac{1}{2}+\beta}\,\alpha + \mathrm{Ha}\,\mathrm{Coth}[\mathrm{Ha}]}}$$

In[95]:=
```
effpggs[Ha, A, GGSmax, Ψ] // FullSimplify
```

Out[95]=

$$\left(e^{-\frac{\sqrt{Ha}}{A}}\left(A+e^{\frac{\sqrt{Ha}}{A}}\left(-A+\sqrt{Ha}\right)\right)\sqrt{\sigma}\,(-1+Ha\,Coth[Ha])\sqrt{A\,c\,\sqrt{Ha}+A\,Ha^{\frac{1}{2}+\beta}\alpha+Ha\,Coth[Ha]}\right)\Big/\left(\sqrt{Ha}\left(1+\sqrt{A\,c\,\sqrt{Ha}+A\,Ha^{\frac{1}{2}+\beta}\alpha+Ha\,Coth[Ha]}\right)\left(\left(-1+A\,c\,\sqrt{Ha}+A\,Ha^{\frac{1}{2}+\beta}\alpha\right)\sqrt{\mu}\sqrt{\frac{\sigma}{\mu}}+Ha\sqrt{\mu}\sqrt{\frac{\sigma}{\mu}}\,Coth[Ha]+\sqrt{\sigma}\left(1+\sqrt{A\,c\,\sqrt{Ha}+A\,Ha^{\frac{1}{2}+\beta}\alpha+Ha\,Coth[Ha]}\right)\right)\right)$$

In[168]:=

**Effpmax[Ha_, A_] :=**

$$\left(\left(1+\frac{A\left(-1+e^{-\frac{\sqrt{Ha}}{A}}\right)}{\sqrt{Ha}}\right)(-1+Ha\,Coth[Ha])\right)\Big/\left(1+\sqrt{A\,c\,\sqrt{Ha}+A\,Ha^{\frac{1}{2}+\beta}\alpha+Ha\,Coth[Ha]}\right)^2$$

This is equivalent to:

In[49]:=

$$\left(1+\frac{A\left(-1+e^{-\frac{\sqrt{Ha}}{A}}\right)}{\sqrt{Ha}}\right)(-1+Ha\,Coth[Ha])\ GGSmax\text{^}2\ //\ FullSimplify$$

Out[49]=

$$\frac{\left(1+\frac{A\left(-1+e^{-\frac{\sqrt{Ha}}{A}}\right)}{\sqrt{Ha}}\right)(-1+Ha\,Coth[Ha])}{\left(1+\sqrt{\frac{A\,(1.-0.956\,Ha)}{\sqrt{Ha}}+Ha\,Coth[Ha]}\right)^2}$$

In[104]:=
```
Plot[{Effpmax[Ha, 0.05], Effpmax[Ha, 0.1], Effpmax[Ha, 0.5], Effpmax[Ha, 0.6]},
 {Ha, 1, 100}, PlotRange → {0, 1}, PlotStyle → "Rainbow", PlotTheme → "ThickLines",
 LabelStyle → {15, GrayLevel[0]}, PlotLegends → {"A=0.05", "A-0.1", "A=0.5", "A=0.6"}]
```

Out[104]=

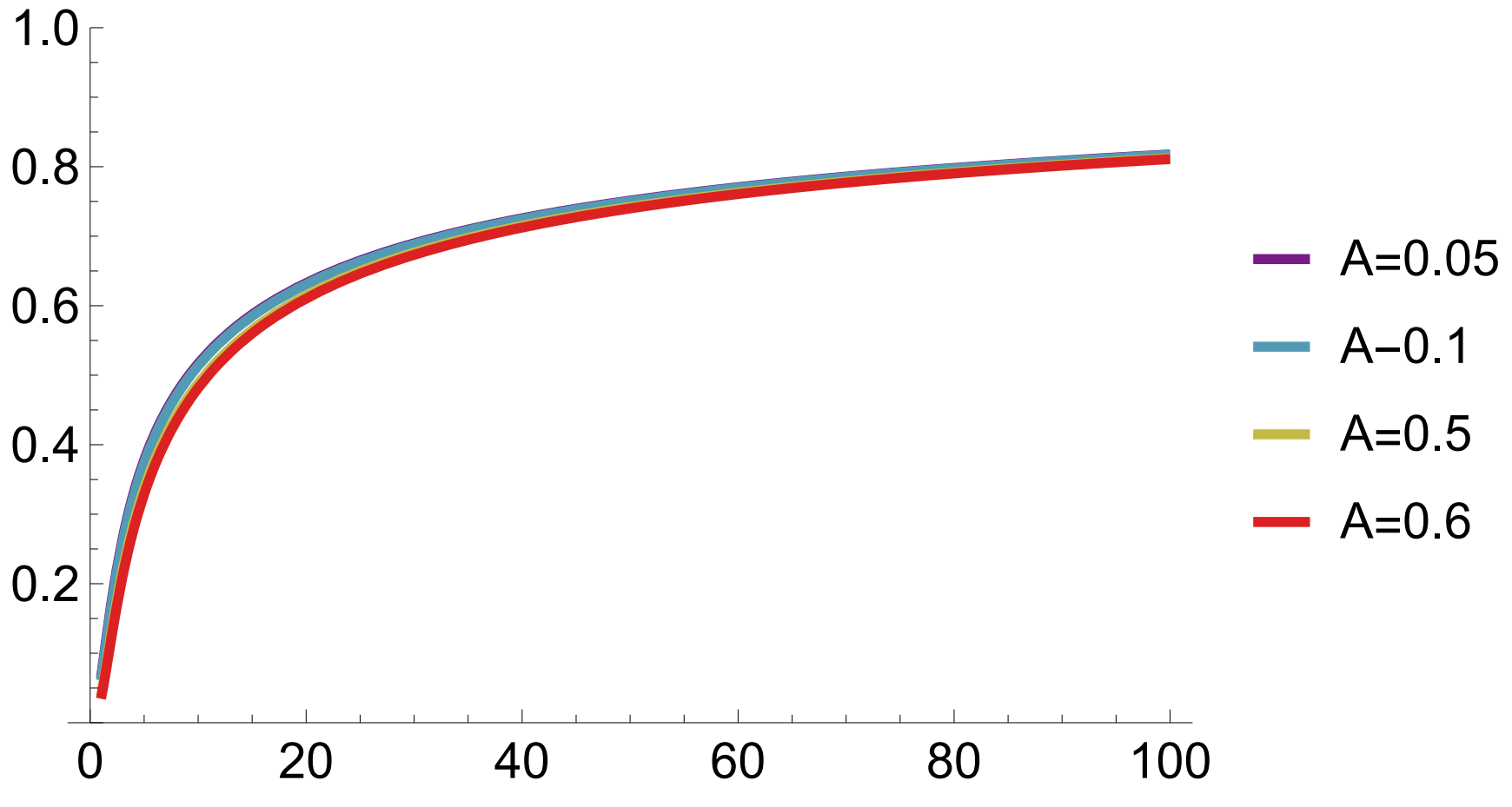

### B2.3 Max Pump Efficiency at a given pressure head

We determine the electrode potential difference that maximizes the efficiency.

.

```
Simplify[(-Q[Ha, A, G, Ψ] *G) / (Ψ *DH[Ha, A, G, Ψ]), Assumptions → σ > 0 && μ > 0 && A > 0 && Ha > 0]
```

Out[ ]=

$$-\left(\left(2 a^2 e^{-\frac{\sqrt{Ha}}{A}} G \left(A\left(-1+e^{\frac{\sqrt{Ha}}{A}}\right) - e^{\frac{\sqrt{Ha}}{A}}\sqrt{Ha}\right)\left(2 a^2 G + A\, Ha \sqrt{\mu\sigma}\,\Psi\right)(-1 + Ha\, \mathrm{Coth}[Ha])\right) \Big/ \left(A\, Ha^{3/2} \mu \Psi \left(-2 a^2 G \left(\sqrt{\frac{\sigma}{\mu}} - A\left(c + Ha^{\beta}\alpha\right)\sqrt{\frac{Ha\,\sigma}{\mu}}\right) - A\, Ha\, \sigma \Psi + 2 a^2 G\, Ha \sqrt{\frac{\sigma}{\mu}}\, \mathrm{Coth}[Ha]\right)\right)\right)$$

In[102]:=

```
Pumpef[Ha_, A_, G_, Ψ_] :=
```

$$-\left(\left(2 a^2 e^{-\frac{\sqrt{Ha}}{A}} G \left(A\left(-1+e^{\frac{\sqrt{Ha}}{A}}\right) - e^{\frac{\sqrt{Ha}}{A}}\sqrt{Ha}\right)\left(2 a^2 G + A\, Ha \sqrt{\mu\sigma}\,\Psi\right)(-1 + Ha\, \mathrm{Coth}[Ha])\right) \Big/ \left(A\, Ha^{3/2} \mu \Psi \left(-2 a^2 G \left(\sqrt{\frac{\sigma}{\mu}} - A\left(c + Ha^{\beta}\alpha\right)\sqrt{\frac{Ha\,\sigma}{\mu}}\right) - A\, Ha\, \sigma \Psi + 2 a^2 G\, Ha \sqrt{\frac{\sigma}{\mu}}\, \mathrm{Coth}[Ha]\right)\right)\right)$$

```
Ψs := -khh G / khe // FullSimplify
```

$$\Psi_s = \frac{-2 a^2 G}{A H a \sqrt{\sigma\mu}}$$

The pump efficiency can be simplified and written in terms of $\Psi/\Psi_s$ , written as ΨΨs:

In[167]:=

```
Pumpeff[Ha_, A_, ΨΨs_] :=
```

$$\left(e^{-\frac{\sqrt{Ha}}{A}}\left(A\left(-1+e^{\frac{\sqrt{Ha}}{A}}\right) - e^{\frac{\sqrt{Ha}}{A}}\sqrt{Ha}\right)(-1 + \Psi\Psi s)(-1 + Ha\,\mathrm{Coth}[Ha])\right) \Big/ \left(Ha^{1/2}\, \Psi\Psi s \left(1 - A\left(c + Ha^{\beta}\alpha\right)\sqrt{Ha} - \Psi\Psi s - Ha\, \mathrm{Coth}[Ha]\right)\right)$$

In[137]:=

```
Plot[{Pumpeff[5, 0.05, ΨΨs], Pumpeff[5, 0.2, ΨΨs], Pumpeff[5, 0.5, ΨΨs],
  Pumpeff[20, 0.05, ΨΨs], Pumpeff[20, 0.2, ΨΨs], Pumpeff[20, 0.5, ΨΨs],
  Pumpeff[500, 0.05, ΨΨs], Pumpeff[500, 0.2, ΨΨs], Pumpeff[500, 0.5, ΨΨs]}, {ΨΨs, 1, 20},
 PlotStyle → {Thick, Dashed, Dotted}, PlotTheme → "Classic", AxesLabel → {"Ψ/Ψ_stall", "ϵ_m"},
 LabelStyle → Directive[ Medium], PlotRange → All, PlotLegends → Automatic]
```

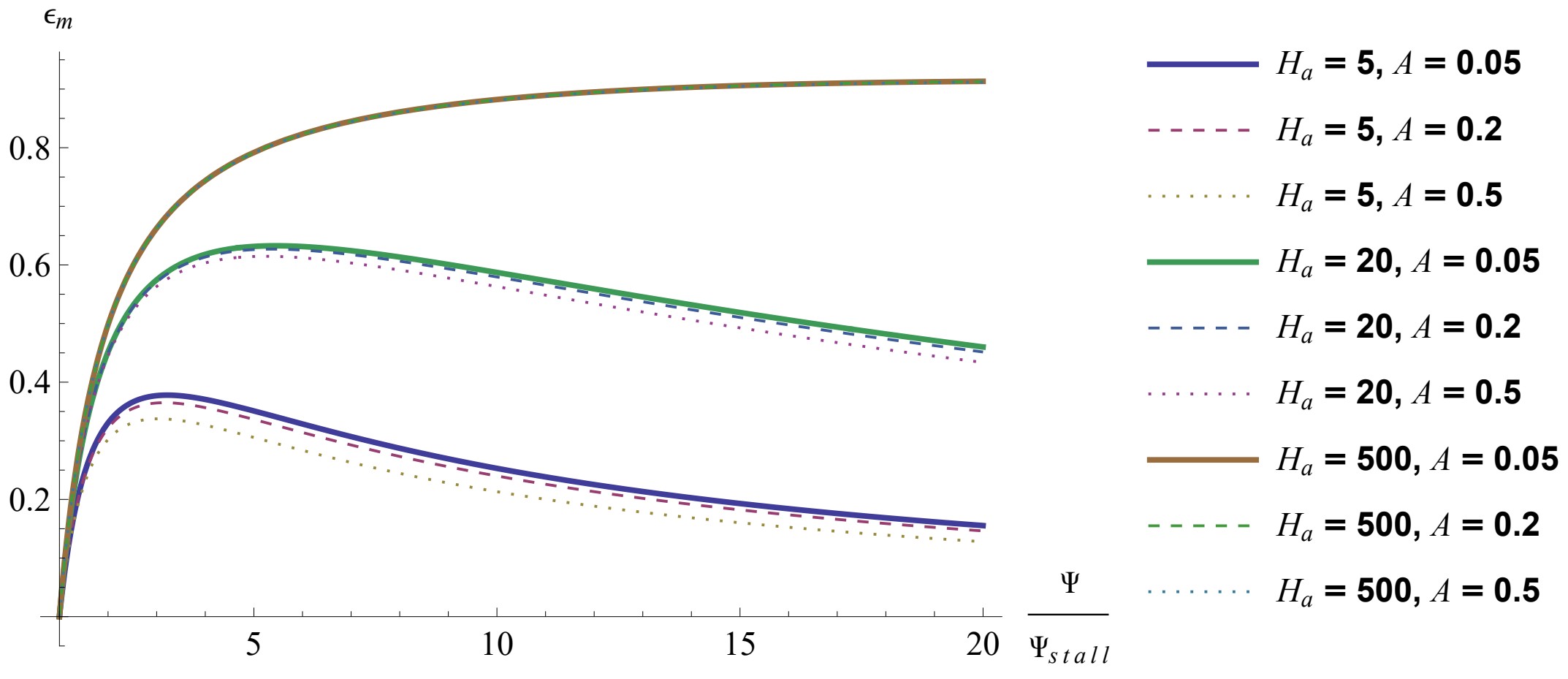

$\Psi_s$ is the electrode potential difference under stall. We determine the optimal $\Psi/\Psi_s$:

This is just the inverse of G/GS max, since G/Gs= $(\Psi/\Psi_s)^{-1}$

In[146]:=

```
G / Gs
```

Out[146]=

$$-\frac{\text{G khh}}{\text{khe }\Psi}$$

In[149]:=

```
(Ψ / Ψs) ^ (-1)
```

Out[149]=

$$-\frac{\text{G khh}}{\text{khe }\Psi}$$

In[ ]:= 

```
Solve[D[Pumpeff[Ha, A, ΨΨs, α, β, c], ΨΨs] == 0, ΨΨs] // FullSimplify
```

Out[ ]=

$$\left\{\left\{\Psi\Psi s \to 1 - \sqrt{\text{A c } \sqrt{\text{Ha}} + \text{A Ha}^{\frac{1}{2}+\beta}\,\alpha + \text{Ha Coth[Ha]}}\right\}, \left\{\Psi\Psi s \to 1 + \sqrt{\text{A c } \sqrt{\text{Ha}} + \text{A Ha}^{\frac{1}{2}+\beta}\,\alpha + \text{Ha Coth[Ha]}}\right\}\right\}$$

Select the positive quantity:

In[155]:=

```
ΨΨsm := 1 + Sqrt[A c Sqrt[Ha] + A Ha^(1/2+β) α + Ha Coth[Ha]];
Pumpeff[Ha, A, ΨΨsm] // FullSimplify
```

Out[156]=

$$-\left(\left(\left(1.\,e^{-\frac{\sqrt{\text{Ha}}}{\text{A}}}\left(-1.\,\text{A} + e^{\frac{\sqrt{\text{Ha}}}{\text{A}}}\left(1.\,\text{A} - 1.\,\sqrt{\text{Ha}}\right)\right)(-1. + \text{Ha Coth[Ha]})\sqrt{\frac{\text{A }(1. - 0.956\text{ Ha})}{\sqrt{\text{Ha}}} + \text{Ha Coth[Ha]}}\right)\Big/\right.\right.$$
$$\left(\left(1. + \sqrt{\frac{\text{A }(1. - 0.956\text{ Ha})}{\sqrt{\text{Ha}}} + \text{Ha Coth[Ha]}}\right)\right.$$
$$\left.\left.\left.\left(\text{A }(1. - 0.956\text{ Ha}) + \sqrt{\text{Ha}}\left(\text{Ha Coth[Ha]} + \sqrt{\frac{\text{A }(1. - 0.956\text{ Ha})}{\sqrt{\text{Ha}}} + \text{Ha Coth[Ha]}}\right)\right)\right)\right)\right)$$

This simplifies to:

In[166]:=

$$\texttt{Pumpeffm[Ha_, A_]} := \left(e^{-\frac{\sqrt{Ha}}{A}} \left(A + e^{\frac{\sqrt{Ha}}{A}} \left(-A + \sqrt{Ha}\right)\right) (-1 + Ha\,\mathrm{Coth}[Ha])\right) \Big/ \left(\sqrt{Ha} \left(1 + A\,c\,\sqrt{Ha} + A\,Ha^{\frac{1}{2}+\beta}\,\alpha + Ha\,\mathrm{Coth}[Ha] + 2\sqrt{A\,c\,\sqrt{Ha} + A\,Ha^{\frac{1}{2}+\beta}\,\alpha + Ha\,\mathrm{Coth}[Ha]}\right)\right)$$

In[163]:=

```
Plot[{Pumpeffm[Ha, 0.05], Pumpeffm[Ha, 0.1], Pumpeffm[Ha, 0.5], Pumpeffm[Ha, 0.6]},
 {Ha, 1, 100}, PlotRange → {0, 1}, PlotStyle → "Rainbow",
 PlotTheme → "ThickLines", PlotLegends → {"A=0.05", "A=0.1", "A=0.5", "A=0.6"},
 AxesLabel → {"Ha", "ϵgΨ,m"}, LabelStyle → Directive[Medium]]
```

Out[163]=

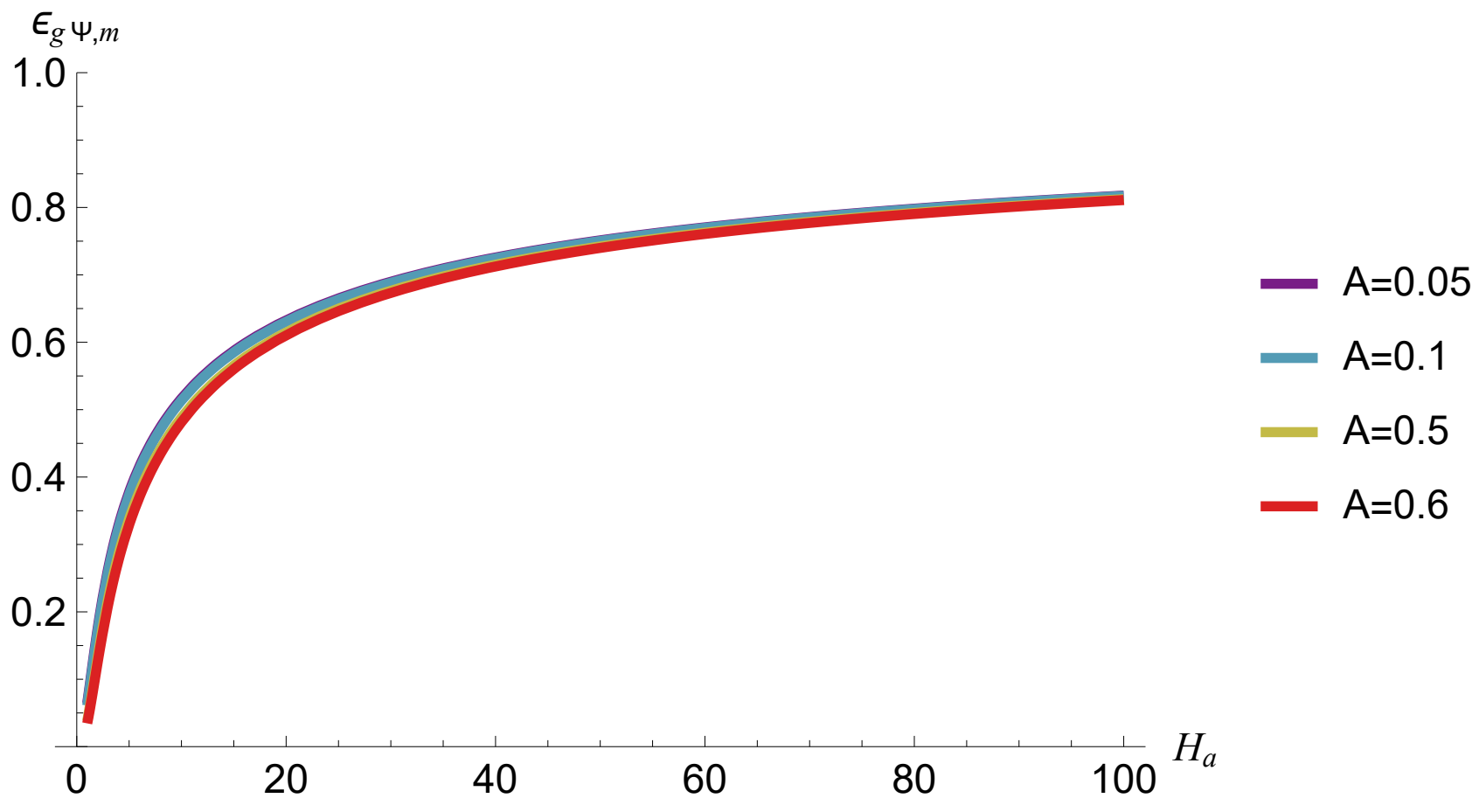

Comparing the obtained max efficiencies between the pump maximized via G/Gs or $\Psi/\Psi_s$:

In[170]:=

```
Clear[α, β, c]
```

In[172]:=

```
Pumpeffm[Ha, A] - Effpmax[Ha, A] // FullSimplify
```

Out[172]=

```
0
```

## B3. MHD sensor calculations

Measurement of electrical signals, such as the potential across the electrodes $\Psi$ and/or the current $\Delta H$ inform on the flow rate ($Q$) and pressure drop in the conduit $G$. Consider, for example, open circuit conditions ($\Delta H$ =0). From the current equation we get $G$ to be:

In[176]:=

```
Gsensor := -KEE[A, Ha] * ψ / KEH[A, Ha]
```

This simplifies to:

In[177]:=

$$\texttt{Gsensor} := \frac{A\,Ha\,\sqrt{\mu\,\sigma}\;\psi}{2\,a^2 \left(-1 + A\,c\,\sqrt{Ha} + A\,Ha^{\frac{1}{2}+\beta}\,\alpha + Ha\,\mathrm{Coth}[Ha]\right)}$$

In[ ]:= `Q[Ha, A, Gsensor, ψ] // FullSimplify`

Out[ ]=

$$\left(2 a^2 e^{-\frac{\sqrt{Ha}}{A}} \left(A + e^{\frac{\sqrt{Ha}}{A}} \left(-A + \sqrt{Ha}\right)\right) \psi \left(-1 + Ha\,Coth[Ha]\right) \left(\left(-1 + A\,c\,\sqrt{Ha} + A\,Ha^{\frac{1}{2}+\beta}\alpha\right)\sqrt{\mu}\sqrt{\sigma} + \sqrt{\mu\,\sigma} + Ha\sqrt{\mu}\sqrt{\sigma}\,Coth[Ha]\right)\right) \Big/ \left(Ha^2 \mu \left(A\left(c + Ha^\beta \alpha\right) + \sqrt{Ha}\,Coth[Ha]\right)\left(-1 + A\,c\,\sqrt{Ha} + A\,Ha^{\frac{1}{2}+\beta}\alpha + Ha\,Coth[Ha]\right)\right)$$

This is equivalent to the following expression:

In[ ]:= `KHE[A, Ha] (1 + (-KHH[A, Ha] * KEE[A, Ha]) / (KHE[A, Ha] * KEH[A, Ha])) ψ // FullSimplify (*verify that answer is the same*)`

Out[ ]=

$$\frac{2 a^2 \left(1 + \frac{A\left(-1 + e^{-\frac{\sqrt{Ha}}{A}}\right)}{\sqrt{Ha}}\right)\sqrt{\sigma}\,\psi\left(-1 + Ha\,Coth[Ha]\right)\left(1 + \frac{\sqrt{\mu}\sqrt{\frac{\sigma}{\mu}}}{\sqrt{\sigma}\left(-1 + A\,c\,\sqrt{Ha} + A\,Ha^{\frac{1}{2}+\beta}\alpha + Ha\,Coth[Ha]\right)}\right)}{Ha^{3/2}\sqrt{\mu}\left(A\left(c + Ha^\beta\alpha\right) + \sqrt{Ha}\,Coth[Ha]\right)}$$

This simplifies to:

In[178]:=

$$Qs[Ha_, A_, \alpha_, \beta_, c_] := \frac{2 a^2 \left(1 + \frac{A\left(-1 + e^{-\frac{\sqrt{Ha}}{A}}\right)}{\sqrt{Ha}}\right)\sqrt{\sigma}\,\psi\left(-1 + Ha\,Coth[Ha]\right)}{Ha\sqrt{\mu}\left(-1 + A\,c\,\sqrt{Ha} + A\,Ha^{\frac{1}{2}+\beta}\alpha + Ha\,Coth[Ha]\right)}$$

In[179]:=

`Limit[Qs[Ha, A, 1, -1, -0.956], Ha → Infinity]`

Out[179]=

$$0.\ \text{if}\ \left(a^2 \mid \frac{1}{\sqrt{\mu}} \mid \sqrt{\sigma} \mid \psi\right) \in \mathbb{R}\ \&\&\ A > 0$$

## B3.1 MHD Sensor Applying the Onsager Casimir Reciprocity

If we use Onsager Relations instead: KEH=-KHE

In[ ]:= `KHE[A, Ha] (1 + (KHH[A, Ha] * KEE[A, Ha]) / KHE[A, Ha] ^2) ψ // FullSimplify`

Out[ ]=

$$\left(2 a^2 e^{-\frac{\sqrt{Ha}}{A}}\sqrt{\sigma}\,\psi\left(A\left(-1 + Ha\,Coth[Ha]\right) + e^{\frac{\sqrt{Ha}}{A}}\left(A + \left(-A\,Ha + Ha^{3/2}\right)Coth[Ha]\right)\right)\right) \Big/ \left(Ha^2\sqrt{\mu}\left(A\left(c + Ha^\beta\alpha\right) + \sqrt{Ha}\,Coth[Ha]\right)\right)$$

In[180]:=

$$ORQS[Ha_, A_, \alpha_, \beta_, c_] := \left(2 a^2 e^{-\frac{\sqrt{Ha}}{A}}\sqrt{\sigma}\,\psi\left(A\left(-1 + Ha\,Coth[Ha]\right) + e^{\frac{\sqrt{Ha}}{A}}\left(A + \left(-A\,Ha + Ha^{3/2}\right)Coth[Ha]\right)\right)\right) \Big/ \left(Ha^2\sqrt{\mu}\left(A\left(c + Ha^\beta\alpha\right) + \sqrt{Ha}\,Coth[Ha]\right)\right)$$

In[ ]:= `Limit[ORQS[Ha, A, 1, -1, -0.956], Ha → Infinity]`

Out[ ]=

$$0.\ \text{if}\ \left(a^2 \mid \frac{1}{\sqrt{\mu}} \mid \sqrt{\sigma} \mid \psi\right) \in \mathbb{R}\ \&\&\ A > 0$$

### B4. MHD Brake

Here we define a brake where the fluid's motion is due to a pressure gradient is in the positive **z**-direction. We apply a negative potential such that a force in the negative **z**-direction is created to slow down the fluid. As long as $Q \geq 0$ , we remain in the break region. Once the flow rate becomes negative, we are in a pump region once again.

$H = K_{E\,E}\,\psi + K_{E\,H}\,G$

$Q = K_{H\,E}\,\Psi + K_{H\,H}\,G$

When $Q \geq 0$, we have:

$$\psi \leq -\frac{K_{H\,H}}{K_{H\,E}}\,G$$

In[181]:=

```
ψbmax[A_, Ha_, G_] := -KHH[A, Ha] / KHE[A, Ha] * G // FullSimplify
```

We can plug that into the H expression:

In[182]:=

```
Hbrake[A_, Ha_, G_] := (KEH[A, Ha] - KEE[A, Ha] * KHH[A, Ha] / KHE[A, Ha]) G // FullSimplify
```

From Physics we know that at Q=0 the forces cancel out such that:

$$\frac{H\,B_0}{2\,a} + G = 0$$

$$\frac{H\,b\,r\,a\,k\,e}{G} = \frac{-2\,a}{B_0}$$

In[ ]:=

```
Simplify[Hbrake[A, Ha, G] / G, Assumptions → μ > 0]
```

Out[ ]=

$$-\frac{2\,a^2\,\sqrt{\frac{\sigma}{\mu}}}{\text{Ha}}$$

Plugging in $H_a = a\,B_0\,\sqrt{\frac{\sigma}{\mu}}$ , this becomes: $\frac{-2\,a}{B_0}$ as desired